\documentclass[11pt,a4paper]{article}
\pdfoutput=1
\usepackage{epsfig,amsfonts,amssymb}

\usepackage{comment}

\usepackage[normalem]{ulem}
\usepackage[utf8]{inputenc}
\usepackage[T1]{fontenc}
\usepackage{lmodern, textcomp}
\usepackage[english]{babel}
\usepackage{csquotes,cancel}

\usepackage[left=1in,right=1in,top=.8in,bottom=.8in]{geometry}

\usepackage{eurosym, xspace}
\usepackage{graphicx, subfig, float}
\usepackage{verbatim}

\usepackage{amsmath, amsfonts, amssymb,upgreek}
\usepackage{cancel}
\usepackage{mathtools}
\usepackage{slashed}

\usepackage[usenames, dvipsnames]{xcolor}
\usepackage[amsmath, hyperref, framed]{}
\usepackage{hhline}
\usepackage{bm}
\usepackage{array, caption}
\usepackage{array,multirow}

\newcolumntype{C}[1]{>{\centering\arraybackslash}m{#1}}

\usepackage[pdftex, breaklinks=true, linktocpage,
	colorlinks=true, urlcolor=blue, linkcolor=blue, citecolor=red
]{hyperref}

\usepackage{cite}
\usepackage{tikz-feynman}
\usepackage{tikz, pgf}
\usetikzlibrary{shapes.misc}
\usetikzlibrary{snakes}
\usetikzlibrary{decorations.pathmorphing}
\usetikzlibrary{shapes}
\usetikzlibrary{fit}

\makeatletter
\newsavebox{\@brx}
\newcommand{\llangle}[1][]{\savebox{\@brx}{\(\m@th{#1\langle}\)}%
  \mathopen{\copy\@brx\kern-0.5\wd\@brx\usebox{\@brx}}}
\newcommand{\rrangle}[1][]{\savebox{\@brx}{\(\m@th{#1\rangle}\)}%
  \mathclose{\copy\@brx\kern-0.5\wd\@brx\usebox{\@brx}}}
\makeatother

\def\ZZZ{{\hbox{ Z\kern-1.6mm Z}}}
\def\RRR{{\hbox{ R\kern-2.4mm R}}}
\def\CCC{{\hbox{ C\kern-2.0mm C}}}
\def\zzz{{\hbox{z\kern-1mm z}}}

\newcommand{\qeq}{{\hbox{=\kern-2.3mm ? \kern.5mm }}}
\renewcommand{\qeq}{=}

\newcommand{\be}{\begin{eqnarray}}
\newcommand{\ee}{\end{eqnarray}}

\newcommand{\vp}{\varphi}

\newcommand{\ben}{\begin{eqnarray}\displaystyle}
\newcommand{\een}{\end{eqnarray}}

\newcommand{\p}{\partial}

\def\one{{\hbox{ 1\kern-.8mm l}}}
\def\zero{{\hbox{ 0\kern-1.5mm 0}}}

\newcommand{\bea}[1]{\begin{eqnarray}\label{#1} }
\newcommand{\eea}{\end{eqnarray}}

\usepackage{bm}

\newcommand\non{\nonumber}

\newcommand\f{\frac}

\def\figone{

\def\JPicScale{0.8}
\ifx\JPicScale\undefined\def\JPicScale{1}\fi
\unitlength \JPicScale mm
\begin{picture}(135,80)(0,0)
\linethickness{0.3mm}
\multiput(40,80)(0.12,-0.18){167}{\line(0,-1){0.18}}
\linethickness{0.3mm}
\multiput(30,70)(0.18,-0.12){167}{\line(1,0){0.18}}
\linethickness{0.3mm}
\put(30,50){\line(1,0){30}}
\linethickness{0.3mm}
\multiput(30,30)(0.18,0.12){167}{\line(1,0){0.18}}
\linethickness{0.3mm}
\multiput(40,20)(0.12,0.18){167}{\line(0,1){0.18}}
\linethickness{0.3mm}
\put(60,50){\line(1,0){40}}
\linethickness{0.3mm}
\multiput(100,50)(0.12,0.18){167}{\line(0,1){0.18}}
\linethickness{0.3mm}
\multiput(100,50)(0.18,0.12){167}{\line(1,0){0.18}}
\linethickness{0.3mm}
\put(100,50){\line(1,0){30}}
\linethickness{0.3mm}
\multiput(100,50)(0.18,-0.12){167}{\line(1,0){0.18}}
\linethickness{0.3mm}
\multiput(100,50)(0.12,-0.18){167}{\line(0,-1){0.18}}

\put(30,80){\makebox(0,0)[cc]{$\zeta Q_B A_1^c$}}

\put(25,70){\makebox(0,0)[cc]{$A_2^c$}}

\put(25,50){\makebox(0,0)[cc]{$A_n^c$}}

\put(20,30){\makebox(0,0)[cc]{$A_1^o$}}

\put(35,20){\makebox(0,0)[cc]{$A_p^o$}}

\put(35,60){\makebox(0,0)[cc]{$\vdots$}}

\put(40,30){\makebox(0,0)[cc]{$\vdots$}}

\put(125,80){\makebox(0,0)[cc]{$B_1^c$}}

\put(135,70){\makebox(0,0)[cc]{$B_2^c$}}

\put(135,50){\makebox(0,0)[cc]{$B_m^c$}}

\put(135,30){\makebox(0,0)[cc]{$B_1^o$}}

\put(120,15){\makebox(0,0)[cc]{$B_q^o$}}

\put(115,35){\makebox(0,0)[cc]{$\vdots$}}

\put(120,57){\makebox(0,0)[cc]{$\vdots$}}

\put(66,47){\makebox(0,0)[cc]{$\psi_r^c\vp_r$}}

\put(93,47){\makebox(0,0)[cc]{$\psi_s^c\vp_s$}}

\end{picture}

}

\def\figtwo{

\def\JPicScale{0.8}
\ifx\JPicScale\undefined\def\JPicScale{1}\fi
\unitlength \JPicScale mm
\begin{picture}(135,80)(0,0)
\linethickness{0.3mm}
\multiput(40,80)(0.12,-0.18){167}{\line(0,-1){0.18}}
\linethickness{0.3mm}
\multiput(30,70)(0.18,-0.12){167}{\line(1,0){0.18}}
\linethickness{0.3mm}
\put(30,50){\line(1,0){30}}
\linethickness{0.3mm}
\multiput(30,30)(0.18,0.12){167}{\line(1,0){0.18}}
\linethickness{0.3mm}
\multiput(40,20)(0.12,0.18){167}{\line(0,1){0.18}}
\linethickness{0.3mm}
\put(60,50){\line(1,0){40}}

\put(30,80){\makebox(0,0)[cc]{$\zeta Q_B A_1^c$}}

\put(25,70){\makebox(0,0)[cc]{$A_2^c$}}

\put(25,50){\makebox(0,0)[cc]{$B_m^c$}}

\put(20,30){\makebox(0,0)[cc]{$A_1^o$}}

\put(35,20){\makebox(0,0)[cc]{$B_q^o$}}

\put(35,60){\makebox(0,0)[cc]{$\vdots$}}

\put(40,30){\makebox(0,0)[cc]{$\vdots$}}

\put(66,47){\makebox(0,0)[cc]{$\psi_r^c\vp_r$}}

\put(93,45){\makebox(0,0)[cc]{$\tilde\psi_s^c\tilde\vp^s$}}

\put(100,50){\makebox(0,0)[cc]{$\times$}}

\end{picture}

}

\definecolor{armygreen}{rgb}{0.29, 0.33, 0.13}

\allowdisplaybreaks

\begin{document}

\baselineskip 24pt

\begin{center}
{\Large \bf Momentum space CFT correlators of non-conserved spinning operators  }

\end{center}

\vskip .6cm
\medskip

\vspace*{4.0ex}

\baselineskip=18pt

\begin{center}

{\large 
\rm Raffaele Marotta$^a$, Kostas Skenderis$^b$ and Mritunjay Verma$^b$ }

\end{center}

\vspace*{4.0ex}

\centerline{\it \small $^a$Istituto Nazionale di Fisica Nucleare (INFN), Sezione di Napoli, }
\centerline{ \it \small Complesso Universitario di Monte S. Angelo ed. 6, via Cintia, 80126, Napoli, Italy.}
\centerline{ \it \small $^b$ Mathematical Sciences and STAG Research Centre, University of Southampton,}
\centerline{\it \small  Highfield, Southampton SO17 1BJ, UK}

\vspace*{1.0ex}
\centerline{\small E-mail:  k.skenderis@soton.ac.uk, raffaele.marotta@na.infn.it, 
m.verma@soton.ac.uk }

\vspace*{5.0ex}

\centerline{\bf Abstract} \bigskip

We analyse the 3-point CFT correlators involving non-conserved spinning operators in momentum space.
We derive a general expression for the conformal Ward identities defining
 the 3-point functions involving two generic spin $s$ non-conserved operators and a spin 1 conserved current. We give explicit expressions for the 3-point function when the two non-conserved operators have spins 1 and 2 and generic conformal dimensions. We also systematically analyse the divergences appearing in these 3-point functions when the conformal dimensions of the two non-conserved operators coincide.
\vfill

\vfill \eject

\baselineskip18pt

\tableofcontents

\section{Introduction } 
\label{s1}
Higher spin interactions are conjectured to be the necessary ingredients to construct UV complete theories of gravity. There are several  observations supporting such a conjecture. High derivative corrections to three-point couplings break causality unless an infinite tower of massive particles of spin larger than 2 is added to the spectrum (CEMZ causality problem) \cite{Camanho:2014apa}. String theories contain an infinite number of massive higher spin excitations and their amplitudes are finite 
 and free from the CEMZ causality problem \cite{DAppollonio:2015fly}.
While the formulation of gauge-invariant higher spin interactions in Minkowski space is problematic (for reviews, see e.g., \cite{Bekaert:2010hw,Ponomarev:2022vjb}), such interacting theories in AdS are conjectured to be dual  to conformal field theories (CFT) containing higher-spin operators. If the higher-spin field in the bulk is a gauge field the dual current is conserved, otherwise it is a non-conserved tensorial operator. Thus, holography may be used to study higher spin interactions in AdS using correlation functions of the dual operators.   

The signature of higher spin interactions in the early universe has also gained interest in the context of holographic cosmology and the cosmological bootstrap. As the isometries of de Sitter match that of the (Euclidean) conformal group in one dimension less, inflationary correlators are constrained by conformal symmetry, see   \cite{Maldacena:2002vr, McFadden:2009fg, Antoniadis:2011ib, Maldacena:2011nz, Bzowski:2011ab, Creminelli:2012ed, Kehagias:2012pd, Bzowski:2012ih, Mata:2012bx, McFadden:2013ria, Ghosh:2014kba, Kundu:2014gxa, Arkani-Hamed:2015bza, Arkani-Hamed:2018kmz, Sleight:2019mgd, Sleight:2019hfp, Baumann:2020dch, Bonifacio:2022vwa} for a sample of works in this direction. 

Motivated by these problems, there has been significant interest in the study of conformal correlators in the momentum space \cite{Coriano:2013jba, Bzowski:2013sza, Bzowski:2015pba, Bzowski:2015yxv, Bzowski:2017poo, Isono:2018rrb, Bzowski:2018fql, Coriano:2018bsy, Albayrak:2018tam, Farrow:2018yni, Isono:2019ihz, Coriano:2019sth, Albayrak:2019yve, Isono:2019wex, Bautista:2019qxj, Gillioz:2019lgs, Bzowski:2019kwd, Baumann:2019oyu, Coriano:2019nkw, Lipstein:2019mpu, Albayrak:2020isk,  Bzowski:2020lip, Jain:2020rmw, Bzowski:2020kfw, Jain:2020puw, Jain:2021wyn, Jain:2021qcl,  Jain:2021vrv, Gillioz:2021sce, Jain:2021whr, Jain:2022ujj, Bzowski:2022rlz, Jain:2022ajd, Caloro:2022zuy}. The analysis in momentum space is quite involved since it requires finding the solution of sets of coupled second and first order differential equations coming from the momentum space conformal Ward identities. It allows, however, to discuss in generality the short distance singularities of correlators, their renormalization and the computation of associated conformal anomalies. Moreover,  for many applications, such as in cosmology,  momentum space is the natural language.

Most of the work on tensorial corretors in momentum space has focussed on CFT correlators involving conserved currents and/or the energy momentum tensor. Even though there are few results about the momentum space correlators of non-conserved operators beyond scalars \cite{Isono:2018rrb,Isono:2019wex
,Isono:2019ihz,Arkani-Hamed:2015bza,Jain:2020rmw}, a systematic study of correlators of such operators is missing. Our goal in this paper will be to initiate such systematic study by considering the 3-point correlators when we have more than one spinning non-conserved operators. In particular, we shall consider the 3-point function of two generic spin $s$ operators and one conserved current.  One reason for this choice is that this 3-point function encodes holographically the leading non-linear coupling of gauge fields with higher-spin fields. We shall find the consequences of the conformal Ward identities when the spin $s$ operators are symmetric and traceless and explicitly solve the equations in terms of triple-K integrals \cite{Bzowski:2013sza} for 3-point function when the non-conserved operators have spins 1 and 2.\footnote{Due to additional constraints between the tensor structures in $d=3$ \cite{Costa:2011mg, Bzowski:2013sza, Bzowski:2017poo}, our expressions for 3-point functions are valid for $d\ge 4$ dimensions when $s\ge 2$.} The restriction to these values is mainly for practical reasons. While the method for solving the equations is qualitatively the same in all cases,  the higher the value of $s$ the higher the number of equations that one needs to solve.

In this paper, we shall consider generic conformal dimensions of the non-conserved operators, with only restriction that the conformal dimensions are above the unitarity bound. With generic dimensions, 
the 2 and 3-point functions considered here are finite\footnote{When the conformal dimensions of the two non-conserved operators are the same, some of the triple K integrals in the expression of 3-point function diverge, but we find that the divergences cancel each other leaving a finite expression.}. We shall analyse in detail the case of both non-conserved operators having spin $1$. In general, for integer conformal dimensions, additional divergences may be present which require the addition of  counterterms. Such cases may be analysed following \cite{Bzowski:2013sza, Bzowski:2015pba, Bzowski:2017poo, Bzowski:2018fql}  and leave such analysis to future work.

The rest of this paper is organised as follows. In section \ref{sec22pt}, we review the 2-point function of an arbitrary spin $s$ operator in the momentum space, exemplifying our general procedure. In section \ref{sec2:identification}, we write the general decomposition of a momentum space 3-point function involving a conserved current and two non-conserved higher spinning operators in terms of form factors, and work out the action of the Ward identities on this 3-point function. This gives rise to coupled differential equations which need to be solved to determine the 3-point functions. In sections \ref{sec3:s0}, \ref{sec4:s1} and \ref{sec5:s2}, we  solve these equations and completely determine the 3-point functions for the cases when the non-conserved operators have spins $0,1$ and $2$, respectively. We end with some discussion in section \ref{s4:dissc}. The appendices contain reviews of some results which are needed for the derivations as well as some details of the main derivations. We use an index-free approach, where spacetime indices are contracted with complex null auxiliary vectors, and we review this formalism in appendix \ref{linearind3456}. Appendix \ref{appenA} contains a summary of the conformal Ward identities both in position and momentum space and appendix \ref{appena.3} reviews the decomposition of correlators involving conserved operators into transverse and non-transverse parts. In appendix \ref{apps1} we review triple-K integrals and in appendix \ref{app:res_3pt} we review relevant position-space results for 3-point functions. In appendix \ref{dregu4} we review the regularisation of divergent triple-K integrals. Appendices 
\ref{appen:s2eq} and \ref{s=2secondary} contain the equations satisfied by the form factors in the case $s=2$ and the solutions of the secondary Ward identities, respectively.

We shall be working in the flat $d$ dimensional space with Euclidean signature and use $\delta^{\mu\nu}$ to raise and lower the indices.

\section{Two-point functions}
\label{sec22pt}
In this section, we consider the 2-point function of a generic spinning operators $\mathcal{O}^{\mu_1\cdots\mu_s}$ which is symmetric and traceless. We start with the 2-point function to introduce our conventions and the method we shall be following for the 3-point function. For earlier works on CFT 2-point function of spinning operators in momentum space using different approaches, see \cite{Arkani-Hamed:2015bza, Doberev}. To avoid dealing with the indices, it is convenient to introduce auxiliary complex vectors $\epsilon^\mu$ and work with the operators in the index free notation \cite{Todorov,Costa:2011mg}
\be
\epsilon\cdot\mathcal{O}_{s} &\equiv& \epsilon^{\mu_1}\cdots \epsilon^{\mu_s}\mathcal{O}_{\mu_1\cdots\mu_s}\, . \label{indexfree}
\ee
Since we are interested in symmetric traceless operators, we shall also need to impose the additional relation $\epsilon^\mu \epsilon_\mu=0$. This is possible since $\epsilon^\mu$ are complex and lie on \textquotedblleft complex null cone\textquotedblright\   in Euclidean space \cite{Todorov}. We review this formalism in appendix \ref{linearind3456}.

Using momentum conservation, the two point function of the spin $s$ operator becomes a function of a single momentum and hence it can be written as 
\be
\mathcal{A}_{2}&\equiv&\llangle[\Bigl]  \epsilon_1\cdot\mathcal{O}_{s}({\bf p})\epsilon_2\cdot\mathcal{O}_{s}(-{\bf p})   \rrangle[\Bigl] \label{2pt5}
\ee
Our notation is explained in appendix \ref{appenA}. The double  bracket denotes that we have stripped off the momentum conserving delta function. 
The conformal dimension of the operator $\mathcal{O}_i$ is $\Delta_i$. Conformal invariance sets $\Delta_1=\Delta_2$ as we shall see below. One may try to Fourier transform the position space expression of the 2-point function to obtain the above 2-point function in the momentum space. While this is possible for generic conformal dimensions, there are short-distance singularities  when the dimensions take specific values and renormalization is required for the 2-point function to have a Fourier transform.  We will work directly in momentum space where such issues can be dealt straightforwardly, and we shall determine the momentum-space 2-point functions by solving the Ward identities in momentum space. 

The strategy is to write down the most general expression that the correlator can take based on Lorentz invariance alone and then impose the conformal Ward identities. The tensor structure 
of the correlator is encoded in monomials constructed out of scalar products of the auxiliary vectors $\epsilon^\mu$ with themselves and the momenta $p^\mu$. The first step is to list the set of such monomials. Then we multiply each such monomial by a form factor, a scalar function of Lorentz invariants constructed out of momenta only and we sum over all of them. The conformal Ward identities would then amount for equations for the scalar functions.

Let us exemplify this for the 2-point function. There are three scalar products one can construct from $p^\mu, \epsilon_1^\mu, \epsilon_2^\mu$: 
\be
w=\epsilon_1\cdot \epsilon_2\quad,\qquad \zeta=\epsilon_1\cdot p \quad,\quad \xi=\epsilon_2 \cdot p \label{der98j}
\ee
The correlator should have $s$ powers of both auxiliary vectors $\epsilon_1$ and $\epsilon_2$ and the list of all possible
such monomials is $w^{s-n}\xi^n\zeta^n$, with $n=0, \ldots s$. Thus the most general expression for the correlator is 
\be
\mathcal{A}_2&=&\sum_{n=0}^s \f{1}{n!} A_n(p) w^{s-n}\xi^n\zeta^n \label{a2p}
\ee
where $A_n(p)$ are the form factors (which are scalar functions of the momentum magnitude, $p=\sqrt{p^\mu p_\mu}$).
 These will be determined using the conformal Ward identities. This requires knowing the action of the generators of the conformal group on \eqref{a2p}. To determine the constraints coming from the Ward identities, we start by noting that for a function of the form $f= f(p, w, \xi,  \zeta )$,
we have 
\be
\f{\p f}{\p p^\mu}&=& \f{p^\mu}{p}  \f{\p f}{\p p} + \epsilon_{1\mu} \f{\p f}{\p \zeta} + \epsilon_{2\mu} \f{\p f}{\p \xi}\label{chain2pt}
\ee
where $p$ denotes the magnitude of momenta $p^\mu$, namely $p=\sqrt{p^\mu p_\mu}$.

Using \eqref{chain2pt}, the dilatation Ward identity, given in equation \eqref{dilwardw}, can be expressed as 
\be
\left(p^\mu\f{\p}{\p p^\mu }+d-\Delta_1-\Delta_2\right)\mathcal{A}_2&=&\left(p\f{\p}{\p p }+\xi\f{\p}{\p \xi}+\zeta\f{\p}{\p \zeta}+d-\Delta_1-\Delta_2\right)\mathcal{A}_2=0\label{wer3}
\ee
Now, substituting \eqref{a2p} in above equation gives 
\be
 \Bigl(p\f{d}{d p }+2n+d-\Delta_1-\Delta_2\Bigl)A_{n}(p) =0
\ee
This has the general solution 
\be
A_n = a_n\; p^{\Delta_1+\Delta_2-d-2n}\label{2.8uy}
\ee
where $a_n$ are some constants which may depend upon $d$ and $\Delta_i$.

Next, we need to determine the action of the special conformal Ward identity. Using \eqref{chain2pt}, the special conformal Ward identities given in equations \eqref{a29} and \eqref{a210} can be expressed as
\be
b\cdot (\mathcal{K}_s+\mathcal{K}_\epsilon)&=&  2(b\cdot p)\biggl[\f{1}{2}K-\f{\zeta}{p}\f{\p^2}{\p\zeta\p p}\biggl]+2(b\cdot\epsilon_1)\biggl[ (\Delta_1-1)\f{\p}{\p\zeta}+w\f{\p^2}{\p w\p\zeta}+\f{\xi}{p}\f{\p^2}{\p w\p p}\biggl]\non\\[.3cm]
&&+2(b\cdot\epsilon_2)\biggl[ (\Delta_1-d)\f{\p}{\p\xi}-w\f{\p^2}{\p w\p\xi}-\f{\zeta}{p}\f{\p^2}{\p w\p p}-\zeta\f{\p^2}{\p\xi\p\zeta}-p\f{\p^2}{\p p\p\xi}-\xi\f{\p^2}{\p\xi^2} \biggl] \label{bkske}
\ee
where, we have contracted with an arbitrary vector $b^\mu$ and defined 
\be
K = 2\Bigl(\Delta_1-\f{d+1}{2}\Bigl)\f{1}{p}\f{d}{d p} -\f{d^2}{d p^2}\label{defKKK}
\ee
The coefficients proportional to $(b\cdot p) , (b\cdot\epsilon_1)$ and $(b\cdot\epsilon_2)$ in \eqref{bkske} are independent as discussed in appendix \ref{linearind3456}. Setting these independent terms to zero, the differential equations for the form factors are obtained to be 
\be
0&=&\biggl[\f{1}{2}K-\f{\zeta}{p}\f{\p^2}{\p\zeta\p p}\biggl]\mathcal{A}_2\label{2eqa}\\[.2cm]
0&=&\biggl[ (\Delta_1-1)\f{\p}{\p\zeta}+w\f{\p^2}{\p w\p\zeta}+\f{\xi}{p}\f{\p^2}{\p w\p p}\biggl]\mathcal{A}_2\label{2eqb}\\[.2cm]
0&=&\biggl[ (\Delta_1-d)\f{\p}{\p\xi}-w\f{\p^2}{\p w\p\xi}-\f{\zeta}{p}\f{\p^2}{\p w\p p}-\zeta\f{\p^2}{\p\xi\p\zeta}-p\f{\p^2}{\p p\p\xi} -\xi\f{\p^2}{\p\xi^2}\biggl]\mathcal{A}_2\label{2eqc}
\ee
These equations can be solved for arbitrary spin $s$. A convenient way to obtain the general result is to proceed recursively. Here, we state the final result for the spin $s$. Using \eqref{a2p} and \eqref{2.8uy}, we find that \eqref{2eqa} is identically satisfied provided we set $\Delta_1=\Delta_2$. Equation \eqref{2eqb} determines $a_n$ in terms of $a_0$ to be 
\be
a_n = (-1)^n\f{c_n(2\Delta_1-d)(2\Delta_1-d-2)(2\Delta_1-d-4)\cdots(2\Delta_1-d-2(n-1))}{(\Delta_1+s-2)(\Delta_1+s-3)\cdots(\Delta_1+s-2-(n-1))}a_0\label{a0uy}
\ee
with 
\be
c_n = s(s-1)(s-2)\cdots(s-n+1)
\ee
The 3rd equation \eqref{2eqc} is now identically satisfied for the above solution. 

Thus, we have shown that the 2-point function for an operator with spin $s$ in momentum space is given by \eqref{a2p}, \eqref{2.8uy} and \eqref{a0uy}. The conformal Ward identities have fixed this up to a single constant $a_0$. E.g., for spin $1$ operator with conformal dimension $\Delta$, we have 
\be
\mathcal{A}_2&=& A_0w +A_1\xi\zeta\;=\;a_0 \epsilon_1^\mu\epsilon_2^\nu \left[\delta_{\mu\nu} -\Bigl(\f{2\Delta-d}{\Delta-1}\Bigl)\f{p_\mu p_\nu}{ p^{2}}\right]p^{2\Delta-d}
\label{2.16}
\ee
Similarly, for spin $2$ operator with conformal dimension $\Delta$, we have 
\be
\mathcal{A}_2&=& A_0w^2 +A_1w\xi\zeta +\f{1}{2}A_2\xi^2\zeta^2\non\\
&=& \epsilon_1^{\mu\sigma}\epsilon_2^{\nu\rho} \left[a_0\delta_{\mu\nu}\delta_{\rho\sigma} +a_1\f{\delta_{\mu\nu}p_\rho p_\sigma}{ p^{2}}+a_2\f{p_{\mu}p_{\nu}p_\rho p_\sigma}{ 2p^{4}}\right]p^{2\Delta-d}
\ee
where we have defined $\epsilon^{\mu\nu}=\epsilon^\mu\epsilon^\nu $ and the relation between the coefficients are given by
\be
a_1 = -\f{2(2\Delta-d)}{\Delta}a_0\qquad;\quad a_2 = \f{2(2\Delta-d)(2\Delta-d-2)}{\Delta(\Delta-1)}a_0
\ee
In the above derivation, we have only used the conformal properties of the operators. Hence, the result is also valid for the conserved spin $s$ currents when we specialise $\Delta$ to $(d+s-2)$ which is the conformal dimension of the conserved current having spin $s$. It is instructive to express the 2-point functions in a form where the conserved nature of the operators are manifest when $\Delta=(d+s-2)$. For $s=1$, we can write
\be
\mathcal{A}_2 &=&a_0 \epsilon_1^\mu\epsilon_2^\nu \left[\pi_{\mu\nu}(p) -\Bigl(\f{\Delta-d+1}{\Delta-1}\Bigl)\f{p_\mu p_\nu}{ p^{2}}\right]p^{2\Delta-d}\label{2ptcons}
\ee
where, $\pi_{\mu\nu}(p)= \delta_{\mu\nu}-\f{p_\mu p_\nu}{ p^{2}}$. For $\Delta=d-1$ which is the dimension of a conserved vector current, the second term in \eqref{2ptcons} vanishes and we get the expected result for the 2-point function of the conserved vector operators (see, e.g., equation (2.12) of \cite{Bzowski:2013sza}). 

In a similar way, we can express the 2-point function of the spin 2 operators as 
\be
\mathcal{A}_2&=& \epsilon_1^{\mu\sigma}\epsilon_2^{\nu\rho} a_0\left[\Pi_{\mu\sigma,\nu\rho}(p)+\f{(d-\Delta)p_\mu p_\nu}{\Delta(\Delta-1)(d-1)p^2}\Bigl( E\,\delta_{\sigma\rho} + F\f{p_\sigma p_\rho}{p^2}  \Bigl) \right]\label{2.20iyt}
\ee
where, $E=2(d-1)(\Delta-1), F= (2\Delta-3d\Delta +d^2+d-2)$ and 
\be
\Pi_{\mu\sigma,\nu\rho}(p) =\f{1}{2}\Bigl[\pi_{\mu\nu}(p)\pi_{\sigma\rho}(p)+\pi_{\mu\rho}(p)\pi_{\nu\sigma}(p)\Bigl]-\f{1}{d-1}\pi_{\mu\sigma}(p)\pi_{\nu\rho}(p).
\ee
In writing equation \eqref{2.20iyt}, we have used the property $\epsilon^\mu\epsilon_\mu=0$ which is satisfied by the auxiliary vectors.
Again, for the conserved spin 2 current, for which $\Delta=d$, the second term in \eqref{2.20iyt} vanishes and we get the expected result for the spin 2 conserved currents (see, e.g., equation (2.11) of \cite{Bzowski:2013sza}).

\section{Ward identity constraints for 3-point function}
\label{sec2:identification}
In this section, we start the analysis of the 3-point function. As mentioned in the introduction, we focus on 3-point functions of one conserved current with spin 1 and two non-conserved spinning operators having spin $s$. We shall write down a general structure of the correlator in terms of scalar form factors and then 
use the conformal Ward identities to determine these form factors.

\subsection{General form of 3-point function}
To avoid dealing with the Lorentz indices, we shall again work in the index free notation given in equation \eqref{indexfree}. Further, due to the translation invariance, we can extract an overall momentum conserving delta function from our correlators and work with the reduced correlators as defined in equation \eqref{stripmomentum}. The 3-point function of one spin-1 conserved current and two spinning non-conserved operators, we are interested in, is
\be
\mathcal{A}_{s_1,J,s_3}&\equiv&\llangle[\Bigl] \epsilon_1\cdot\mathcal{O}_{s_1}({\bf p}_1)\epsilon_2\cdot J({\bf p}_2) \epsilon_3\cdot\mathcal{O}_{s_3}({\bf p}_3)   \rrangle[\Bigl] ~.\label{3ptre}
\ee
In the above expression, the spin and conformal dimension of the operator $\mathcal{O}_i$ are $s_i$ and $\Delta_i$ respectively. The conformal dimension of the conserved current $J^\mu$ is $\Delta_2=d-1$. Due to the momentum conservation $p_1^\mu+p_2^\mu+p_3^\mu=0$ and we  shall eliminate $p_3^\mu$ in favour of $p_1^\mu$ and $p_2^\mu$. The scalar form factors depend on three scalar variables that we choose to be  the magnitude of the momenta $p_1, p_2, p_3$\footnote{The inner product between different momenta can be expressed in terms of the magnitudes as 
\be
p_i\cdot p_j &=&  \f{1}{2}\Bigl[p_k^2-p_i^2-p_j^2\Bigl]\quad;\qquad i,j,k =1,2,3
\ee
where $p_i$ denotes the magnitude of the $i^{th}$ momenta $p_i^\mu$.}.

Next, we need to write the above 3-point function in terms of form factors. Following \cite{Bzowski:2013sza}, we split the correlator as sum of transverse and longitudinal parts (see appendix \ref{appena.3}). Using equation \eqref{a117t}, we can express the 3-point function \eqref{3ptre} as
\be
\mathcal{A}_{s_1,J,s_3}(p_1,p_2)&=& \llangle[\Bigl] \epsilon_1\cdot\mathcal{O}({\bf{p}_1}) \epsilon_2\cdot j({\bf p}_2) \epsilon_3\cdot\mathcal{O}({\bf p}_3)\rrangle[\Bigl]\;\;+\;\;\f{\epsilon_2\cdot p_2}{p_2^2}\llangle[\Bigl]  \epsilon_1\cdot\mathcal{O}({\bf p}_1)p_{2\nu} J^\nu({\bf p}_2) \epsilon_3\cdot\mathcal{O}({\bf p}_3)\rrangle[\Bigl]\non\\
&\equiv&\mathcal{A}^\perp +\mathcal{A}^{||}\, , \label{3.21}
\ee
where $\mathcal{A}^\perp$ and $\mathcal{A}^{||}$ denote the transverse and longitudinal parts, respectively, and  $j^\mu$ denotes the transverse current defined as $j^\mu = J^\nu\,\pi_{2\nu}^{~\;\mu}$, with 
\be
\pi_2^{\mu\nu} =\delta^{\mu\nu}-\f{p_2^\mu p_2^\nu}{p_2^2}\qquad;\qquad p_2^\mu\; \pi_{2\,\mu\nu} =0~,\label{3.22}
\ee
The transverse current $j^\mu$ satisfies $p_\mu j^\mu=0$. The second term in RHS of \eqref{3.21} can be reduced to sum of two 2-point functions by using the conservation Ward identity \eqref{B.128} and hence it is easily determined from the knowledge of 2-point functions obtained in the previous section. This means that we only need to focus on the transverse part of the correlator. 
The transverse part should be linear in the project operator $\pi_2$ with one of its indices contracted by $\epsilon_2$. Given that we have eliminated $p_3^\mu$ in terms of $p_1^\mu$ and $p^\mu_2$ and that $p_2 \cdot \pi_2=0$, there are three possibilities 
for the contraction of the the second index of $\pi_2$ and the 3-point function must be of the form:
\be
\mathcal{A}^\perp({\bf p}_1,{\bf p}_2) &=&(\epsilon_2\cdot \pi_2\cdot p_1)A+ (\epsilon_2\cdot \pi_2\cdot \epsilon_1)B_1 +(\epsilon_2\cdot \pi_2\cdot \epsilon_3)B_2
\label{ansatzs}
\ee
We now need to determine the form of $A, B_1, B_2$. We have already accounted for the $\epsilon_2$ dependence, so 
the possible scalar products between the auxiliary vectors $\epsilon_1, \epsilon_3$ and the momenta are:
\be
z=\epsilon_1\cdot \epsilon_3\quad,\qquad \xi_1=\epsilon_3\cdot p_2\quad,\quad \xi_2=\epsilon_3\cdot (p_1+p_2)\quad,\quad \zeta_1=\epsilon_1\cdot p_2\quad,\quad \zeta_2=\epsilon_1\cdot p_1\label{fgtr546}
\ee
This choice is motivated by the simple transformation property under the exchange $({\bf p}_1 \leftrightarrow {\bf p}_3, \epsilon_1  \leftrightarrow \epsilon_3)$, 
\begin{equation}
z \leftrightarrow z, \qquad \xi_1 \leftrightarrow \zeta_1, \qquad \xi_2 \leftrightarrow - \zeta_2\, .
\end{equation}
which we will use later on in section \ref{sec:symm_prop}.
The correlator is proportional to $s_1$ powers of $\epsilon_1$ and $s_2$ powers of $\epsilon_3$. In this paper we focus on the case $s_1=s_2=s$,  so $A$ should contain 
$s$ powers of the auxiliary vectors $\epsilon_1^\mu$ and $\epsilon_3^\mu$,  $B_1$ should have $s$ powers of $\epsilon_3^\mu$ and $s-1$ powers of $\epsilon_1^\mu$, and $B_2$ should have $s$ powers of $\epsilon_1^\mu$ but only $s-1$ powers of $\epsilon_3^\mu$. Thus the general form of $A, B_1$ and $B_2$ is
\be
A({\bf p}_1,{\bf p}_2)&=&\sum_{n=0}^s \sum_{p,q=0}^n \f{1}{n!p!q!} A_n^{(p,q)}(p_1,p_2,p_3) z^{s-n}\xi_1^p\xi_2^{n-p}\zeta_1^q\zeta_2^{n-q}   \label{25rt1}\\[.3cm]
B_1({\bf p}_1,{\bf p}_2)&=&\sum_{n=0}^{s-1} \sum_{p=0}^{n+1}\sum_{q=0}^{n} \f{1}{n!p!q!} B_{1;n}^{(p,q)}(p_1,p_2,p_3) z^{s-1-n}\xi_1^p\xi_2^{n-p+1}\zeta_1^q\zeta_2^{n-q}   
 \label{25rt2}\\[.3cm]
 B_2({\bf p}_1,{\bf p}_2)&=&\sum_{n=0}^{s-1} \sum_{p=0}^{n}\sum_{q=0}^{n+1} \f{1}{n!p!q!} B_{2;n}^{(p,q)}(p_1,p_2,p_3) z^{s-1-n}\xi_1^p\xi_2^{n-p}\zeta_1^q\zeta_2^{n-q+1}  \label{25rt3}
\ee
where the form factors $A_n^{(p,q)},\, B_{1;n}^{(p,q)}$ and $B_{2;n}^{(p,q)}$ depend on the magnitudes of  the three momenta. The problem of determining the 3-point function has been reduced to finding the form factors $A_{n}^{(p,q)}, B_{1;n}^{(p,q)}$ and $B_{2;n}^{(p,q)}$. We shall fix these form factors by applying the conformal Ward identities on the transverse part of 3-point function given in equation \eqref{ansatzs}.

\subsection{Action of Ward identities }
To determine the form factors, the first step is to obtain the equations satisfied by them. To obtain these equations, we start by noting that the momentum space CFT Ward identities, as reviewed in appendix \ref{appenAb}, involve derivatives with respect to the components of the individual momenta. However, when acting on 3-point functions, they can be converted into derivatives with respect to the magnitudes of the momenta. Further, the 3-point function \eqref{3ptre} also depends upon the auxiliary vectors
$\epsilon_1, \epsilon_3$ via $\zeta_i, \xi_i$ and $z$. It is useful to derive a chain rule which takes into account these variables. For a general function of the form $f= f(p_{1},p_2,p_3
, z, \xi_{1},\xi_2, \zeta_{1},\zeta_2 )$, we have 
\be
\f{\p f}{\p p_1^\mu}&=& \f{(p_1+p_2)_{\mu}}{p_3} \f{\p f}{\p p_3} +\f{p_1^\mu}{p_1}  \f{\p f}{\p p_1} + \epsilon_{1\mu} \f{\p f}{\p \zeta_2} + \epsilon_{3\mu} \f{\p f}{\p \xi_2} \non\\[.3cm]
\f{\p f}{\p p_2^\mu}&=& \f{(p_1+p_2)_{\mu}}{p_3} \f{\p f}{\p p_3} +\f{p_2^\mu}{p_2}  \f{\p f}{\p p_2}  +\epsilon_{1\mu} \f{\p f}{\p \zeta_1} + \epsilon_{3\mu} \f{\p f}{\p \xi_2} + \epsilon_{3\mu} \f{\p f}{\p \xi_1} \label{chainwer}
\ee
Using this, the action of dilatation operator \eqref{dilwardw} can be expressed in the form
\be
\left(\sum_{i=1}^{2}p^\mu_i\f{\p}{\p p_i^\mu }+2d-\Delta_t\right)\mathcal{A}^\perp\;=\; \left(\sum_{i=1}^{3}p_i\f{\p}{\p p_i }+\xi_1\f{\p}{\p \xi_1}+\xi_2\f{\p}{\p \xi_2}+\zeta_1\f{\p}{\p \zeta_1}+\zeta_2\f{\p}{\p \zeta_2}+2d-\Delta_t\right)\mathcal{A}^\perp=0\non
\ee
where, we have defined $\Delta_t\equiv\Delta_1+\Delta_2+\Delta_3$.

The above expression is valid when acting on the 3-point correlator. It is useful to consider the action of the dilatation operator on the functions $A, B_1$ and $B_2$ given in equation \eqref{ansatzs}. Due to the linear independence of three terms in RHS of \eqref{ansatzs}, the action of dilatation Ward identity on each of the terms in RHS is separately zero. Using this fact, a little algebra gives the action of the dilatation on the functions $A$ and $B_i$ to be 
\be
\left(\sum_{i=1}^{2}p^\mu_i\f{\p}{\p p_i^\mu }+2d-\Delta_t\right)T\;=\; \Bigl( \sum_{i=1}^{3}p_i\f{\p}{\p p_i }+\xi_1\f{\p}{\p \xi_1}+\xi_2\f{\p}{\p \xi_2}+\zeta_1\f{\p}{\p \zeta_1}+\zeta_2\f{\p}{\p \zeta_2}+2d-\Delta_t+a\Bigl)T=0\non\\
\label{dilght}
\ee
where $T$ can be $A,B_1$ or $B_2$ and $a=1$ for $A$ and $0$ for $B_1$ and $B_2$.

Next we consider the action of the special conformal Ward identity. The corresponding differential operator is $\mathcal{K}^\mu=\mathcal{K}^\mu_s+\mathcal{K}^\mu_\epsilon$, where we have denoted the scalar and spin parts of the special conformal operator by $\mathcal{K}_s^\mu$ and $\mathcal{K}_\epsilon^\mu$ respectively 
\be
\mathcal{K}^\mu_s&=&\sum_{i=1}^{2}\biggl[2(\Delta_i-d)\f{\p}{\p p_{i\mu}}-2p_i^\nu\f{\p}{\p p_i^\nu }\f{\p}{\p p_{i\mu}}+p^\mu_{i} \f{\p}{\p p_i^\nu }\f{\p}{\p p_{i\nu} }\biggl]\non\\[.3cm]
\mathcal{K}^\mu_\epsilon&=& 2\sum_{i=1}^2 S_i^{\mu\nu} \f{\p}{\p p_i^\nu} \qquad;\qquad S^{\mu\nu}_i =\epsilon_i^\mu\f{\p}{\p \epsilon_{i\nu}}-\epsilon_i^\nu\f{\p}{\p \epsilon_{i\mu}}\label{a210a}
\ee
Since these operators involve a vector index, it is convenient to contract them with an arbitrary vector $b_\mu$ and work with the resulting scalar operator. Using the chain rules given in \eqref{chainwer}, we find
\be
b\cdot \mathcal{K}_s\; (T)
&=&(b\cdot p_1)\biggl[P_1^{(0,a)}+2\epsilon_1\cdot\epsilon_3\f{\p^2}{\p\xi_2\p\zeta_2}\biggl]T+(b\cdot p_2)\biggl[P_2^{(0,a)}+2z\f{\p}{\p\zeta_1}\Bigl(\f{\p}{\p\xi_1}+\f{\p}{\p\xi_2}\Bigl)\biggl]T\non\\[.28cm]
&&-2(b\cdot\epsilon_1)\biggl[ (\Delta_3-1-a)\f{\p}{\p\zeta_2}+(R+d-\Delta_2)\left(\f{\p}{\p\zeta_1} -\f{\p}{\p\zeta_2} \right)\biggl]T\non\\[.3cm]
&&-2(b\cdot\epsilon_3)\biggl[ (\Delta_3-1-a)\f{\p}{\p\xi_2}+(R+d-\Delta_2)\f{\p}{\p\xi_1} \biggl]T
\label{bks}
\ee
where $T$ can be $A, B_1$ or $B_2$ and 
\be
P_1^{(\alpha,\beta)} &\equiv& K_1-K_3-2\biggl(\alpha-\beta-\xi_2\f{\p}{\p\xi_2}-(\zeta_1+\zeta_2)\f{\p}{\p\zeta_2}\biggl)\f{1}{p_3}\f{\p}{\p p_3}
\non\\[.2cm]
P_2^{(\alpha,\beta)}&\equiv&K_2-K_3-2\biggl(\alpha-\beta-\xi_2\Bigl(\f{\p}{\p\xi_1}+\f{\p}{\p\xi_2}\Bigl)-(\zeta_1+\zeta_2)\f{\p}{\p\zeta_1}\biggl)\f{1}{p_3}\f{\p}{\p p_3}\non\\[.3cm]
R&\equiv&p_2\f{\p}{\p p_2}+\f{1}{2}\f{p_2^2+p_3^2-p_1^2}{p_3}\f{\p}{\p p_3} +\xi_1\Bigl(\f{\p}{\p\xi_1} +\f{\p}{\p\xi_2} \Bigl)+\zeta_1\f{\p}{\p\zeta_1}\non\\[.3cm]
K_i &\equiv&  -\f{\p^2}{\p p_i^2}\;+\;2\biggl(\Delta_i-\f{d+1}{2}\biggl)\f{1}{p_i}\f{\p}{\p p_i}
\ee
Similarly, the action of the spin part can be expressed as 
\be
b\cdot \mathcal{K}_\epsilon\;(T)
&=&-2(b\cdot p_1)E_1\f{\p}{\p\zeta_2}T -2(b\cdot p_2)E_1\f{\p}{\p\zeta_1}T +2(b\cdot\epsilon_1)\left[F_1\left(\f{\p}{\p\zeta_1}-\f{\p}{\p\zeta_2}\right) +F_2\f{\p}{\p\zeta_2}+E_2\f{\p}{\p z}  \right]T\non\\[.3cm]
&&-2(b\cdot\epsilon_3)\Bigl(E_1\f{\p}{\p z}  \Bigl)T
\ee
where again $T=A,B_1,B_2$ and we have defined 
\be
E_1&=& z\frac{\partial }{\partial \xi _2 }+\frac{\zeta _2}{p_1}\frac{\partial }{\partial p_1}+\frac{ (\zeta_1+\zeta _2)}{p_3} \frac{\partial }{\partial p_3}\non\\[.3cm]
E_2&=&z\f{\p}{\p \zeta_2}+  \f{\xi_2-\xi_1}{p_1}\f{\p}{\p p_1} + \f{\xi_2}{p_3}\f{\p}{\p p_3} \non\\[.3cm]
F_1&=& \f{1}{2}\f{p_3^2-p_2^2-p_1^2}{p_1}\f{\p}{\p p_1}+\f{1}{2}\f{p_3^2+p_2^2-p_1^2}{p_3}\f{\p}{\p p_3} +\xi_1\f{\p}{\p\xi_2}+\zeta_1\f{\p}{\p\zeta_2}\non\\[.3cm]
F_2&=&\f{1}{2}\f{p_3^2-p_2^2+p_1^2}{p_1}\f{\p}{\p p_1}+{p_3}\f{\p}{\p p_3} +\xi_2\f{\p}{\p\xi_2}+(\zeta_1+\zeta_2)\f{\p}{\p\zeta_2}+d-1
\ee
Finally, we are ready to write down the action of the special conformal Ward identity on the full 3-point function. This is easily done by noting the following product rule for two arbitrary functions $f_1$ and $f_2$
\be
 \mathcal{K}_\epsilon^\mu (f_1f_2)  &=& f_1 \mathcal{K}_\epsilon^\mu f_2+ f_2 \mathcal{K}_\epsilon^\mu f_1+2\sum_{i=1}^{2}\left(\epsilon_i^\mu\f{\p f_2}{\p \epsilon_{i\nu}}\f{\p f_1}{\p p_{i}^\nu }+\epsilon_i^\mu\f{\p f_1}{\p \epsilon_{i\nu}}\f{\p f_2}{\p p_{i}^\nu }-\epsilon_i^\nu\f{\p f_2}{\p \epsilon_{i\mu}}\f{\p f_1}{\p p_{i}^\nu }-\epsilon_i^\nu\f{\p f_1}{\p \epsilon_{i\mu}}\f{\p f_2}{\p p_{i}^\nu }\right)\non\\[.3cm]
 \mathcal{K}_s^\mu (f_1f_2)
  &=& f_1 \mathcal{K}_s^\mu f_2+ f_2 \mathcal{K}_s^\mu f_1+2\sum_{i=1}^{2}p_{i\mu} \f{\p f_1}{\p p_i^\nu}\f{\p f_2}{\p p_{i\nu}}\; -\; 2\sum_{i=1}^{2}p_{i}^{\nu} \Bigl(\f{\p f_1}{\p p_i^\nu}\f{\p f_2}{\p p^\mu_{i}}+\f{\p f_1}{\p p_i^\mu}\f{\p f_2}{\p p^\nu_{i}}\Bigl)
\ee
Using this, we can determine the equations resulting from the action of the special conformal Ward identity. Its' action on the full 3-point function \eqref{3.21}  implies
\be
\hspace*{-.3in}b\cdot (\mathcal{K}_s+\mathcal{K}_\epsilon)\mathcal{A}^\perp + b\cdot (\mathcal{K}_s+\mathcal{K}_\epsilon)\mathcal{A}^\parallel=0\label{consu8}
\ee
After a long but straightforward calculation,  the action on the transverse part of the correlator can be found to be 
\be
&&\hspace*{-.3in}b\cdot (\mathcal{K}_s+\mathcal{K}_\epsilon)\mathcal{A}^\perp \non\\[.3cm]
&=&(\epsilon_2\cdot \pi_2\cdot p_1)\biggl[b\cdot (\mathcal{K}_s+\mathcal{K}_\epsilon)A-\f{2b\cdot (p_1+p_2)}{p_3}\f{\p A}{\p p_3}+\f{2(b\cdot \epsilon_1)}{p_1}\f{\p B_1}{\p p_1}-\f{2(b\cdot \epsilon_3)}{p_3}\Bigl(\f{\p B_2}{\p p_3}+\f{\p A}{\p \xi_2}\Bigl)\biggl]\non\\
&&+ (\epsilon_2\cdot \pi_2\cdot \epsilon_1)\biggl[b\cdot (\mathcal{K}_s+\mathcal{K}_\epsilon)B_1 -2b\cdot (p_1+p_2) \f{\p A}{\p\zeta_1}-2(b\cdot\epsilon_1)\Bigl(\f{\p B_1}{\p\zeta_1}-\f{\p B_1}{\p\zeta_2}\Bigl)-2(b\cdot\epsilon_3)\Bigl(\f{\p B_2}{\p\zeta_1}+\f{\p A}{\p z}\Bigl)\biggl]\non\\
&&+ (\epsilon_2\cdot \pi_2\cdot \epsilon_3)\biggl[b\cdot (\mathcal{K}_s+\mathcal{K}_\epsilon)B_2 -2(b\cdot p_1) \f{\p A}{\p\xi_1}+2(b\cdot\epsilon_1)\Bigl(\f{\p A}{\p z}-\f{\p B_1}{\p\xi_1}\Bigl)-2(b\cdot\epsilon_3)\Bigl(\f{\p B_2}{\p\xi_1}+\f{\p B_2}{\p \xi_2}\Bigl)\biggl]\non\\
&&+{(\epsilon_2\cdot\pi_2\cdot b)}\Biggl[\biggl(2\Delta_1-2d+\f{\Delta_2-1}{p_2^2}(p_1^2+p_2^2-p_3^2)-2p_1\f{\p}{\p p_1}+2(\xi_2- \xi_1) \f{\p}{\p\xi_1}+2\zeta_2\f{\p}{\p\zeta_1}\non\\
&&+\f{(p_3^2-p_1^2-p_2^2)}{p_2}\f{\p}{\p p_2}  \biggl)A+\f{2\zeta_1}{p_2^2} \biggl(1-\Delta_2+p_2\f{\p}{\p p_2}-\f{p_2^2}{p_1}\f{\zeta_2}{\zeta_1}\f{\p}{\p p_1}+\f{zp_2^2}{\zeta_1}\f{\p}{\p\xi_1}\biggl)B_1\non\\
&&+\f{2\xi_1}{p_2^2} \biggl(1-\Delta_2+p_2\f{\p}{\p p_2}+\f{p_2^2}{p_3}\f{\xi_2}{\xi_1}\f{\p}{\p p_3}+\f{zp_2^2}{\xi_1}\f{\p}{\p\zeta_1}\biggl)B_2\Biggl]+L\label{kskeps}
\ee
where $L$ denotes a longitudinal contribution given by 
\be
L &\equiv & -\f{2{(\epsilon_2\cdot p_2)}}{p_2^2}(\Delta_2-d+1)\Biggl[(b\cdot p_1)A+(b\cdot \epsilon_1)B_1+(b\cdot \epsilon_3)B_2\non\\
&&-(b\cdot p_2) \biggl(\f{p_3^2-p_1^2-p_2^2}{2p_2^2}A+\f{\zeta_1}{p_2^2} B_1+\f{\xi_1}{p_2^2} B_2\biggl)\Biggl]
\ee
For the conserved current $\Delta_2=d-1$, this longitudinal contribution vanishes identically. This particular result can also be derived using the commutation relations of the generators of the special conformal transformations with the momentum. To see this, we note that the longitudinal contribution can be determined from the $\epsilon$ stripped correlation function, namely ${\cal A}^\perp _\mu$, 
through the identity
\begin{eqnarray}
 (\mathcal{K}_s+\mathcal{K}_\epsilon)_\nu  \,\epsilon_2^\mu\, {\cal A}^\perp _\mu  &=&\epsilon_{2\sigma} \Big[2\left(\delta^\sigma_\nu \frac{\partial}{\partial p_2^\mu}-\delta_{\nu\mu} \, \frac{\partial}{\partial p_{2\sigma}}\right) +  \delta_\mu^\sigma (\mathcal{K}_s+\mathcal{K}_\epsilon)_\nu\Big]  \,{{\cal A}^\perp} ^\mu\nonumber\\
&\equiv &\epsilon_{2\sigma}\Big[
 (\pi_2 \cdot p_1)^\sigma\, I^{ (A)}_\nu+ (\pi_2\cdot \epsilon_1)^\sigma\,I^{(B_1)}_\nu+(\pi_2\cdot \epsilon_3)^\sigma\,I^{(B_2)}_\nu+\frac{p_2^\sigma}{p_2^2} \,I_\nu^\parallel\Big]\label{alter56y}
\end{eqnarray}
where in the last line we have written the most general expression that one gets from the action of the generator of the  special conformal transformations on the correlator. The longitudinal contribution $I_\nu^\parallel$ can be obtained by replacing $\epsilon_{2\mu}\rightarrow p_{2\mu}$ since the first 3 terms in the second line of \eqref{alter56y} vanish when $\epsilon_{2\mu} = p_{2\mu}$, and we get
\begin{eqnarray}
 I^\parallel_\nu&=& \left[2\left( p_{2\nu} \frac{\partial}{\partial p_2^\mu} - \delta_{\nu\mu}\, p_2\cdot \frac{\partial}{\partial p_2}\right)+ p_{2\,\mu} \, (\mathcal{K}_s+\mathcal{K}_\epsilon)_\nu\right]\,{{\cal A}^\perp} ^\mu \label{3.42}
\end{eqnarray}
 Now, the commutator between the generator of the special conformal transformations and the momentum $p_2$ is given by
\begin{eqnarray}
\bigl[p_2^\mu,\,({\cal K}_s)_\nu\bigl]&=& 2\delta^\mu_\nu\left(d -\Delta_2-1+p_2\cdot\frac{\partial}{\partial p_2}\right) +2\frac{\partial}{\partial p_2^\nu}p_2^\mu-2p_{2\nu} \frac{\partial}{\partial p_{2\mu}} \label{2.62}
\end{eqnarray}
Furthermore, 
\begin{equation} \label{pke}
p_{2\,\mu} (\mathcal{K}_\epsilon)_\nu\mathcal{A}^{\perp\mu}= 2 p_{2\,\mu} \sum_{i=1}^2 S_{i\ \nu}^{\kappa} \f{\p}{\p p_i^\kappa} \mathcal{A}^{\perp\mu}=0
\end{equation} 
The term with $i=1$ gives zero because $p_{2\,\mu}$ commutes with this term and $p_{2\mu}\mathcal{A}^{\perp\mu}=0$, while the term with $i=2$ is zero because $\mathcal{A}^{\perp\mu}$ (by definition) does not depend on $\epsilon_2$.
Using these results in \eqref{3.42}, we find
\begin{equation}
 I^\parallel_\nu= 2 (d -\Delta_2-1) \mathcal{A}^{\perp}_{\nu} =0,
\end{equation}
since the dimension of the conserved current is $\Delta_2=d-1$.

The action of the special conformal Ward identity on the longitudinal part of the 3-point function is given by
\be
b\cdot(\mathcal{K}_s+\mathcal{K}_\epsilon)\mathcal{A}^{||}&=&b\cdot(\mathcal{K}_s+\mathcal{K}_\epsilon )\left[\f{\epsilon_2\cdot p_2}{p_2^2}\llangle [\Bigl] \epsilon_1\cdot\mathcal{O}({\bf p}_1)p_{2\nu} J^\nu({\bf p}_2) \epsilon_3\cdot\mathcal{O}({\bf p}_3)\rrangle[\Bigl]\right]\non\\[.3cm]
&=&\f{\epsilon_2\cdot p_2}{p_2^2} b\cdot(\mathcal{K}_s+\mathcal{K}_\epsilon+2\f{\p}{\p p_2})H\;+\;\f{2(\Delta_2-1)}{p_2^2}(\epsilon_2\cdot \pi_2\cdot b) H\label{3.44}
\ee
where $H\equiv \llangle[\bigl]  \epsilon_1\cdot\mathcal{O}({\bf p}_1)p_{2\nu} J^\nu({\bf p}_2) \epsilon_3\cdot\mathcal{O}({\bf p}_3)\rrangle[\bigl]$.

By using the conservation Ward identity \eqref{B.128}, the correlator $H$ can be written as the sum of two 2-point functions. However, the operators ${\cal{K}}_s$ and $\cal{K}_\epsilon$ are the ones that act on 3-point functions. Thus, we have
\be
&&b\cdot\Bigl(\mathcal{K}_s+\mathcal{K}_\epsilon+2\f{\p}{\p p_2}\Bigl)_{\mbox{3-pt}}H\non\\
&=&  -b\cdot\Bigl(\mathcal{K}_s+\mathcal{K}_\epsilon+2\f{\p}{\p p_2}\Bigl)_{\mbox{3-pt}}\Big[g_1\llangle[\bigl] \epsilon_1 \cdot \mathcal{O}_1 (-{\bf p}_3)\,\epsilon_3\cdot \mathcal{O}_3({\bf p}_3)\rrangle[\bigl]+g_3\llangle[\bigl]  \epsilon_1\cdot \mathcal{O}_1({\bf p}_1)\,\epsilon_3 \cdot  \mathcal{O}_3(-{\bf p}_1)\rrangle[\bigl]\Big]\non
\ee
It is easy to see that when acting on the 2nd two point function in the above expression, the operator $b\cdot(\mathcal{K}_s+\mathcal{K}_\epsilon+2\f{\p}{\p p_2})$ reduces to the one corresponding to 2-point function and hence vanishes. For the 1st 2-point function, we can use momentum conservation to write $p_3 =-p_1-p_2$. A simple calculation now shows
\be
&&b\cdot\Bigl(\mathcal{K}_s+\mathcal{K}_\epsilon+2\f{\p}{\p p_2}\Bigl)_{\mbox{3-pt}}\Big[\llangle[\bigl] \epsilon_1 \cdot \mathcal{O}_1 (-{\bf p}_3={\bf p}_1+{\bf p}_2)\,\epsilon_3\cdot \mathcal{O}_3({\bf p}_3=-{\bf p}_1-{\bf p}_2)\rrangle[\bigl]\Big]\non\\
&=&b\cdot\Bigl(\mathcal{K}_s+\mathcal{K}_\epsilon\Bigl)_{\mbox{2-pt}}\Big[\llangle[\bigl] \epsilon_1 \cdot \mathcal{O}_1 (-{\bf p}_3)\,\epsilon_3\cdot \mathcal{O}_3({\bf p}_3)\rrangle[\bigl]\Big]
\ee
The operator in the second line contains derivatives w.r.t. the $p_3$ variable only. Hence, we again have the Ward identity corresponding to a 2-point function acting on the 2-point function and hence it vanishes. This means that the first term in RHS of equation \eqref{3.44} does not contribute and we are left with
\be
&&\hspace*{-.47in}b\cdot(\mathcal{K}_s+\mathcal{K}_\epsilon)\mathcal{A}^{||}\non\\[.2cm]
&=&\;-\f{2(\Delta_2-1)}{p_2^2}(\epsilon_2\cdot \pi_2\cdot b) \Big[g_1\llangle[\bigl] \epsilon_1 \cdot \mathcal{O}_1 (-{\bf p}_3)\,\epsilon_3\cdot \mathcal{O}_3({\bf p}_3)\rrangle[\bigl]+g_3\llangle[\bigl]  \epsilon_1\cdot \mathcal{O}_1({\bf p}_1)\,\epsilon_3 \cdot  \mathcal{O}_3(-{\bf p}_1)\rrangle[\bigl]\Big]
\label{3.45}
\end{eqnarray}
where $g_1$ and $g_3$ denote the gauge couplings of the operators $\mathcal{O}_1$ and $\mathcal{O}_3$ respectively (see appendix \ref{longitudinal} for details).  
The contribution \eqref{3.45} vanishes when $\Delta_1\not=\Delta_3$ (since the 2-point vanishes) and it contributes to
the equations originating from the term proportional to $(\epsilon_2\cdot \pi_2\cdot b)$  when $\Delta_1=\Delta_3$.

\subsection{ Ward identity equations}
\label{prisecwi}
The action of special conformal Ward identity on the full 3-point function is given by equations \eqref{consu8}, \eqref{kskeps} and \eqref{3.45}. In the transverse part \eqref{kskeps}, the coefficients of the independent tensor structures such as  $(\epsilon_2\cdot\pi_2\cdot p_1)(b\cdot p_1), (\epsilon_2\cdot\pi_2\cdot \epsilon_1)(b\cdot p_2)$ {\it etc.} are all independent. This means that we can set the coefficients of these different tensor structures to zero. However, the coefficient of $(\epsilon_2\cdot\pi_2\cdot b)$ has to be combined with the terms involving the two point functions according to equation \eqref{3.45}. It is useful to divide the equations based on the order of the differential equations. Those differential equations which involve second derivative with respect to momenta are called primary Ward identities and those involving only a single momentum derivative are called secondary Ward identities \cite{Bzowski:2013sza}. The secondary Ward identities can be homogeneous or inhomogeneous. As discussed above, the inhomogeneous Ward secondary equations originate from the coefficient of $(\epsilon_2\cdot\pi_2\cdot b)$ (though, as we will see below there are also secondary Ward equations originating from the coefficient of $(\epsilon_2\cdot\pi_2\cdot b)$ that are still homogeneous). The primary Ward identities are used to fix the form of 3-point function up to free parameters. The homogeneous secondary Ward identities constrain these parameters, and the inhomogeneous secondary Ward identities relate the parameters to the normalization coefficient of the two-point function. The equations are  given explicitly below.

\subsubsection{Primary Ward identities}
As mentioned above, the equations involving second derivative with respect to momenta are called primary Ward identities. These are the equations associated with $(b\cdot p_1)$ and $(b\cdot p_2)$ and are given by

\vspace*{.1in}\noindent{\underline{$\epsilon_2\cdot\pi_2\cdot p_1$ }}
\be
(b\cdot p_1)&:&\quad 0\;=\;    \f{1}{2}P_1^{(0,1)}A+z\f{\p^2A}{\p\xi_2\p\zeta_2} -E_1\f{\p A}{\p\zeta_2}-\f{1}{p_3}\f{\p A}{\p p_3}\non\\[.4cm]
(b\cdot p_2)&:&\quad  0\;=\;  \f{1}{2}P_2^{(0,1)}A+z\f{\p}{\p\zeta_1}\Bigl(\f{\p}{\p\xi_1}+\f{\p}{\p\xi_2}\Bigl)A -E_1\f{\p A}{\p\zeta_1}-\f{1}{p_3}\f{\p A}{\p p_3} 
\ee
{\underline{$\epsilon_2\cdot\pi_2\cdot \epsilon_1$ }}
\be
(b\cdot p_1)&:&\quad 0\;=\;   \f{1}{2}P_1^{(0,0)}B_1+z\f{\p^2B_1}{\p\xi_2\p\zeta_2'}-\f{\p A}{\p\zeta_1} -E_1\f{\p B_1}{\p\zeta_2}\non\\[.4cm]
(b\cdot p_2)&:&\quad   0\;=\;  \f{1}{2}P_2^{(0,0)}B_1+z\f{\p}{\p\zeta_1}\Bigl(\f{\p}{\p\xi_1}+\f{\p}{\p\xi_2}\Bigl)B_1-\f{\p A}{\p\zeta_1} -E_1\f{\p B_1}{\p\zeta_1} 
\ee
{\underline{$\epsilon_2\cdot\pi_2\cdot \epsilon_3$ }}
\be
(b\cdot p_1)&:&\quad  0\;=\;   \f{1}{2}P_1^{(0,0)}B_2+z\f{\p^2B_2}{\p\xi_2\p\zeta_2}-\f{\p A}{\p\xi_1} -E_1\f{\p B_2}{\p\zeta_2}\non\\[.4cm]
(b\cdot p_2)&:&\quad  0\;=\;   \f{1}{2}P_2^{(0,0)}B_2+z\f{\p}{\p\zeta_1}\Bigl(\f{\p}{\p\xi_1}+\f{\p}{\p\xi_2}\Bigl)B_2 -E_1\f{\p}{\p\zeta_1}B_2 
\ee

\subsubsection{Secondary Ward Identities}
\label{tse4}

The equations involving a single derivative with respect to momenta are called the secondary Ward identities. They are associated with $(b\cdot\epsilon_1), (b\cdot\epsilon_3)$ and $(\epsilon_2\cdot \pi_2\cdot b)$ and are given by

\vspace*{.1in}\noindent{\underline{$\epsilon_2\cdot\pi_2\cdot p_1$ }}
\be
(b\cdot \epsilon_1)&:& \quad 0\;=\;  -\biggl[ (\Delta_3-2-F_2)\f{\p}{\p\zeta_2}+(R-F_1+d-\Delta_2)\Bigl(\f{\p}{\p\zeta_1}-\f{\p}{\p\zeta_2}\Bigl)-E_2\f{\p}{\p z} \biggl]A+\f{1}{p_1}\f{\p B_1}{\p p_1}\non\\[.4cm]
(b\cdot \epsilon_3)&:&\quad 0\;=\;  \biggl[ (\Delta_3-1)\f{\p}{\p\xi_2}+(R+d-\Delta_2)\f{\p}{\p\xi_1}+E_1\f{\p }{\p z} \biggl]A+\f{1}{p_3}\f{\p B_2}{\p p_3}\\ [.1cm]\non
\ee
{\underline{$\epsilon_2\cdot\pi_2\cdot \epsilon_1$ }}
\be
(b\cdot \epsilon_1)&:&\quad 0\;=\;  \biggl[ (\Delta_3-1-F_2)\f{\p}{\p\zeta_2}+(R-F_1+d+1-\Delta_2)\Bigl(\f{\p}{\p\zeta_1}-\f{\p}{\p\zeta_2}\Bigl) -E_2\f{\p}{\p z}  \biggl]B_1\non\\[.4cm]
(b\cdot \epsilon_3)&:&\quad 0\;=\; \biggl[ (\Delta_3-1)\f{\p}{\p\xi_2}+(R+d-\Delta_2)\f{\p}{\p\xi_1}+E_1\f{\p}{\p z} \biggl]B_1+ \f{\p A}{\p z}+\f{\p B_2}{\p \zeta_1}\\ [.1cm]\non
\ee
{\underline{$\epsilon_2\cdot\pi_2\cdot \epsilon_3$ }}
\be
(b\cdot \epsilon_1)&:&\quad 0\;=\;  \Bigl[ (\Delta_3-1-F_2)\f{\p}{\p\zeta_2}+(R-F_1+d-\Delta_2)\Bigl(\f{\p}{\p\zeta_1}-\f{\p}{\p\zeta_2}\Bigl) -E_2\f{\p}{\p z} \Bigl]B_2-\f{\p A}{\p z}+\f{\p B_1}{\p\xi_1}\non\\[.4cm]
(b\cdot \epsilon_3)&:&\quad 0\;=\; \biggl[ \Delta_3\f{\p}{\p\xi_2}+(R+d+1-\Delta_2)\f{\p}{\p\xi_1} +E_1\f{\p }{\p z}\biggl]B_2
\ee

The final secondary equation comes from the coefficient of $(\epsilon_2\cdot \pi_2\cdot b)$. This is first order non-homogeneous differential equation having two point functions as sources and is given by

\vspace*{.2in}\noindent{\underline{$\epsilon_2\cdot\pi_2\cdot b$ }}
\be
&&\Big[(\Delta_1-\Delta_3) -p_1\frac{\partial}{\partial p_1} +p_3\frac{\partial}{\partial p_3} -\frac{(p_1^2-p_3^2)}{p_2} \frac{\partial}{\partial p_2}-(\xi_1-\xi_2)\partial_{\xi_1}+\xi_2(\partial_{\xi_2}+\partial_{\xi_1})+\zeta_2(\partial_{\zeta_1}-\partial_{\zeta_2})\nonumber\\
&& +(\zeta_2+\zeta_1)\partial_{\zeta_1}\Big]A+\frac{2 (1-\Delta_2)}{p_2^2} \Big( \frac{1}{2}(p_3^2- p_1^2)A +\zeta_1B_1+\xi_1B_2\Big)+2\Big[\frac{\zeta_1}{p_2}\frac{\partial}{\partial p_2}-\frac{\zeta_2}{p_1}\frac{\partial}{\partial p_1}+z\partial_{\xi_1}\Big]B_1\nonumber\\
&&+2\Big[ \frac{\xi_1}{p_2} \frac{\partial}{\partial p_2} +\frac{\xi_2}{p_3} \frac{\partial}{\partial p_3}+ z\partial_{\zeta_1} \Big]B_2\non\\[.3cm]
&&=\f{2(\Delta_2-1)}{p_2^2}\delta_{\Delta_1,\Delta_3}\sum_{n=0}^s\frac{1}{n!} z^{s-n}\,a_n\Big[g_1\,\xi_2^n\,(\zeta_1+\zeta_2)^n\,p_3^{2\Delta_1 -d-2n}+g_3\,(\xi_2-\xi_1)^n\,\zeta_2^n\,p_1^{2\Delta_1 -d-2n}\Big]\non\\
\label{3.52}
\ee
In writing the left hand side of the above equation, we have used the dilatation Ward identity \eqref{dilght}. For the right hand side, we have used the result for two point function given in previous section. 
The $\delta_{\Delta_1,\Delta_3}$ ensures that the right hand side only contributes when the two point function is non zero. We note that on the right side of \eqref{3.52}, the combination $\zeta_1^n\,\xi_1^n$ never appears. The equations corresponding to this combination are homogeneous equations which, together with other homogeneous secondary equations,  restrict the parameters appearing in the solutions of primary equations.

\subsection{Solution of the dilation Ward identities}

The dilatation Ward identity implies that the form factors are homogeneous functions of momenta.
To see this, we start with the form factors $A_n^{(p,q)}$ writing 
\be
A&=&\sum_{n=0}^sz^{s-n}T^{(n)}\qquad;\qquad T^{(n)}\equiv\sum_{p,q=0}^n \f{1}{n!p!q!} A_n^{(p,q)} \xi_1^p\xi_2^{n-p}\zeta_1^q\zeta_2^{n-q}
\ee
Since the dilatation Ward identity does not involve powers and derivative with respect to $z$, the $T^{(n)}$ is also annihilated by the dilatation ward identity, \eqref{dilght}. Now, a small calculation gives
\be
\Bigl(\xi_1\f{\p}{\p \xi_1}+\xi_2\f{\p}{\p \xi_2}+\zeta_1\f{\p}{\p \zeta_1}+\zeta_2\f{\p}{\p \zeta_2}\Bigl)T^{(n)}&=&  2nT^{(n)}
\ee
Using this, we can get rid of $\xi_i$ and $\zeta_i$ variables in \eqref{dilght} and obtain
\be
\Bigl( \sum_{i=1}^{3}p_i\f{\p}{\p p_i }+2d+2n-\Delta_t+1\Bigl)A_{n}^{(p,q)}=0\label{andilr5}
\ee
The above equation says that $A_n^{(p,q)}$ is a homogeneous function of degree $-(2 d + 2n -\Delta_t +1)$. 

Anticipating the results to follow, we express $A_n^{(p,q)}$ in terms of triple K integrals (see appendix \ref{apps1} for details on triple K integrals). Triple K integrals satisfy the above equation (see \eqref{jn1}), and they indeed turn out to satisfy all conformal Ward identities. In particular, we shall find that the primary Ward identities are satisfied by the functions of the form
\be
A_n^{(p,q)} = a_n^{(p,q)}J_{N\{k_1,k_2,k_3\}}\qquad;\qquad N=2n+1+k_t\quad;\qquad k_t = k_1+k_2+k_3\label{andil}
\ee
The $k_1,k_2$ and $k_3$ are some integers. The above expression can be written more explicitly as 
\be
A_n^{(p,q)}&=&a_n^{(p,q)}J_{2n+1+k_t\{k_i\}}\non\\
&=&a_n^{(p,q)} \int_0^\infty dx\, x^{\f{d}{2}+2n+k_t} \prod_{j=1}^3 p_j^{\Delta_j-\f{d}{2}+k_j} K_{\Delta_j-\f{d}{2}+k_j} (xp_j)
\ee
The expressions for $B_{1,n}^{(p,q)}$ and $B_{2,n}^{(p,q)}$ will also be given by similar expressions
\be
B_{1;n}^{(p,q)}=b_{1;n}^{(p,q)}J_{N\{k_1,k_2,k_3\}}\qquad,\qquad B_{2;n}^{(p,q)}=b_{2;n}^{(p,q)}J_{N\{k_1,k_2,k_3\}}\qquad,\qquad N=2n+1+k_t\label{andil1}
\ee
The undetermined quantities in \eqref{andil} and \eqref{andil1} are $k_i, a_n^{(p,q)}, b_{1;n}^{(p,q)}$ and $b_{2;n}^{(p,q)}$. These will be fixed by making use of the special conformal Ward identity. In particular, the primary Ward identities fix the integer parameters $k_i$. The secondary Ward identities imply some relations between the coefficients $a_n^{(p,q)}, b_{1;n}^{(p,q)}$ and $b_{2;n}^{(p,q)}$.

It turns out that the primary and secondary Ward identities are very involved and cannot be solved easily for arbitrary spin $s$. In the next sections, we shall solve these equations for spins $0,1$ and $2$. However, before that, we describe some useful symmetry properties of the form factors which turn out to be useful in finding the solution and providing consistency checks when $\Delta_1=\Delta_3$.

\subsection{Symmetry properties of form factors} \label{sec:symm_prop}

In our discussion so far, we considered the general case where $s_1 \neq s_2$. When $s_1 = s_2$ (with the operators not necessarily the same) the equations exhibit a simple transformation property under the exchange $(\epsilon_1,p_1)\leftrightarrow (\epsilon_3,p_3)$, which amounts to
\be
z \leftrightarrow z, \qquad \xi_1 \leftrightarrow \zeta_1, \qquad \xi_2 \leftrightarrow - \zeta_2\, .
\ee
It is easy to see that the equations coming from the Ward identities remain invariant under this exchange if the functions $A, B_1$ and $B_2$ transform as 
 \be
A(p_1,p_2,p_3,\xi_1,\xi_2,\zeta_1,\zeta_2,z)&\rightarrow & \mp A(p_3,p_2,p_1,\zeta_1,-\zeta_2,\xi_1,-\xi_2,z) \non\\[.3cm]
B_{1}(p_1,p_2,p_3,\xi_1,\xi_2,\zeta_1,\zeta_2,z)&\rightarrow&\pm  B_{2}(p_3,p_2,p_1,\zeta_1,-\zeta_2,\xi_1,-\xi_2,z) \non\\[.3cm]
B_{2}(p_1,p_2,p_3,\xi_1,\xi_2,\zeta_1,\zeta_2,z)&\rightarrow& \pm B_{1}(p_3,p_2,p_1,\zeta_1,-\zeta_2,\xi_1,-\xi_2,z)~. \label{5.68}
\ee
Useful identities for showing the symmetries  under $(\epsilon_1,p_1)\leftrightarrow (\epsilon_3,p_3)$ exchange are
\begin{eqnarray}
(2+d-\Delta_2-\Delta_3+R-F_1+F_2)\frac{\partial}{\partial \zeta_2}&=&(\Delta_1-a) \frac{\partial}{\partial \zeta_2},
\end{eqnarray}
where we have used equation \eqref{dilght}, and the transformation properties
\begin{eqnarray}
E_1    \overset{1\leftrightarrow 3}{\longleftrightarrow}
-E_2 \qquad;\qquad R-F_1\overset{1\leftrightarrow 3}{\longleftrightarrow}R
\end{eqnarray}
The Ward identity equation \eqref{3.52}, which receives contribution from 2-point function, needs special attention. This equation remains invariant under the above exchange only if $g_1\leftrightarrow \pm g_3$ (where the signs correspond to the signs in \eqref{5.68}). 

When $\Delta_1=\Delta_3$, the above transformation property imply useful constraints on the form factors. In this case, the transverse part of 3-point function \eqref{ansatzs}, will be either symmetric or antisymmetric under the above exchange. For the upper sign in \eqref{5.68}, the equation \eqref{ansatzs} remains invariant, whereas for the lower sign, \eqref{ansatzs} picks up an overall minus sign. This property gives constraints on the form factors. By using the series expansions \eqref{25rt1}, \eqref{25rt2} and \eqref{25rt3}, we find
\be
A_n^{(p,q)}(p_1,p_2,p_3)&=& \mp (-1)^{p+q}A_n^{(q,p)}(p_3,p_2,p_1) \non\\[.3cm]
B_{1,n}^{(p,q)}(p_1,p_2,p_3)&=& \pm (-1)^{p+q}B_{2,n}^{(q,p)}(p_3,p_2,p_1) \label{symmcons}
\ee 
Combining the above symmetry constraints with \eqref{andil} and \eqref{andil1}, we find
\be
A_n^{(p,q)}\quad&:&\quad a_n^{(p,q)}J_{2n+1+k_t\{k_1,k_2,k_3\}}=\mp (-1)^{p+q}a_n^{(q,p)}J_{2n+1+k_t\{k_3,k_2,k_1\}}\non\\[.3cm]
B_{1;n}^{(p,q)},B_{2;n}^{(p,q)}\quad&:&\quad b_{1;n}^{(p,q)}J_{2n+1+k_t\{k_1,k_2,k_3\}}=\pm (-1)^{p+q}b_{2;n}^{(q,p)}J_{2n+1+k_t\{k_3,k_2,k_1\}}\label{symconstrs}
\ee
These relations give consistency checks for the solutions of the coefficients $a_n^{(p,q)}, b_{1;n}^{(p,q)} $ and $b_{2;n}^{(p,q)}$. Further, we can also use these to fix $B_1$ from the knowledge of $B_2$ or vice versa. 

In the following, in the case of non-conserved spin-1 fields, we will get only  the solution of the Ward identities that is antisymmetric in the exchange of $(\epsilon_1,p_1)\leftrightarrow (\epsilon_3,p_3)$. This is also consistent with the results obtained in position space where the three point correlation function with identical spin-1 operators $O_1^\mu=O_3^\mu$ vanishes  due to the exchange symmetry of the two vectors.

\section{Solution for $s=0$ }
\label{sec3:s0}
This case has been considered in \cite{Bzowski:2018fql} and will be briefly reviewed here. When the 1st and 3rd operators are scalars, the second and third terms in RHS of \eqref{ansatzs} are absent and the correlator is given by
\be
\mathcal{A}_{0,J,0}^\perp &=&(\epsilon_2\cdot \pi_2\cdot p_1)A
\label{ansatzs1}
\ee
The function $A$ now involves only one form factor
\be
A&=&A_0^{(0,0)}(p_1,p_2,p_3)
\ee
In this case, the primary Ward identities take a very simple form
\be
 (K_1-K_3)A_0^{(0,0)}=0\qquad;\qquad (K_2-K_3)A_0^{(0,0)}=0\, .
\ee
Together with the dilatation W ard identity \eqref{dilght}, and using equations \eqref{jn1} and \eqref{jn4}, we find that the above equations have unique solution
\be
A_0^{(0,0)} = a_0^{(0,0)}\; J_{1\{0,0,0\}}\;.
\ee
We still need to satisfy the secondary Ward identity which, in this case, comes from equation \eqref{3.52} with $n=0$ and is given by the following single equation
\be
&&\hspace*{-.47in}\biggl[\Delta_1-\Delta_3+\f{\Delta_2-1}{p_2^2}(p_1^2-p_3^2)-p_1\f{\p}{\p p_1}+p_3\f{\p}{\p p_3} +\f{(p_3^2-p_1^2)}{p_2}\f{\p}{\p p_2}  \biggl]A_0^{(0,0)} \nonumber\\[.3cm]
&=&\frac{2(\Delta_2-1)}{p_2^2} \,a_0\,\delta_{\Delta_1,\Delta_3} \Big[g_1\,p_3^{2\Delta_1 -d}+g_3\,p_1^{2\Delta_1 -d}\Big] \label{0spinh}
\ee
If $\Delta_1 \not=\Delta_3$, then the right hand side vanishes and the only way to satisfy the above equation is to set $a_0^{(0,0)}=0$. This implies that the 3-point function vanishes identically. This is consistent with the well known result in position space. For $\Delta_1=\Delta_3=\Delta$, the right hand side of \eqref{0spinh} is non zero.  In this case, the equation can be conveniently analysed by taking the limit $p_3\rightarrow 0$ in both sides. Making use of \eqref{jn2}, \eqref{wotder1}, \eqref{wotder} and \eqref{zep} gives the relation between the 2 and 3-point functions to be
\begin{eqnarray}
a_0=- 2^{\frac{d}{2}-4}\,\frac{2\Delta -d}{g_3(d-2)}\, \Gamma\left(\frac{d}{2}\right)\,\Gamma\left(\frac{d-2\Delta}{2}\right)\,\Gamma\left(\frac{2\Delta -d}{2}\right)\, a_0^{(0,0)}
\end{eqnarray}
where we have used $\Delta_2=d-1$. In getting this relation, we have used the standard identity $\Gamma\left(x\right)\Gamma\left(-x\right) =- \frac{ \pi }{x}\; \mbox{Cosec}\left(\pi x \right)$.

 For $\Delta_1=\Delta_3$, the left side of \eqref{0spinh} is antisymmetric in the exchange $p_1\leftrightarrow  p_3$. The right hand side must also have this property. This is possible only if $g_1=-g_3$. This condition must be satisfied for the 3-point function to be non zero and can be ensured if the 1st and 3rd operators are complex conjugates of each other. In other words, the 3 point function will vanish for uncharged scalar operators.

\section{Solution for $s=1$ }
\label{sec4:s1}
If the two non-conserved operators have spin $1$, the 3-point function is given by
\be
\mathcal{A}_{1,J,1}^\perp &=&(\epsilon_2\cdot \pi_2\cdot p_1)A+ (\epsilon_2\cdot \pi_2\cdot \epsilon_1)B_1 +(\epsilon_2\cdot \pi_2\cdot \epsilon_3)B_2
\ee
where,
\be
A&=& A_0^{(0,0)} z +A_1^{(0,0)}\xi_2\zeta_2+A_1^{(0,1)}\xi_2\zeta_1+A_1^{(1,0)}\xi_1\zeta_2+A_1^{(1,1)}\xi_1\zeta_1\non\\[.3cm]
B_1&=&B_{1;0}^{(0,0)}\xi_2+B_{1;0}^{(1,0)}\xi_1\non\\[.3cm]
B_2&=&B_{2;0}^{(0,0)}\zeta_2+B_{2;0}^{(0,1)}\zeta_1
\ee
Thus, in this case, there are a total of 9 form factors and hence 9 undetermined coefficients. However, not all of these are independent. As we shall see below, this 3-point function is characterised by only 2 independent constants for $\Delta_1\not=\Delta_3$ and by 3 independent constants for $\Delta_1=\Delta_3$.

\subsection{Equations for form factors}
The equations for the form factors can be obtained by substituting the expansions \eqref{25rt1}-\eqref{25rt3} in the equations given in section \ref{prisecwi} for $s=1$. 
\subsubsection{Primary Ward Identities}

For the case we are considering in this section, the terms can be proportional to either $z$ or $z^0$. At $O(z)$, we have following equations
\be
(\epsilon_2\cdot\pi_2\cdot p_1)(b\cdot p_1)&:&  0\;=\;  (K_1-K_3)A^{(0,0)}_0\label{bp1a1s1}\\[.4cm]
(\epsilon_2\cdot\pi_2\cdot p_1)(b\cdot p_2)&:& 0\;=\;
 (K_2-K_3)A_0^{(0,0)} +2A_1^{(1,1)}\label{f1p2zs1}
\ee
At $O(z^{0})$, we have following equations
\be
(\epsilon_2\cdot\pi_2\cdot p_1)(b\cdot p_1)&:&0\;=\;     (K_1-K_3)A_{1}^{(0,0)}-\f{2}{p_1}\f{\p A_1^{(0,0)}}{\p p_1} +\f{2}{p_3}\f{\p A_1^{(0,0)}}{\p p_3}\non\\[.3cm]
&& 0\;=\; (K_1-K_3)A_{1}^{(0,1)}+\f{2}{p_3}\f{\p A_1^{(0,1)}}{\p p_3} \non\\[.3cm]
&& 0\;=\;  (K_1-K_3)A_{1}^{(1,0)}-\f{2}{p_1}\f{\p A_1^{(1,0)}}{\p p_1}\label{gty7h}\\[.3cm]
&&  0\;=\; (K_1-K_3)A_{1}^{(1,1)}\label{f1p1s1}
\ee
\be
(\epsilon_2\cdot\pi_2\cdot p_1)(b\cdot p_2)&:&  0\;=\;  (K_2-K_3)A_{1}^{(0,0)} -\f{2}{p_1}\f{\p A_1^{(0,1)}}{\p p_1}+\f{2}{p_3}\f{\p (A_1^{(0,0)}+A_1^{(1,0)})}{\p p_3} \non\\[.3cm]
&&  0\;=\;  (K_2-K_3)A_{1}^{(0,1)}+\f{2}{p_3}\f{\p (A_1^{(0,1)}+A_1^{(1,1)})}{\p p_3}  \non\\[.3cm]
&&  0\;=\; (K_2-K_3)A_{1}^{(1,0)}-\f{2}{p_1}\f{\p A_1^{(1,1)}}{\p p_1}\label{gty8h}\\[.3cm]
&& 0\;=\;  (K_2-K_3)A_{1}^{(1,1)}\label{f1p2s1}
\ee
\be
(\epsilon_2\cdot\pi_2\cdot \epsilon_1)(b\cdot p_1)&:& 0\;=\; 
(K_1-K_3)B_{1,0}^{(1;0)}-2A_1^{(1,1)}\non\\[.4cm]
&& 0\;=\; (K_1-K_3)B_{1;0}^{(0,0)}+\f{2}{p_3}\f{\p B_{1,0}^{(0,0)}}{\p p_3}-2A_1^{(0,1)}\label{f2p1s1}\\[.7cm]
(\epsilon_2\cdot\pi_2\cdot \epsilon_1)(b\cdot p_2)&:&
 0\;=\;  (K_2-K_3)B_{1,0}^{(1;0)} -2A_1^{(1,1)}\non\\[.3cm]
&&  0\;=\; (K_2-K_3)B_{1;0}^{(0,0)}+\f{2}{p_3}\f{\p (B_{1,0}^{(0,0)}+B_{1;0}^{(1,0)})}{\p p_3} -2A_1^{(0,1)}\label{f2p2s1}
\ee
\be
(\epsilon_2\cdot\pi_2\cdot \epsilon_3)(b\cdot p_1)&:& 0\;=\; 
(K_1-K_3)B_{2,0}^{(0;1)}-2A_1^{(1,1)}\non\\[.3cm]
&& 0\;=\; (K_1-K_3)B_{2;0}^{(0,0)}-2A_1^{(1,0)}-\f{2}{p_1}\f{\p B_{2,0}^{(0,0)}}{\p p_1}\label{f3p1s1}\\[.4cm]
(\epsilon_2\cdot\pi_2\cdot \epsilon_3)(b\cdot p_2)&:& 0\;=\;  (K_2-K_3)B_{2;0}^{(0,0)}-\f{2}{p_1}\f{\p B_{2,0}^{(0,1)}}{\p p_1} \non\\[.3cm]
 &&  0\;=\; (K_2-K_3)B_{2;0}^{(0,1)}\label{f3p2s1}
\ee
\subsubsection{Secondary Ward Identities}
We first consider the equations coming from $b\cdot \epsilon_1$ and $b\cdot\epsilon_3$. At $O(z)$, there are no equations. On the other hand, at $O(z^0)$, we have following 6 equations

\vspace*{.3cm}\noindent{$(\epsilon_2\cdot\pi_2\cdot p_1)(b\cdot \epsilon_1)$ }
\be
0&=& (2+2d-\Delta_2-\Delta_3)A_1^{(0,0)} +(\Delta_2-d)A_1^{(0,1)} +\f{1}{p_1}\f{\p A_0^{(0,0)}}{\p p_1}+\f{1}{p_3} \f{\p A_0^{(0,0)}}{\p p_3}\non\\
&&+\f{p_1\cdot p_2}{p_1}\f{\p A_1^{(0,1)}}{\p p_1}+\f{1}{p_1}\f{\p B_{1,0}^{(0,0)}}{\p p_1}-p_2\f{\p A_1^{(0,1)}}{\p p_2} +p_1\f{\p A_1^{(0,0)}}{\p p_1}+p_2\f{\p A_1^{(0,0)}}{\p p_2}+p_3\f{\p A_1^{(0,0)}}{\p p_3}\non\\[.6cm]
0&=&(2+2d-\Delta_2-\Delta_3)A_1^{(1,0)} +(\Delta_2-d-1)A_1^{(1,1)} -\f{1}{p_1}\f{\p A_0^{(0,0)}}{\p p_1}+\f{p_1\cdot p_2}{p_1}\f{\p A_1^{(1,1)}}{\p p_1} \non\\
&&+\f{1}{p_1}\f{\p B_{1,0}^{(1,0)}}{\p p_1}-p_2\f{\p A_1^{(1,1)}}{\p p_2} +p_1\f{\p A_1^{(1,0)}}{\p p_1}+p_2\f{\p A_1^{(1,0)}}{\p p_2}+p_3\f{\p A_1^{(1,0)}}{\p p_3}\non
\ee
{$(\epsilon_2\cdot\pi_2\cdot p_1)(b\cdot \epsilon_3)$ }
\be
 0&=& (\Delta_3-1)A_1^{(0,0)} +(d-\Delta_2)A_1^{(1,0)} +\f{1}{p_1}\f{\p A_0^{(0,0)}}{\p p_1} -\f{p_2\cdot p_3}{p_3}\f{\p A_1^{(1,0)}}{\p p_3}\non\\
&&+\f{1}{p_3}\f{\p B_{2,0}^{(0,0)}}{\p p_3}+p_2\f{\p A_1^{(1,0)}}{\p p_2} +\f{1}{p_3} \f{\p A_0^{(0,0)}}{\p p_3}\non\\[.6cm]
0&=&(\Delta_3-1)A_1^{(0,1)} +(d-\Delta_2+1)A_1^{(1,1)} +\f{1}{p_3}\f{\p A_0^{(0,0)}}{\p p_3}-\f{p_2\cdot p_3}{p_3}\f{\p A_1^{(1,1)}}{\p p_3} \non\\
&&+\f{1}{p_3}\f{\p B_{2,0}^{(0,1)}}{\p p_3}+p_2\f{\p A_1^{(1,1)}}{\p p_2}\non
\ee
{$(\epsilon_2\cdot\pi_2\cdot \epsilon_1)(b\cdot\epsilon_3)$ }
\be
0&=&-\f{p_1^2-p^2_2- p_3^2}{2p_3}\f{\p B_{1,0}^{(1,0)}}{\p p_3} +p_2\f{\p B_{1,0}^{(1,0)}}{\p p_2}+A_0^{(0,0)}+(\Delta_3-1)B_{1,0}^{(0,0)}+(d-\Delta_2)B_{1,0}^{(1,0)}+B_{2,0}^{(0,1)}\non
\ee
{$(\epsilon_2\cdot\pi_2\cdot \epsilon_3)(b\cdot\epsilon_1)$ }
\be
0&=&\f{p_1\cdot p_2}{p_1}\f{\p B_{2,0}^{(0,1)}}{\p p_1}  +p_1\f{\p B_{2,0}^{(0,0)}}{\p p_1}+p_2\f{\p (B_{2,0}^{(0,0)}-B_{2,0}^{(0,1)})}{\p p_2}+p_3\f{\p B_{2,0}^{(0,0)}}{\p p_3}\non\\[.3cm]
&&+A_0^{(0,0)}+(2d-\Delta_2-\Delta_3)B_{2,0}^{(0,0)}+(\Delta_2-d)B_{2,0}^{(0,1)}-B_{1,0}^{(1,0)}\non
\ee
Finally, we consider the secondary Ward identities coming from the coefficient {$(\epsilon_2\cdot\pi_2\cdot b)$ }. There are a total of 5 equations coming from this. Four of these equations are inhomogeneous and one is homogeneous.
At $O(z)$, we have a single equation
\be
&&\f{(p_3^2-p_1^2-p_2^2)}{p_2}\f{\p A_0^{(0,0)}}{\p p_2}+\f{(\Delta_2-1)(p_1^2+p_2^2-p_3^2)}{p_2^2}A_0^{(0,0)}
 -2p_1\f{\p A_0^{(0,0)}}{\p p_1} +2(\Delta_1-d)A_0^{(0,0)}  +2(B_{1,0}^{(1,0)}+B_{2,0}^{(0,1)})\nonumber\\
&&=\delta_{\Delta_1,\Delta_3} \frac{2(d-2)}{p_2^2}\,\left(g_1\,p_1^{2\Delta-d} +g_3\,p_3^{2\Delta -d}\right)  \label{2.3.7}
\ee
where we have used equation \eqref{3.52} and the expression of the two point function for spin-1 operators given in equation \eqref{2.16}.

At $O(z^0)$, we have following 4 equations  
\be
&&\f{(p_3^2-p_2^2-p_1^2)}{p_2}\f{\p A_1^{(1,1)}}{\p p_2}+\f{2}{p_2} \f{\p}{\p p_2}(B_{1,0}^{(1,0)}+B_{2,0}^{(0,1)}) + \f{(\Delta_2-1)(p_1^2+p_2^2-p_3^2)}{p_2^2}A_1^{(1,1)}\non \\[.2cm]
&&-2p_1\f{\p A_1^{(1,1)}}{\p p_1}+2(\Delta_1-d-1) A_1^{(1,1)} - \f{2(\Delta_2-1)}{p_2^2} (B_{1,0}^{(1,0)}+B_{2,0}^{(0,1)}) \;\;=\;\;0
\label{5.83}
\end{eqnarray}
\begin{eqnarray}
&&\f{(p_3^2-p_2^2-p_1^2)}{p_2}\f{\p A_1^{(0,1)}}{\p p_2}+\f{2}{p_2} \f{\p}{\p p_2}(B_{1,0}^{(0,0)})+\f{2}{p_3} \f{\p}{\p p_3}(B_{2,0}^{(0,1)}) + \f{(\Delta_2-1)(p_1^2+p_2^2-p_3^2)}{p_2^2}A_1^{(0,1)} \non\\[.2cm]
&&-2p_1\f{\p A_1^{(0,1)}}{\p p_1}+2(\Delta_1-d) A_1^{(0,1)} - \f{2(\Delta_2-1)}{p_2^2}B_{1,0}^{(0,0)}+2A_1^{(1,1)} \non\\[.4cm]
&&=\delta_{\Delta_1,\Delta_3}\, \frac{2(d-2)(d-2\Delta_1)}{p_2^2(\Delta_1-1)}\,a_0\,p_3^{2\Delta_1 -d-2}
\end{eqnarray}
\begin{eqnarray}
&&\f{(p_3^2-p_2^2-p_1^2)}{p_2}\f{\p A_1^{(1,0)}}{\p p_2}+\f{2}{p_2} \f{\p}{\p p_2}(B_{2,0}^{(0,0)})-\f{2}{p_1} \f{\p}{\p p_1}(B_{1,0}^{(1,0)}) + \f{(\Delta_2-1)(p_1^2+p_2^2-p_3^2)}{p_2^2}A_1^{(1,0)} \non\\[.2cm]
&&-2p_1\f{\p A_1^{(1,0)}}{\p p_1}+2(\Delta_1-d-2) A_1^{(1,0)} - \f{2(\Delta_2-1)}{p_2^2}B_{2,0}^{(0,0)}+2A_1^{(1,1)} \non\\[.4cm]
&&=-\delta_{\Delta_1,\Delta_3}\, \frac{2(d-2)(d-2\Delta_1)}{p_2^2\,(\Delta_1-1)}\,a_0\,p_1^{2\Delta_1-d-2}
\end{eqnarray}
\begin{eqnarray}
&&\f{(p_3^2-p_2^2-p_1^2)}{p_2}\f{\p A_1^{(0,0)}}{\p p_2}-\f{2}{p_1} \f{\p}{\p p_1}(B_{1,0}^{(0,0)})+\f{2}{p_3} \f{\p}{\p p_3}(B_{2,0}^{(0,0)}) + \f{(\Delta_2-1)(p_1^2+p_2^2-p_3^2)}{p_2^2}A_1^{(0,0)} \non\\[.2cm]
&&-2p_1\f{\p A_1^{(0,0)}}{\p p_1}+2(\Delta_1-d-1) A_1^{(0,0)} +2A_1^{(0,1)}+2A_1^{(1,0)}\nonumber\\
&&= \delta_{\Delta_1,\Delta_3} \frac{2(d-2)(d-2\Delta_1)}{p_2^2 (\Delta_1-1)}\, a_0\,\left(g_1\,p_1^{2\Delta_1-d-2}+g_3\,p_3^{2\Delta_1 -d -2}\right)
\ee
 The equation \eqref{5.83} remains a homogeneous differential equation since it comes from the tensor structure involving $\xi_1\,\zeta_1$. Hence, as discussed below equation \eqref{3.52}, it does not have a source term coming from the two-point function. All the other equations coming from $(\epsilon_2\cdot\pi_2\cdot b)$ are inhomogeneous.

We now turn to solving these equations. 

\subsection{Solution of Primary Ward Identities}

The primary Ward identities can be solved recursively by making use of \eqref{jn2} and \eqref{jn4}. We shall illustrate the procedure by solving for $A_1^{(1,1)}, A_0^{(0,0)}$ and $A_1^{(1,0)}$. Using equations \eqref{f1p1s1}, \eqref{f1p2s1} and \eqref{jn4}, we find that $A_1^{(1,1)}$ must have the following form
\be
A_1^{(1,1)} = a_1^{(1,1)} J_{N\{0,0,0\}}
\ee
where, $a_1^{(1,1)}$ is some constant. The $N$ is fixed by demanding the correct scaling property. Comparing \eqref{andilr5} and \eqref{jn1} and noting the above solution, we find $N=3$. Next, we consider equations \eqref{bp1a1s1} and \eqref{f1p2zs1} for $A_0^{(0,0)}$, with  \eqref{f1p2zs1} being inhomogeneous. Again using \eqref{jn4}, the solution of the homogeneous equation has the form $a_0^{(0,0)}J_{N\{0,0,0\}}$ where $N$ is again fixed by demanding the correct scaling property to be equal to $1$. To obtain the general solution of the inhomogeneous equation \eqref{f1p2zs1}, we substitute the solution for $A_1^{(1,1)}$ in \eqref{f1p2zs1} to get
\be
(K_2-K_3)A_0^{(0,0)}=-2a_1^{(1,1)}J_{3\{0,0,0\}}
\ee
Comparing the above equation with \eqref{jn4}, we find that a particular solution is given by $-a_1^{(1,1)}J_{2\{0,1,0\}}$. Therefore, the general solution of \eqref{f1p2zs1} is given by
\be
A_0^{(0,0)}= -a_{1}^{(1,1)} J_{2\{0,1,0\}}+ a_0^{(0,0)}J_{1\{0,0,0\}} 
\ee
Next, we consider equations \eqref{gty7h} and \eqref{gty8h} for $A_1^{(1,0)}$. The general solution of the system of homogeneous 
equations corresponding to \eqref{gty7h} and \eqref{gty8h}  is 
\begin{equation}
a_{1}^{(1,0)} J_{2\{-1,0,0\}}, 
\end{equation}
where $a_{1}^{(1,0)}$ is a constant.
To obtain a particular solution, we substitute the solution of $A_1^{(1,1)}$ in \eqref{gty8h} and use \eqref{jn2} to get
\be
(K_2-K_3)A_1^{(1,0)} = 2a_1^{(1,1)}J_{4\{-1,0,0\}} 
\ee
Again using \eqref{jn4}, we find that the general solution of the above equation is
\be
A_1^{(1,0)} = -a_{1}^{(1,1)} J_{3\{-1,1,0\}}+  a_{1}^{(1,0)} J_{2\{-1,0,0\}}
\ee
This completes the analysis for $A_1^{(1,1)}, A_0^{(0,0)}$ and $A_1^{(1,0)}$. The other equations can be analysed in similar manner. The final solution is given by
\be
 A_{1}^{(1,1)} &=& a_1^{(1,1)}J_{3\{0,0,0\}}\non\\[.2cm]
   A_{1}^{(1,0)} &=&-a_{1}^{(1,1)} J_{3\{-1,1,0\}}+  a_{1}^{(1,0)} J_{2\{-1,0,0\}} \non\\[.2cm]
  A_{1}^{(0,1)} &=& a_{1}^{(1,1)} J_{3\{0,1,-1\}}+ a_{1}^{(0,1)} J_{2\{0,0,-1\}}  \non\\[.2cm]
 A_{0}^{(0,0)} &=& -a_{1}^{(1,1)} J_{2\{0,1,0\}}+ a_0^{(0,0)}J_{1\{0,0,0\}}  \non\\[.2cm]
  A_{1}^{(0,0)} &=& -a_{1}^{(1,1)} J_{3\{-1,2,-1\}}+ a_1^{(0,0)}J_{1\{-1,0,-1\}}+(a_1^{(1,0)} -a_1^{(0,1)})J_{2\{-1,1,-1\}}    \non\\[.2cm]
   B_{1;0}^{(0,0)} &=&  -a_{1}^{(1,1)} J_{2\{0,1,0\}}+ b_{1,0}^{(0,0)}J_{1\{0,1,-1\}}+(b_{1,0}^{(0,0)}-b_{1,0}^{(1,0)})J_{1\{1,0,-1\}} +(b_{1,0}^{(0,0)}-b_{1,0}^{(1,0)} - a_{1}^{(0,1)})J_{1\{0,0,0\}} \non\\[.3cm]
       B_{1;0}^{(1,0)} &=& -a_{1}^{(1,1)}J_{2\{0,0,1\}}+ b_{1;0}^{(1,0)}J_{1\{0,0,0\}} \;\non\\[.3cm]
  B_{2;0}^{(0,0)} &=&  -a_{1}^{(1,1)} J_{2\{0,1,0\}}+ b_{2,0}^{(0,0)}J_{1\{-1,1,0\}}+ ( b_{2;0}^{(0,0)}+ b_{2;0}^{(0,1)})J_{1\{-1,0,1\}} +(b_{2,0}^{(0,0)}+ b_{2,0}^{(0,1)} + a_{1}^{(1,0)})J_{1\{0,0,0\}} \non\\[.3cm]
  B_{2;0}^{(0,1)} &=& a_{1}^{(1,1)}J_{2\{1,0,0\}}+ b_{2;0}^{(0,1)}J_{1\{0,0,0\}} \label{2316a}
\ee
From the above result, we see that each form factor is determined up to one free parameter. Thus, there are a total of 9 independent parameters in the solution of primary Ward identities. However, from the position space analysis\cite{Costa:2011mg}, it is known that the 3-point function involving a conserved current and two spin 1 operators has only 2 independent parameters (for $\Delta_1\not=\Delta_3$) or 3 independent parameters (for $\Delta_1=\Delta_3$). This means that there must be some additional relations between the above 9 parameters. We shall determine these relations by making use of the secondary identities.

\subsection{Solution of Secondary Ward Identities}

In order to solve the secondary equations, we first eliminate the momenta in the denominators by multiplying the equation with a suitable power of momenta. Further, the secondary equations only involve single derivatives of momenta. These derivatives may be eliminated by making use of identity \eqref{jn2}. The terms of the form $p_i^2J_{N\{k_1,k_2,k_3\}}$ may then be eliminated using identity \eqref{wotder}. This leaves us with a linear combination of different triple K integrals. A convenient way to solve this equation is to use the identity \eqref{wotder1} to convert all the triple K integrals of the form $J_{N'\{k_1,k_2,k_3\}}$ to the triple K integrals of the form $J_{N\{k_1,k_2,k_3\}}$, with $N$ being the highest value among all $N'$ appearing in the equation. Doing this gives algebraic equations involving the 9 undetermined coefficients and hence fixes the coefficients. This procedure may be used to solve all the homogeneous secondary equations.
However, this procedure is not sufficient for solving the inhomogeneous secondary equations. These are most conveniently solved by taking one of the momenta to be zero \cite{Bzowski:2013sza}. In fact, analysing the equations in this limit is another way to also solve the homogeneous secondary equations. When one of the momenta is send to zero, say, $p_3\rightarrow0$, then the equations take a simple form. The relevant identity to take this limit is given in equation \eqref{zep}. In this limit the secondary equations give algebraic equations involving the undetermined parameters. We have explicitly verified that the two procedures give the same results for the homogeneous equations once the regularization procedure is carefully implemented, as we shall see later. 

Now, there are two cases depending upon the conformal dimensions of the 1st and 3rd operators. We shall consider these cases separately in what follows. When $\Delta_1\neq \Delta_3$ all the secondary Ward identities are homogeneous differential equations. When $\Delta_1=\Delta_3$, instead, there are only seven homogeneous differential equations that have to be solved separately from the 4 inhomogeneous Ward identities. As mentioned earlier, the homogeneous equations give relations between the coefficients appearing in the solution of primary equations. On the other hand, the inhomogeneous equations relate the 2-point coefficient with the 3-point coefficients.

\subsubsection{Case 1: $\Delta_1\not=\Delta_3$}
For the case, $\Delta_1\not=\Delta_3$, the 11 secondary equations can be solved to express seven out of the nine coefficients in terms of two undetermined coefficients as
\be
 a_{0}^{(0,0)} &=& -(2-d)a_{1}^{(1,1)} \non\\[.2cm]
  a_{1}^{(0,0)} &=&- \f{(d-2) \left(\Delta _1+\Delta _3-2\right) \left(d+\Delta _1-\Delta _3-4\right)}{2(\Delta_1-1)}a_{1}^{(1,1)} - \f{(d-2)(\Delta_1+\Delta_3-2)}{(\Delta_1-1)(\Delta_3-1)}b_{1;0}^{(1,0)}\non\\[.2cm]
  a_{1}^{(0,1)} &=&-(d-2) a_{1}^{(1,1)}-\frac{1}{\Delta _3-1}b_{1;0}^{(1,0)}\non\\[.2cm]
    a_{1}^{(1,0)} &=&-\f{(d-2)(2-\Delta_1-\Delta_3)+(\Delta_3-\Delta_1)(\Delta_1+\Delta_3)}{2(\Delta_1-1)}a_{1}^{(1,1)}     +\f{1}{(\Delta_1-1)}b_{1;0}^{(1,0)}\non\\[.3cm]
b_{1;0}^{(0,0)} &=& -\frac{ \left(d-\Delta _1-\Delta _3\right)}{2 \left(\Delta _3-1\right)}b_{1;0}^{(1,0)}\non\\[.2cm]
b_{2;0}^{(0,0)} &=&\frac{\left(\Delta _1-\Delta _3\right) \left(-d+\Delta _1+\Delta _3\right) \left(-d+\Delta _1+\Delta _3+2\right)}{4 \left(\Delta _1-1\right)}a_{1}^{(1,1)}-\frac{ \left(d-\Delta _1-\Delta _3\right)}{2 \left(\Delta _1-1\right)}b_{1;0}^{(1,0)} \non\\[.2cm]
b_{2;0}^{(0,1)} &=& \frac{1}{2} \left(\Delta _1-\Delta _3\right)  \left(d-\Delta _1-\Delta _3-2\right)a_{1}^{(1,1)}-b_{1;0}^{(1,0)} 
\ee
The above solution shows that the 3-point function involving a conserved current and two spin 1 fields with different conformal dimension has two free parameters. We have chosen $a_{1}^{(1,1)} $ and $b_{1;0}^{(1,0)} $ as independent pararmeters. However, clearly, it is also possible to choose any other two coefficients (out of 9) as free parameters.

In this case, namely for $\Delta_1\not=\Delta_3$, none of the triple K integrals appearing in the solution \eqref{2316a} of primary identities diverge for generic non integer values of $\Delta_1$ and $\Delta_3$. This can be checked using the divergence condition given in \eqref{g31}. If we use the zero momentum limit $p_3\rightarrow0$ to solve the secondary equations, we find the same solution as above.

\subsubsection{Case 2: $\Delta_1=\Delta_3=\Delta$} 

In this case, some of the triple K integrals appearing in the solution of primary equations, namely $J_{1\{0,1,-1\}},J_{1\{1,0,-1\}}, J_{1\{-1,1,0\}} $ and $J_{1\{-1,0,1\}}$ are divergent. The condition \eqref{g31} is satisfied for $k=0$ for some choice of signs for each of these triple K integrals. Hence, we need to regularise these integrals by shifting the parameters $\Delta$ and $d$ as described in appendix \ref{regtripleK}. We shall analyse the regularisation of these integrals in more detail below. However, assuming that regularisation has been done, we can use the same approach as above to solve all the homogeneous equations without the two points as source term. Using identities \eqref{jn2}, \eqref{jn4} and \eqref{wotder1}, we find the following solution of the homogeneous secondary equations
\be
 a_{0}^{(0,0)} &=& (d-2)\Delta a_{1}^{(1,1)}-(\Delta-1)a_1^{(1,0)} +b_{1;0}^{(1,0)} \non\\[.2cm]
     a_{1}^{(0,0)} &=&2(d-2)\Delta a_{1}^{(1,1)}  -(2\Delta+d-4)a_1^{(1,0)}   +\f{(2\Delta-d)}{(\Delta-1)}b_{1;0}^{(1,0)}\non\\[.2cm]
  a_{1}^{(0,1)} &=&-a_{1}^{(1,0)}\non\\[.3cm]
    b_{1;0}^{(0,0)}&=&\f{(2\Delta-d)}{2(\Delta-1)}b_{1;0}^{(1,0)}\non\\[.2cm]
b_{2;0}^{(0,0)}&=&b_{1;0}^{(0,0)}=\f{(2\Delta-d)}{2(\Delta-1)}b_{1;0}^{(1,0)}\non\\[.2cm]
b_{2;0}^{(0,1)}&=&-b_{1;0}^{(1,0)}  \label{2349afnewd}
\ee
Now, there are three independent parameters for our 3-point function as expected. We have chosen $a_1^{(1,1)}, a_1^{(1,0)}$ and $b_{1;0}^{(1,0)}$ as 3 independent parameters. However, clearly, we can also choose some other combination of three parameters instead of these. We also note that the above solution satisfies the relation \eqref{symconstrs} with the lower sign. This means that this 3-point function is anti-symmetric under the exchange of $(\epsilon_1\,p_1) \leftrightarrow (\epsilon_3,\,p_3)$. We do not find any solution which is symmetric under this exchange. This is in agreement with the position space results, where the 3-point function of same operators vanishes if it is symmetric under this exchange \cite{Costa:2011mg}.

Again, we can also solve the homogeneous secondary equations by analysing them in the limit $p_3\rightarrow0$ making use of equation \eqref{zep}. In this case, by carefully applying  the regularization procedure, as discussed in 
detail in the next section, we get the same result as before. The difference between the two approaches is that in the first approach the homogeneous equations are solved to all order in the regulator $\epsilon$ and the limit $\epsilon\rightarrow 0$ gives a finite non divergent answer. In the kinematic region  $p_3\rightarrow 0$, instead, the solution is obtained as a perturbative series in the regulator. The divergent terms have to be vanishing, while the finite $O(\epsilon^0)$ terms are determined by solving the homogeneous Ward identities  at this order in the $\epsilon$-expansion. We discuss this second approach in the next section. However, before that, we consider the inhomogeneous secondary equations which can be solved by taking one of the momentum to be zero. We shall consider the kinematic region $p_3\rightarrow 0$. Due to the symmetry of the solution,  we can find a non zero result only when $g_1=-g_3\equiv g$. Again, this implies that the spin 1 operators must be charged. Now, there are four inhomogeneous secondary differential equations to solve but they give only one condition which turns out to be
  \begin{eqnarray}
 a_0&=& - 2^{\frac{d}{2} -4}\, \frac{(d-2\Delta)}{g_3(d-2)}\,\Gamma \left(\frac{ d-2\Delta}{2}\right)\Gamma \left(\frac{ 2\Delta-d}{2}\right)\,\Gamma\left(\frac{d}{2}\right)\nonumber\\[.2cm]
 &&\left[(\Delta-1)\left(a_1^{(0,1)}+(d-2) a_1^{(1,1)}\right)+\alpha_{1,0}^{(1,0)}
 \right]\label{gtr5d}
\end{eqnarray}
where $\alpha_{1,0}^{(1,0)}$ is defined in equation \eqref{fgt6} below. To leading order in the regulator, we can replace $\alpha_{1,0}^{(1,0)}$ with $b_{1,0}^{(1,0)}$ in \eqref{gtr5d}. In getting the above relation, we have also  regularized the two point function with the prescription  given in equation \eqref{shiftpara} before taking the zero momentum limit.
Thus, as expected, the solution of the inhomogeneous secondary Ward identities relate the coefficient of the two-point correlation function with those of the three points without giving any further condition.

\subsection{Regularisation for $\Delta_1=\Delta_3=\Delta$}
\label{regularization}
For non-integer dimensions $\Delta_1$ and $\Delta_3$, which is the case we consider in this paper, most of the triple K integrals appearing in the solution \eqref{2316a} are well behaved and give finite results (possibly after an analytic continuation). However, some of the triple K integrals satisfy the divergence condition \eqref{g31} for $\Delta_1=\Delta_3$. As mentioned earlier, these triple K integrals are $J_{1\{0,1,-1\}},J_{1\{1,0,-1\}}, J_{1\{-1,1,0\}} $ and $J_{1\{-1,0,1\}}$ whose regularisation has been considered in appendix \ref{dregu4}. In general, for integer dimensions, the condition \eqref{g31} may also be satisfied for other triple K integrals. However, in this paper, we only focus on non-integer dimensions. 

It turns out that even though the above triple K integrals diverge individually, the divergences cancel for the combination in which they appear in the 3-point function. To see this, we start by noting that these divergent triple K integrals appear in $B_{1;0}^{(0,0)}$ and $B_{2;0}^{(0,0)}$. 
Hence, we need to focus on these form factors. Using the relations given in \eqref{2349afnewd}, the combination involving the divergent triple K integrals which appear in these form factors can be written as 
\be
D_1&=& b_{1;0}^{(1,0)}\biggl[\f{(2\Delta-d)}{2(\Delta-1)}J_{1\{0,1,-1\}}+\f{(2-d)}{2(\Delta-1)}J_{1\{1,0,-1\}}\biggl]\label{combinwdf4a}\\[.3cm]
  D_2&=&b_{1;0}^{(1,0)}\biggl[\f{(2\Delta-d)}{2(\Delta-1)}J_{1\{-1,1,0\}}+\f{(2-d)}{2(\Delta-1)}J_{1\{-1,0,1\}}\biggl]\label{combinwdf4}
\ee
The $D_1$ and $D_2$ correspond to the middle two terms in $B_{1;0}^{(0,0)}$ and $B_{2;0}^{(0,0)}$ respectively in equation \eqref{2316a}. 

Now, as discussed in appendix \ref{regtripleK}, the regularisation is done by shifting the parameters $d$ and $\Delta_i$ as
\be
 \tilde d = d+2u\epsilon \quad,\quad \tilde\Delta_1 =\Delta_1 +(u+v_1)\epsilon\quad,\quad \tilde\Delta_2 =\Delta_2 +(u+v_2)\epsilon\quad,\quad \tilde\Delta_3 =\Delta_3 +(u+v_3)\epsilon
 \label{shiftpara}
\ee
In our case, $\Delta_2$ corresponds to gauge field. To preserve the gauge invariance in the regulated theory, we need to set $v_2=u$. Further, we also have $\Delta_1=\Delta_3=\Delta$. To maintain this condition also in the regulated theory, we need to have $v_1=v_3$.

Since only $\Delta$ appears in equation \eqref{combinwdf4} and there is no $\Delta_1$ or $\Delta_3$ in the prefactors multiplying the triple K integrals, we shall use $v_1=v_3$ from the beginning. From the regularised expressions given in appendix \ref{dregu4}, we see that the individual triple K integrals will have poles in $(u-v_2)$. However, as we shall see below, these poles disappear when we consider the full combination of the divergent integrals. Hence, we can take the limit $u\rightarrow v_2$ at the end. 

We start with $D_1$. Shifting the parameters $d$ and $\Delta$ and using equation \eqref{kcrt1}, we find that the poles in $\epsilon$ cancel between the two terms of \eqref{combinwdf4a} and we get 
\be
D_1 = 2^{\f{d}{2}-3}b_{1;0}^{(1,0)}\left(\f{d-2\Delta}{\Delta-1}\right)\Gamma \left(\frac{d}{2}-\Delta +1\right) \Gamma \left(\Delta-\frac{d}{2}\right) \Gamma \left(\frac{d}{2}-1\right)p_3^{2\Delta-d-2}  +O(\epsilon)\label{463r}
\ee
We also see that the dependence on the regularisation parameters $u$ and $v_2$ has also disappeared from the $O(\epsilon^0)$ terms. 

In a similar way, shifting the parameters $d$ and $\Delta$ in \eqref{combinwdf4} and using equation \eqref{kcrt2}, the divergent part of $D_2$ is regularised to be
\be
D_2 = 2^{\f{d}{2}-3}b_{1;0}^{(1,0)}\left(\f{d-2\Delta}{\Delta-1}\right)\Gamma \left(\frac{d}{2}-\Delta +1\right) \Gamma \left(\Delta-\frac{d}{2}\right) \Gamma \left(\frac{d}{2}-1\right)p_1^{2\Delta-d-2}  +O(\epsilon)\label{464r}
\ee
Again, the dependence on the regularisation parameters $u$ and $v_2$ has disappeared from the $O(\epsilon^0)$ terms. 

All the other triple K integrals are finite.  Hence, shifting the parameters in all the other terms of the 3-point function will not give rise to any poles in $\epsilon$ or $(u-v_2)$.  Now, since, the divergences have been cancelled, we can now set $\epsilon =0$. The final regularised expression of the form factors is thus given by the same solution as in \eqref{2316a} and \eqref{2349afnewd} except that the naive divergent part of the middle two terms in the expressions of $B_{1;0}^{(0,0)}$ and $B_{2;0}^{(0,0)}$ should be replaced by their regularised expressions, namely, $D_1$ and $D_2$ given in equations \eqref{463r} and \eqref{464r} respectively (with $\epsilon $ set to zero).

Next, we show that solving the secondary equations by sending one of the momenta to zero is consistent with the results given in equation \eqref{2349afnewd}. In solving the secondary identities in this kinematic region, the first step is to note that the divergent triple K integrals in our case give terms of order $O(\f{1}{\epsilon})$ after regularisation. To cancel these divergences, we need to shift the coefficients multiplying these triple K integrals by $\epsilon$. This means that, in our case, we need to write
 \begin{eqnarray}
 &&{b}^{(0,0)}_{1;0}= \alpha_{1;0}^{(0,0)}+\epsilon {\beta}_{1;0}^{(0,0)}\qquad;\qquad b_{1;0}^{(1,0)}=\alpha_{1;0}^{(1,0)} +\epsilon\beta_{1;0}^{(1,0)}\nonumber\\[.3cm]
&&{b}_{2 ;0}^{(0,0)} = {\alpha}_{2;0}^{(0,0)}+\epsilon{\beta}^{(0,0)}_{2;0}\qquad;\qquad b_{2;0}^{(0,1)}={\alpha}_{2;0}^{(0,1)}+\epsilon {\beta}_{2;0}^{(0,1)}\label{fgt6}
\end{eqnarray}
We do not need to shift the other coefficients since the triple K integrals multiplying them are not divergent. Now, substituting \eqref{fgt6} in the secondary equation and taking the limit $p_3\rightarrow 0$, we find that $a$-coefficients are given by the same expression as in \eqref{2349afnewd}. The $O(\epsilon^0)$ coefficients in \eqref{fgt6}, as explained before, is fixed by imposing the  cancellation of the  triple-K integral divergences and hence, we have
\be
\alpha_{1;0}^{(0,0)} =  -\frac{(d-2\Delta)}{2(\Delta-1)} \alpha_{1;0}^{(1,0)}\qquad;\qquad \alpha_{2;0}^{(0,0)} =  \frac{(d-2\Delta)}{2(\Delta-1)} \alpha_{2;0}^{(0,1)}
\ee
For the other coefficients, the secondary equations in $p_3\rightarrow0$ limit give
\begin{eqnarray}
\alpha_{2;0}^{(0,1)}= -\alpha_{1;0}^{(1,0)} ~~;~~ \beta_{2,0}^{(0,0)} =\beta_{2,0}^{(0,1)} \frac{d-2\Delta}{2(\Delta -1)} +\beta_{1;0}^{(1,0)} \frac{(u-v_2)}{2(\Delta-1)}\,
\end{eqnarray}
To preserve the gauge invariance in the regulated theory, we need to set $u=v_2$. Furthermore, by solving the equations in the limit $p_1\rightarrow 0$ or requiring the antisymmetry of the solution in the exchange of $(\epsilon_1,\,p_1) \leftrightarrow (\epsilon_3,\,p_3)$,  we find
\begin{eqnarray}
\beta_{1;0}^{(0,0)}= \beta_{2,0}^{(0,0)}\qquad;\qquad \beta_{1;0}^{(1,0)}= -\beta_{2;0}^{(0,1)}
\end{eqnarray}
Using the above results, we can recover the results for $b$-coefficients given in \eqref{2349afnewd}. Indeed, writing $\alpha_{1;0}^{(1,0)}=b_{1,0}^{(1,0)}-\epsilon \beta_{1,0}^{(1,0)}$, we get
 \begin{eqnarray}
 b_{2;0}^{(0,0)}&=&-\frac{(d-2\Delta)}{2(\Delta-1)}\left( b_{1;0}^{(1,0)}-\epsilon\beta_{1;0}^{(1,0)}-\epsilon \beta_{2,0}^{(0,1)}\right)=-\frac{(d-2\Delta)}{2(\Delta-1)}\, b_{1;0}^{(1,0)}\non\\[.2cm]
 b_{2;0}^{(0,1)}&=&  -b_{1;0}^{(0,1)}+\epsilon\left(  \beta_{1,0}^{(1,0)}+ \beta_{2;0}^{(0,1)}\right)= -b_{1;0}^{(0,1)}\nonumber\\[.2cm]
 b_{1;0}^{(0,0)}&=& -\frac{(d-2\Delta)}{2(\Delta-1)}\left[b_{1;0}^{(1,0)} -\epsilon\left( \beta_{1;0}^{(1,0)}+\beta_{2;0}^{(0,1)}\right)\right]= -\frac{(d-2\Delta)}{2(\Delta-1)}b_{1;0}^{(1,0)} 
 \end{eqnarray}
These results are in agreement with equation \eqref{2349afnewd}.

\section{Solution for $ s=2$}
\label{sec5:s2}
For $s_1=s_3=2$, the 3 point function is given by
\be
\mathcal{A}_{2,J,2}^\perp &=&(\epsilon_2\cdot \pi_2\cdot p_1)A+ (\epsilon_2\cdot \pi_2\cdot \epsilon_1)B_1 +(\epsilon_2\cdot \pi_2\cdot \epsilon_3)B_2
\ee
where, the functions $A, B_1$ and $B_2$ are now parametrized as
\be
A&=&z^2 A_0^{(0,0)}+\zeta _2 \xi _2 z A_1^{(0,0)}+ \frac{1}{2} \zeta _2^2 \xi _2^2 A_2^{(0,0)}+\zeta _1 \xi _2 z A_1^{(0,1)}+\zeta _2 \xi _1 z A_1^{(1,0)}+\zeta _1 \xi _1 z A_1^{(1,1)}\non\\
&&+\frac{1}{2} \zeta _1 \zeta _2 \xi _2^2 A_2^{(0,1)}+\frac{1}{4} \zeta _1^2 \xi _2^2 A_2^{(0,2)}+\frac{1}{2} \zeta _2^2 \xi _1 \xi _2 A_2^{(1,0)}+\frac{1}{2} \zeta _1 \zeta _2 \xi _1 \xi _2 A_2^{(1,1)}+\frac{1}{4} \zeta _1^2 \xi _1 \xi _2 A_2^{(1,2)}\non\\
&&+\frac{1}{4} \zeta _2^2 \xi _1^2 A_2^{(2,0)}+\frac{1}{4} \zeta _1 \zeta _2 \xi _1^2 A_2^{(2,1)}+\frac{1}{8} \zeta _1^2 \xi _1^2 A_2^{(2,2)}\non\\[.3cm]
B_1&=&\xi _1 z {B}_{1;0}^{(1,0)}+\xi _2 z {B}_{1;0}^{(0,0)}+\zeta _2 \xi _2 \xi _1 {B}_{1;1}^{(1,0)}+\zeta _1 \xi _2 \xi _1 {B}_{1;1}^{(1,1)}+\zeta _2 \xi _2^2 {B}_{1;1}^{(0,0)}+\zeta _1 \xi _2^2 {B}_{1;1}^{(0,1)}\non\\
&&+\frac{1}{2} \zeta _2 \xi _1^2 {B}_{1;1}^{(2,0)}+\frac{1}{2} \zeta _1 \xi _1^2 {B}_{1;1}^{(2,1)}\non\\[.3cm]
B_2&=&\zeta _1 z {B}_{2;0}^{(0,1)}+\zeta _2 z {B}_{2;0}^{(0,0)}
+\zeta _2 \zeta _1 \xi _2 {B}_{2;1}^{(0,1)}+\zeta _2 \zeta _1 \xi _1 {B}_{2;1}^{(1,1)}+\zeta _2^2 \xi _2 {B}_{2;1}^{(0,0)}+\zeta _2^2 \xi _1 {B}_{2;1}^{(1,0)}\non\\
&&+\frac{1}{2} \zeta _1^2 \xi _2 {B}_{2;1}^{(0,2)}+\frac{1}{2} \zeta _1^2 \xi _1 {B}_{2;1}^{(1,2)}\label{8103e}
\ee
Thus, in this case, there are a total of 30 form factors and hence 30 undetermined coefficients. However, again all of these are not independent. This 3-point function is characterised by only 4 independent parameters for $\Delta_1\not=\Delta_3$ and by 5 independent parameters for $\Delta_1=\Delta_3$.

\subsection{Equations for form factors}
In this case, we need to solve 60 primary equations and 46 secondary equations. As before, the secondary equations can be divided in homogeneous and inhomogeneous equations. It turns out that 37 out of 46 are homogeneous differential equations and 9 are  inhomogeneous equations connecting the 3-point function to the 2-point function.

Again, these equations can be obtained by substituting the ansatz \eqref{8103e} in the equations given in section \ref{prisecwi} and setting the coefficients of independent tensor structures involving $\xi_i$ and $\zeta_i$ to zero. The explicit form of the these equations is not very illuminating which can be easily derived using some computer program. We give the 60 primary equations in appendix \ref{appen:s2eq}. 
\subsection{Solution of Primary Equations}

The solution of the primary Ward identities is given by
\be
A_2^{(2,2)} &=& a_2^{(2,2)}J_{5\{0,0,0\}}\non\\[.2cm]
A_2^{(2,1)}&=&  a_2^{(2,1)} J_{4\{-1,0,0\}}-a_2^{(2,2)} J_{5\{-1,1,0\}}\non\\[.2cm]
A_2^{(1,2)}&=&  a_2^{(1,2)} J_{4\{0,0,-1\}}+a_2^{(2,2)} J_{5\{0,1,-1\}}\non\\[.2cm]
A_2^{(2,0)}&=&  a_2^{(2,0)} J_{3\{-2,0,0\}}-a_2^{(2,1)} J_{4\{-2,1,0\}}+\f{1}{2}a_2^{(2,2)} J_{5\{-2,2,0\}}\non\\[.2cm]
A_2^{(0,2)}&=&  a_2^{(0,2)} J_{3\{0,0,-2\}}+a_2^{(1,2)} J_{4\{0,1,-2\}}+\f{1}{2}a_2^{(2,2)} J_{5\{0,2,-2\}}\non\\[.2cm]
 A_{2}^{(1,1)} &=& a_2^{(1,1)}J_{3\{-1,0,-1\}}-a_{2}^{(2,2)} J_{5\{-1,2,-1\}}+   \bigl(a_{2}^{(2,1)}-a_2^{(1,2)}\bigl)J_{4\{-1,1,-1\}}    \non\\[.2cm]
 A_{2}^{(1,0)} &=& \Bigl(a_2^{(2,0)}-a_2^{(1,1)}\Bigl)J_{3\{-2,1,-1\}}+  \f{1}{2}a_2^{(2,2)}J_{5\{-2,3,-1\}}  +a_2^{(1,0)}J_{2\{-2,0,-1\}}    \non\\
 &&+\Bigl(\f{1}{2}a_{2}^{(1,2)}-a_{2}^{(2,1)}\Bigl) J_{4\{-2,2,-1\}}\non\\[.2cm]
 A_{2}^{(0,1)} &=& \Bigl(a_2^{(1,1)}-a_2^{(0,2)}\Bigl)J_{3\{-1,1,-2\}}-  \f{1}{2}a_2^{(2,2)}J_{5\{-1,3,-2\}}  +a_2^{(0,1)}J_{2\{-1,0,-2\}}    \non\\
 &&+\Bigl(\f{1}{2}a_{2}^{(2,1)}-a_{2}^{(1,2)}\Bigl) J_{4\{-1,2,-2\}}\non\\[.2cm]
  A_{2}^{(0,0)} &=&a_2^{(0,0)}J_{1\{-2,0,-2\}} +\f{1}{4}a_2^{(2,2)}J_{5\{-2,4,-2\}} +\Bigl(\f{1}{2}a_{2}^{(0,2)}+\f{1}{2}a_{2}^{(2,0)}-a_{2}^{(1,1)}\Bigl) J_{3\{-2,2,-2\}}\non\\[.2cm]
 && +\f{1}{2}\bigl( a_2^{(1,2)}- a_2^{(2,1)}\bigl)J_{4\{-2,3,-2\}}   + \Bigl(a_2^{(1,0)}-a_2^{(0,1)}\Bigl)J_{2\{-2,1,-2\}} 
 \ee
  \be
A_1^{(1,1)}&=&  a_1^{(1,1)} J_{3\{0,0,0\}}-\f{1}{2}a_2^{(2,2)} J_{4\{0,1,0\}}\non\\
 A_{1}^{(1,0)} &=&a_{1}^{(1,0)} J_{2\{-1,0,0\}}-\Bigl(a_{1}^{(1,1)}+\f{1}{2}a_2^{(2,1)}\Bigl) J_{3\{-1,1,0\}}+\f{1}{2} a_2^{(2,2)}J_{4\{-1,2,0\}}   \non\\[.2cm]
 A_{1}^{(0,1)} &=& a_{1}^{(0,1)} J_{2\{0,0,-1\}}+\Bigl( a_{1}^{(1,1)}-\f{1}{2}a_2^{(1,2)} \Bigl)J_{3\{0,1,-1\}}-\f{1}{2} a_2^{(2,2)}J_{4\{0,2,-1\}}   \non\\[.2cm]
  A_{1}^{(0,0)} &=& -\Bigl(a_{1}^{(1,1)}+\f{1}{2}a_2^{(2,1)}-\f{1}{2}a_2^{(1,2)} \Bigl)J_{3\{-1,2,-1\}}+  \bigl(a_{1}^{(1,0)}-a_1^{(0,1)}-\f{1}{2}a_2^{(1,1)}\bigl)J_{2\{-1,1,-1\}} \non\\
        &&+a_1^{(0,0)}J_{1\{-1,0,-1\}}+\f{1}{2}a_2^{(2,2)}J_{4\{-1,3,-1\}}   \non\\[.2cm]
A_0^{(0,0)}&=&a_0^{(0,0)} J_{1\{0,0,0\}}-a_1^{(1,1)} J_{2\{0,1,0\}}+\f{1}{4}a_2^{(2,2)} J_{3\{0,2,0\}}
\ee

\be
  B_{2;0}^{(0,1)} &=&b_{2;0}^{(0,1)}J_{1\{0,0,0\}} + \Bigl(\f{1}{2}a_2^{(2,1)}-b_{2;1}^{(1,1)}-b_{2;1}^{(1,2)}\Bigl)J_{2\{0,1,0\}}+\Bigl(\f{1}{2}a_2^{(2,1)}-b_{2;1}^{(1,1)}\Bigl)J_{2\{0,0,1\}}\non\\[.2cm]
  &&- \f{1}{2}a_{2}^{(2,2)}J_{3\{1,1,0\}}+ \Bigl(\f{1}{2}a_2^{(2,1)}-b_{2;1}^{(1,1)}+a_{1}^{(1,1)}\Bigl)J_{2\{1,0,0\}} \non\\[.3cm]
  B_{2;0}^{(0,0)} &=&b_{2;0}^{(0,0)}J_{0\{-1,0,0\}} -\Bigl(a_{1}^{(1,1)}+\f{1}{2}a_{2}^{(2,1)} \Bigl)J_{2\{0,1,0\}}+b_{2;0}^{(0,1)}J_{1\{-1,0,1\}}+  \bigl(b_{2;0}^{(0,1)}+a_1^{(1,0)}\bigl)J_{1\{0,0,0\}} \non\\
   &&+\f{1}{2}a_2^{(2,2)}J_{3\{0,2,0\}} +b_{2;1}^{(1,2)}J_{2\{-1,2,0\}}   \non\\[.2cm] 
  B_{1;0}^{(1,0)} &=&b_{1;0}^{(1,0)}J_{1\{0,0,0\}} + \Bigl(\f{1}{2}a_2^{(1,2)}+b_{1;1}^{(1,1)}-b_{1;1}^{(2,1)}\Bigl)J_{2\{0,1,0\}}+\Bigl(\f{1}{2}a_2^{(1,2)}+b_{1;1}^{(1,1)}\Bigl)J_{2\{1,0,0\}}\non\\[.2cm]
  &&+\f{1}{2}a_{2}^{(2,2)}J_{3\{0,1,1\}}+ \Bigl(\f{1}{2}a_2^{(1,2)}+b_{1;1}^{(1,1)}-a_{1}^{(1,1)}\Bigl)J_{2\{0,0,1\}} \non\\[.3cm]
  B_{1;0}^{(0,0)} &=&b_{1;0}^{(0,0)}J_{0\{0,0,-1\}} -\Bigl(a_{1}^{(1,1)}-\f{1}{2}a_{2}^{(1,2)} \Bigl)J_{2\{0,1,0\}}-  \bigl(b_{1;0}^{(1,0)}+a_1^{(0,1)}\bigl)J_{1\{0,0,0\}} \non\\
   &&-b_{1;0}^{(1,0)}J_{1\{1,0,-1\}}+\f{1}{2}a_2^{(2,2)}J_{3\{0,2,0\}} -b_{1;1}^{(2,1)}J_{2\{0,2,-1\}}   
   \ee

   \be
   B_{2;1}^{(1,2)} &=& b_{2;1}^{(1,2)}J_{3\{0,0,0\}}+ \f{1}{2}a_{2}^{(2,2)}J_{4\{1,0,0\}} \non\\
   B_{2;1}^{(1,1)} &=&b_{2;1}^{(1,1)}J_{3\{0,0,0\}} -\Bigl(b_{2;1}^{(1,2)}-b_{2;1}^{(1,1)}+\f{1}{2}a_2^{(2,1)} \Bigl)J_{3\{-1,1,0\}}+  \bigl(b_{2;1}^{(1,1)}-\f{1}{2}a_2^{(2,1)}\bigl)J_{3\{-1,0,1\}} \non\\
   &&-\f{1}{2}a_2^{(2,2)}J_{4\{0,1,0\}}   \non\\[.2cm] 
   B_{2;1}^{(1,0)} &=&b_{2;1}^{(1,0)}J_{1\{-2,0,0\}}+\f{1}{2}b_{2;1}^{(1,1)}J_{3\{0,0,0\}} +\f{1}{2}a_2^{(2,1)} J_{3\{-2,1,1\}}+  \f{1}{2}b_{2;1}^{(1,1)}J_{3\{-2,0,2\}} +b_{2;1}^{(1,1)}J_{3\{-1,0,1\}} \non\\[.2cm]
   &&+\Bigl(\f{1}{2}a_{2}^{(2,1)}-\f{1}{2}b_{2;1}^{(1,1)}+\f{1}{2}b_{2;1}^{(1,2)}\Bigl)J_{3\{-2,2,0\}} +\f{1}{4}a_{2}^{(2,2)}J_{4\{-1,2,0\}}+\f{1}{2}a_{2}^{(2,0)}J_{2\{-1,0,0\}} \non\\[.2cm] 
    B_{2;1}^{(0,1)} &=&b_{2;1}^{(0,1)}J_{1\{-1,0,-1\}}-\f{1}{2}a_{2}^{(2,2)}J_{4\{0,2,-1\}} +\f{1}{2}a_2^{(1,1)} J_{2\{0,0,-1\}}+  \Bigl(b_{2;1}^{(1,2)}+b_{2;1}^{(1,1)}-\f{1}{2}a_{2}^{(1,2)} \Bigl)J_{3\{0,1,-1\}} \non\\[.2cm]
   &&+\Bigl(b_{2;1}^{(1,1)}-\f{1}{2}a_{2}^{(2,1)}\Bigl)J_{3\{-1,2,-1\}} +\Bigl(b_{2;1}^{(1,2)}+b_{2;1}^{(1,1)}-\f{1}{2}a_{2}^{(2,1)}\Bigl)J_{3\{-1,1,0\}}-b_{2;1}^{(0,2)}J_{2\{-1,1,-1\}} \non\\[.2cm] 
    B_{2;1}^{(0,2)} &=&b_{2;1}^{(0,2)}J_{2\{0,0,-1\}} -b_{2;1}^{(1,2)}J_{3\{0,0,0\}} +  \f{1}{2}a_{2}^{(2,2)}J_{4\{1,1,-1\}} +\Bigl(\f{1}{2}a_{2}^{(1,2)}-b_{2;1}^{(1,2)}\Bigl)J_{3\{1,0,-1\}} 
   \ee

  \be
  B_{2;1}^{(0,0)} &=&b_{2;1}^{(0,0)}J_{0\{-2,0,-1\}}-\f{1}{2}a_{2}^{(1,0)}J_{1\{-2,0,0\}} +  \Bigl(\f{1}{4}a_{2}^{(1,2)}-\f{1}{2}b_{2;1}^{(1,1)}-\f{1}{2}b_{2;1}^{(1,2)} \Bigl)J_{3\{-1,2,-1\}} \non\\[.2cm]
   &&+\Bigl(\f{1}{2}a_2^{(2,0)} -\f{1}{2}a_2^{(1,1)}\Bigl)J_{2\{-1,1,-1\}}+\Bigl(\f{1}{2}a_{2}^{(2,1)}-\f{1}{2}b_{2;1}^{(1,1)}-\f{1}{2}b_{2;1}^{(1,2)} \Bigl)J_{3\{-2,2,0\}}+\f{1}{4}a_{2}^{(2,2)}J_{4\{-1,3,-1\}}\non\\
   &&+\Bigl(b_{2;1}^{(1,0)}-\f{1}{2}a_{2}^{(1,0)}-b_{2;1}^{(0,1)}\Bigl)J_{1\{-2,1,-1\}} +\f{1}{2}b_{2;1}^{(0,2)}J_{2\{-2,2,-1\}}+\Bigl(\f{1}{2}a_{2}^{(2,1)}-\f{2}{3}b_{2;1}^{(1,1)}\Bigl)J_{3\{-2,3,-1\}}\non\\
   &&-\f{1}{2}b_{2;1}^{(1,1)}J_{3\{0,0,0\}} -\f{1}{2}b_{2;1}^{(1,1)}J_{3\{-1,0,1\}} -\f{1}{6}b_{2;1}^{(1,1)}J_{3\{1,0,-1\}}-\f{1}{6}b_{2;1}^{(1,1)}J_{3\{-2,0,2\}}  
  \ee

   \be
   B_{1;1}^{(2,1)} &=& b_{1;1}^{(2,1)}J_{3\{0,0,0\}}- \f{1}{2}a_{2}^{(2,2)}J_{4\{0,0,1\}} \non\\
   B_{1;1}^{(1,1)} &=&b_{1;1}^{(1,1)}J_{3\{0,0,0\}} +\Bigl(b_{1;1}^{(2,1)}+b_{1;1}^{(1,1)}+\f{1}{2}a_2^{(1,2)} \Bigl)J_{3\{0,1,-1\}}+  \bigl(b_{1;1}^{(1,1)}+\f{1}{2}a_2^{(1,2)}\bigl)J_{3\{1,0,-1\}} \non\\
   &&-\f{1}{2}a_2^{(2,2)}J_{4\{0,1,0\}}   \non\\[.2cm] 
   B_{1;1}^{(0,1)} &=&b_{1;1}^{(0,1)}J_{1\{0,0,-2\}}-\f{1}{2}b_{1;1}^{(1,1)}J_{3\{0,0,0\}} +\f{1}{2}a_2^{(1,2)} J_{3\{1,1,-2\}}-  \f{1}{2}b_{1;1}^{(1,1)}J_{3\{2,0,-2\}} -b_{1;1}^{(1,1)}J_{3\{1,0,-1\}} \non\\[.2cm]
   &&+\Bigl(\f{1}{2}a_{2}^{(1,2)}+\f{1}{2}b_{1;1}^{(1,1)}+\f{1}{2}b_{1;1}^{(2,1)}\Bigl)J_{3\{0,2,-2\}} -\f{1}{4}a_{2}^{(2,2)}J_{4\{0,2,-1\}}-\f{1}{2}a_{2}^{(0,2)}J_{2\{0,0,-1\}} \non\\[.2cm] 
    B_{1;1}^{(1,0)} &=&b_{1;1}^{(1,0)}J_{1\{-1,0,-1\}}+\f{1}{2}a_{2}^{(2,2)}J_{4\{-1,2,0\}} -\f{1}{2}a_2^{(1,1)} J_{2\{-1,0,0\}}+  \Bigl(b_{1;1}^{(2,1)}-b_{1;1}^{(1,1)}-\f{1}{2}a_{2}^{(2,1)} \Bigl)J_{3\{-1,1,0\}} \non\\[.2cm]
   &&-\Bigl(b_{1;1}^{(1,1)}+\f{1}{2}a_{2}^{(1,2)}\Bigl)J_{3\{-1,2,-1\}} +\Bigl(b_{1;1}^{(2,1)}-b_{1;1}^{(1,1)}-\f{1}{2}a_{2}^{(1,2)}\Bigl)J_{3\{0,1,-1\}}+b_{1;1}^{(2,0)}J_{2\{-1,1,-1\}} \non\\[.2cm] 
     B_{1;1}^{(2,0)} &=&b_{1;1}^{(2,0)}J_{2\{-1,0,0\}} +b_{1;1}^{(2,1)}J_{3\{0,0,0\}} +  \f{1}{2}a_{2}^{(2,2)}J_{4\{-1,1,1\}} +\Bigl(-\f{1}{2}a_{2}^{(2,1)}+b_{1;1}^{(2,1)}\Bigl)J_{3\{-1,0,1\}} 
   \ee

 \be
  B_{1;1}^{(0,0)} &=&b_{1;1}^{(0,0)}J_{0\{-1,0,-2\}}+\f{1}{2}a_{2}^{(0,1)}J_{1\{0,0,-2\}} -  \Bigl(\f{1}{4}a_{2}^{(2,1)}+\f{1}{2}b_{1;1}^{(1,1)}-\f{1}{2}b_{1;1}^{(2,1)} \Bigl)J_{3\{-1,2,-1\}} \non\\[.2cm]
   &&+\Bigl(\f{1}{2}a_2^{(0,2)} -\f{1}{2}a_2^{(1,1)}\Bigl)J_{2\{-1,1,-1\}}+\Bigl(-\f{1}{2}a_{2}^{(1,2)}-\f{1}{2}b_{1;1}^{(1,1)}+\f{1}{2}b_{1;1}^{(2,1)} \Bigl)J_{3\{0,2,-2\}}+\f{1}{4}a_{2}^{(2,2)}J_{4\{-1,3,-1\}}\non\\
   &&+\Bigl(-b_{1;1}^{(0,1)}+\f{1}{2}a_{2}^{(0,1)}+b_{1;1}^{(1,0)}\Bigl)J_{1\{-1,1,-2\}} +\f{1}{2}b_{1;1}^{(2,0)}J_{2\{-1,2,-2\}}-\Bigl(\f{1}{2}a_{2}^{(1,2)}+\f{2}{3}b_{1;1}^{(1,1)}\Bigl)J_{3\{-1,3,-2\}}\non\\
   &&-\f{1}{2}b_{1;1}^{(1,1)}J_{3\{0,0,0\}} -\f{1}{2}b_{1;1}^{(1,1)}J_{3\{1,0,-1\}} -\f{1}{6}b_{1;1}^{(1,1)}J_{3\{-1,0,1\}}-\f{1}{6}b_{1;1}^{(1,1)}J_{3\{2,0,-2\}}  
  \ee

\subsection{Solution of Secondary Equations}
When $\Delta_1\not=\Delta_3$, the solution of the secondary equations is given in appendix \ref{s=2secondary}. Here, we consider the case where $\Delta_1=\Delta_3=\Delta$ for which the secondary Ward identities imply the following constraints
\be
a_{1}^{(0,1)}&=&\frac{1}{8} (d+2) d^2 (\Delta +1) a_2^{(2,2)}+\left(-\frac{3 d^2}{4}-\frac{1}{2} d (\Delta +2)+\Delta +1\right) a_1^{(1,1)}+\left(\frac{d}{2 \Delta }+\frac{1}{\Delta }\right) a_0^{(0,0)}\non\\
&&+\frac{1}{8} (d-2) \Delta  a_2^{(1,1)}-d \Delta  b_{1;1}^{(1,1)}\non\\[.3cm]
a_1^{(0,0)}&=&\frac{(\Delta -1) \left(d^2-2 d+4 \Delta \right) a_0^{(0,0)}}{\Delta ^2}+\frac{d^2 \left(d^2-4\right) \left(\Delta ^2-1\right) a_2^{(2,2)}}{4 \Delta }+\frac{1}{4} (d-2)^2 (\Delta -1) a_2^{(1,1)}\non\\
&&-\frac{(d-2) (\Delta -1) \left(3 d^2+2 d (\Delta +1)-4 \Delta \right) a_1^{(1,1)}}{2 \Delta }-2 d (d-2) (\Delta -1) b_{1;1}^{(1,1)}
\label{6.109}
\ee

\be
a_{2}^{(1,2)}&=&\frac{d^2 (\Delta +1) a_2^{(2,2)}}{4 \Delta }-\frac{(3 d+2 \Delta -2) a_1^{(1,1)}}{2 \Delta }+\frac{a_0^{(0,0)}}{\Delta ^2}+\frac{1}{4} a_2^{(1,1)}-2 b_{1;1}^{(1,1)}\non\\[.4cm]
a_{2}^{(0,2)}&=&\frac{d \left(d^2+2 d-4\right) (\Delta +1) a_2^{(2,2)}}{4 (\Delta -1)}-\frac{\left(3 d^2+2 d (\Delta +1)-8 (\Delta +1)\right) a_1^{(1,1)}}{2 (\Delta -1)}+\frac{d a_0^{(0,0)}}{(\Delta -1) \Delta }\non\\
&&+\frac{(d-4) \Delta  a_2^{(1,1)}}{4 (\Delta -1)}-\frac{2 d \Delta  b_{1;1}^{(1,1)}}{\Delta -1}\non\\[.4cm]
a_{2}^{(0,1)}&=&\frac{\left(3 d^2+2 d (2 \Delta -5)+8 \Delta \right) }{2 \Delta ^2} a_0^{(0,0)}+\frac{(d+2) d^2 (\Delta +1) (3 d+4 \Delta -8) a_2^{(2,2)}}{8 \Delta }\non\\
&&+\left(-\frac{9 d^3}{4 \Delta }+d^2 \left(\frac{5}{\Delta }-\frac{9}{2}\right)+d \left(-2 \Delta +\frac{3}{\Delta }+5\right)+4 (\Delta -1)\right) a_1^{(1,1)}\non\\
&&+\frac{1}{8} \Bigl(3 d^2+4 d \Delta -18 d-8 \Delta +16\Bigl) a_2^{(1,1)}+d (-3 d-4 \Delta +6) b_{1;1}^{(1,1)}\non\\[.6cm]
a_2^{(0,0)}&=&\frac{\left(d^2-4\right) d^2 \left(\Delta ^2-\Delta -2\right) (d+4 \Delta -5) a_2^{(2,2)}}{4 (\Delta -1) \Delta }-\frac{2 (d-2) d (\Delta -2) (d+4 \Delta -4) b_{1;1}^{(1,1)}}{\Delta -1}\non\\
&&-\frac{(d-2) (\Delta -2) \left(14 d^2 (\Delta -1)+3 d^3+8 d \left(\Delta ^2-\Delta -1\right)-16 (\Delta -1) \Delta \right) a_1^{(1,1)}}{2 (\Delta -1) \Delta }\non\\
&&+\frac{(\Delta -2) \left(4 d^2 (\Delta -2)+d^3+d (12-8 \Delta )+8 (\Delta -1) \Delta \right) a_0^{(0,0)}}{(\Delta -1) \Delta ^2}\non\\
&&+\frac{(d-2) (\Delta -2) \left(d^2+4 d (\Delta -2)-8 \Delta +8\right) a_2^{(1,1)}}{4 (\Delta -1)}
\label{6.10}
\ee

\be
a_{1}^{(1,0)}=-a_1^{(0,1)}\quad;\quad a_{2}^{(2,1)}&=&-a_{2}^{(1,2)}\quad;\quad a_{2}^{(2,0)}=a_2^{(0,2)}\quad;\quad a_{2}^{(1,0)}=-a_2^{(0,1)}
\ee

\be
b_{1;0}^{(1,0)}&=& \left(1-\f{d}{2}-\f{1}{\Delta}\right)a_0^{(0,0)} + \f{1}{4}\Bigl(2d(3+\Delta^2)+3d^2\Delta-4(\Delta-1)^2\Bigl)a_1^{(1,1)} -\f{1}{8}\Delta \left(2+(d-2)\Delta\right)a_2^{(1,1)}\non\\
&&-\frac{1}{8} d^2 (\Delta +1) ((d+2) \Delta +2) {a}_2^{(2,2)}+d \Delta ^2 {b}_{1;1}^{(1,1)}\non\\[.6cm]
b_{1;0}^{(0,0)}&=&\frac{1}{8} d^2 \left(\Delta ^2-1\right) \left(d^2-2 d \Delta -4 (\Delta +2)\right) a_2^{(2,2)}-\frac{1}{8} (\Delta -1) \Delta  \left(-d^2+2 d (\Delta +2)-4 \Delta \right) a_2^{(1,1)}\non\\
&&+\frac{1}{4} (\Delta -1) \left(d^2 (4 \Delta +6)-3 d^3+4 d (\Delta ^2+2 \Delta +2)-8 (\Delta -1) \Delta \right) a_1^{(1,1)}\non\\
&&+\frac{(\Delta -1) \left(d^2-2 d (\Delta +2)+4 \Delta \right) a_0^{(0,0)}}{2 \Delta }+d \Delta  \left(-d \Delta +d+2 \Delta ^2-2\right) b_{1;1}^{(1,1)}
\label{6.11}
\ee

\be
b_{1;1}^{(2,1)}&=&\frac{1}{4} (3 d+2 \Delta +6) a_1^{(1,1)}-\frac{1}{8} d (d+4) (\Delta +1) a_2^{(2,2)}-\frac{a_0^{(0,0)}}{2 \Delta }-\frac{1}{8} \Delta  a_2^{(1,1)}+\Delta  b_{1;1}^{(1,1)}\non\\[.4cm]
b_{1;1}^{(2,0)}&=&\left(-\frac{3 d^2}{4}+d (3-2 \Delta )-\Delta ^2+1\right) a_1^{(1,1)}+\frac{1}{8} d (\Delta +1) \left(d^2+2 d (\Delta -1)+8 (\Delta -1)\right) a_2^{(2,2)}\non\\
&&+\left(\frac{d}{2 \Delta }-\frac{3}{\Delta }+1\right) a_0^{(0,0)}+\frac{1}{8} \Delta  (d+2 \Delta -6) a_2^{(1,1)}-\Delta  (d+2 \Delta -2) b_{1;1}^{(1,1)}\non\\[.4cm]
b_{1;1}^{(1,0)}&=&-\frac{(d+2) d^2 \left(\Delta ^2-1\right) (d+2 \Delta -4) a_2^{(2,2)}}{8 \Delta }+\left(\frac{d^2}{2 \Delta ^2}-\frac{d^2}{2 \Delta }-\frac{3 d}{\Delta ^2}+\frac{4 d}{\Delta }-d-\frac{2}{\Delta }+2\right) a_0^{(0,0)}\non\\
&&+\frac{(\Delta -1) \left(4 d^2 (2 \Delta -3)+3 d^3+4 d \left(\Delta ^2-3 \Delta -1\right)-8 (\Delta -1) \Delta \right) }{4 \Delta }a_1^{(1,1)}\non\\
&&-\frac{1}{8} (\Delta -1) \left(d^2+2 d (\Delta -3)-4 \Delta \right) a_2^{(1,1)}+d (\Delta -1) (d+2 \Delta -2) b_{1;1}^{(1,1)}\non\\[.4cm]
b_{1;1}^{(0,1)}&=&\frac{1}{16} (d-2) d^3 (\Delta +1) a_2^{(2,2)}+\left(\frac{d^2}{4 \Delta }-\frac{3 d}{2 \Delta }+\frac{2}{\Delta }\right) a_0^{(0,0)}+\frac{1}{16} \left(d^2-6 d+8\right) \Delta  a_2^{(1,1)}\non\\[.2cm]
&&-\frac{1}{8} (d-2) \left(3 d^2+2 d (\Delta -3)-8 \Delta +8\right) a_1^{(1,1)}+\frac{1}{2} \Delta  \left(-d^2+2 d+4 \Delta -4\right) b_{1;1}^{(1,1)}\non\\[.5cm]
b_{1;1}^{(0,0)}&=&\frac{(d+2) d^2 \left(\Delta ^2-\Delta -2\right) \left(d^2+4 d (\Delta -3)-16 \Delta +24\right) a_2^{(2,2)}}{16 \Delta }\non\\
&&+\frac{d (\Delta -2) \left(d^2+2 d (2 \Delta -7)-24 \Delta +32\right) a_0^{(0,0)}}{4 \Delta ^2}\non\\
&&-\f{1}{8\Delta}\biggl[2 d^3 \left(7 \Delta ^2-32 \Delta +36\right)+4 d^2 \left(2 \Delta ^3-23 \Delta ^2+53 \Delta -30\right)+3 d^4 (\Delta -2)\non\\
&&-16 d (\Delta -2)^2 (3 \Delta +1)+64 \Delta  \left(\Delta ^2-3 \Delta +2\right) \biggl]a_1^{(1,1)}\non\\
&&+\frac{1}{16} (d-4) (\Delta -2) \left(d^2+2 d (2 \Delta -5)-8 \Delta +8\right) a_2^{(1,1)}\non\\
&&+\frac{1}{6} (\Delta -2) \left(-6 d^2 (2 \Delta -5)-3 d^3+48 d (\Delta -1)+8 (\Delta -1) \Delta \right) b_{1;1}^{(1,1)}
\label{6.12}
\ee
The corresponding coefficients for $B_2$ are given in terms of the coefficients of $B_1$ as
\be
b_{2;0}^{(0,1)}&=&-b_{1;0}^{(1,0)}\qquad;\quad b_{2;0}^{(0,0)}=b_{1;0}^{(0,0)}\non\\[.4cm]
b_{2;1}^{(1,2)}&=&-b_{1;1}^{(2,1)}\qquad;\qquad b_{2;1}^{(0,2)}=b_{1;1}^{(2,0)}\qquad;\qquad b_{2;1}^{(1,1)}=b_{1;1}^{(1,1)}\non\\[.4cm]
b_{2;1}^{(1,0)}&=&-b_{1;1}^{(0,1)}\qquad;\qquad b_{2;1}^{(0,1)}=-b_{1;1}^{(1,0)}\qquad;\qquad b_{2;1}^{(0,0)}=b_{1;1}^{(0,0)}
\ee
The above results can be obtained by either using the identities \eqref{jn2}, \eqref{jn4} and \eqref{wotder1} or analysing the homogeneous Ward identities in the limit $p_3 \rightarrow 0$. The regularisation of the equations needs to be done, as before, according to equation \eqref{shiftpara} and the procedure described in section \ref{regularization}.  

Finally, we need to consider 9 non homogeneous secondary Ward identities which relate the coefficients appearing in 2 and 3 point functions. These equations can be used to find the relation between two and three point coefficients to obtain
 \begin{eqnarray}
 a_0= \,2^{\frac{d}{2} -6}\,\frac{(d-2\Delta)}{g_3\,(d-2)} \,\Gamma\left(\frac{2\Delta-d}{2}\right)\Gamma\left( \frac{d-2\Delta}{2} \right)\,\Gamma\left(\frac{d}{2}\right)\,\left[4 a_0^{(0,0)}
+(d-2) (-4a_1^{(1,1)} +d a_2^{(2,2)})\right]
\end{eqnarray} 
This completes the analysis for the 3-point function of the two spin 2 non-conserved operators and a gauge field.

\section{Discussion}
\label{s4:dissc}
In this paper,  we have derived the consequences of the conformal Ward identities for three-point correlation functions of a conserved current and two generic spin-$s$ operators in the momentum space. We have solved the resulting set of differential equations in the case when spin $s$ is 1 and 2  for operators having the same or different conformal dimensions. The correlator involving  spin-1 operators depends on 9 form factors that are constrained by 28 coupled differential equations. For spin-2, the number of form factors grows to 30  which are fixed by a system of 106 differential equations. The solution has been found analytically by using the properties of the triple-K integrals. The extension of the calculation to higher spins is more tedious but can be done with the help of computer softwares such as Mathematica. There are numerous natural extensions of this work, including the computation of three-point correlation functions of the stress-energy tensor with higher spin non-conserved operators and the computation of four-point functions. 

Apart from the general utility of the results derived here in physics situations where CFTs with suitable spectrum are relevant, 
our results should also have applications to the AdS/CFT correspondence and the study of signatures 
of higher spin fields in early universe cosmology.  One of our motivations for this work was to use CFT and holography to understand interactions of higher-spin fields with gauge fields and gravitons, both in AdS and dS and also in flat space using an appropriate flat space limit, and results in this direction will be presented in  \cite{AdS}.

\bigskip

{\bf Acknowledgement:} 
{We thank Massimo Taronna and Charlotte Sleight for collaboration during initial stages of this work.  We also thank Paul McFadden and Adam Bzowski for discussions. MV and KS are supported in part by the STFC consolidated grant ST/T000775/1 ``New Frontiers in Particle Physics, Cosmology and Gravity''. }

\appendix

\section{Auxiliary vectors}
\label{linearind3456}

In this appendix we recall some basic facts about these auxiliary vectors following \cite{Todorov, Costa:2011mg}.  Suppose we are dealing with an $SO(d)$ symmetric traceless tensor $f_{\mu_1\dots \mu_l}$. It is in one-to-one correspondence with a homogeneous polynomials $f(z)$
\begin{eqnarray}
f_{\mu_1\dots \mu_l} \;\;\longleftrightarrow  \;\;f(z)\;=\;f_{\mu_1\dots \mu_l}\,z^{\mu_1}\,z^{\mu_2}\,\dots z^{\mu_l}\label{onetoone}
\end{eqnarray}
The vectors $z^\mu$ are complex and define a \textquotedblleft complex null cone\textquotedblright  $\; \mathbb{K}_d$ in the Euclidean space $R^d$ \cite{Todorov}
\begin{eqnarray}
\mathbb{K}_d=\Big\{z\in \mathbb{C}^d, \,z^2\equiv (z^1)^2+\dots +(z^d)^2=0\Big\}\label{onetoonea}
\end{eqnarray}
In CFT, symmetric traceless tensors describe spin $l$ primary fields and they can be represented by homogeneous polynomial defined on the complex light cone in one-to-one correspondence as in equation \eqref{onetoone} and \eqref{onetoonea}. From these equations, it then also follows that spin $l$ primary fields  can be represented by different homogeneous polynomials, $f(z)$ and $\tilde{f}(z)$ differing by terms proportional to $z^2$
 \begin{eqnarray}
f_{\mu_1\dots \mu_l} \leftrightarrow  \tilde{f}_l(z)\Big|_{z^2=0}~~;~~f_l(z)-\tilde{f}_l(z) = z^2\,g_{l-2}(z)\label{1.6a}
\end{eqnarray}
This is easily seen by introducing the projector for symmetric, traceless tensors $\pi^{\mu_1\dots\mu_l}_{\nu_1\dots\nu_l}$ and observing that\begin{eqnarray}
f_{\mu_1\dots \mu_l} \pi^{\mu_1\dots\mu_l}_{\nu_1\dots\nu_l}=f_{\nu_1\dots \nu_l}\, ; \qquad \qquad\tilde{f}_{\mu_1\dots \mu_l} \pi^{\mu_1\dots\mu_l}_{\nu_1\dots\nu_l}=f_{\nu_1\dots \nu_l}\, .
\end{eqnarray}
It is useful to note that the above projector operator can also be expressed as a differential operator
\begin{eqnarray}
\pi^{\mu_1\dots\mu_l}_{\nu_1\dots \nu_l} =\frac{1}{l!(h-1)_{l}}D_{\mu_1}\dots  D_{\nu_l} z^{\mu_1}\dots z^{\mu_l}\, ;\quad\quad h =\f{d}{2}\,  ,
\end{eqnarray}
where $(a)_l=\Gamma(a+l)/\Gamma(a)$ and
\begin{eqnarray}
D_\mu&=&\left( h -1+z\cdot \frac{\partial}{\partial z}\right)\frac{\partial}{\partial z^\mu}
-\frac{1}{2} z_\mu \frac{\partial^2}{\partial z\cdot \partial z}\, .  \label{1.8a}
\end{eqnarray}
Next, we define the notion of interior differential operator. An interior differential operator $Q$ on the complex light cone $\mathbb{K}_d$  satisfies the identity
\begin{eqnarray}
(Q \,z^2 \,f(z))\Big|_{z^2}\;\;=\;\;0 \quad\mbox{for every polynomial $f(z)$}
\end{eqnarray}
Such an operator preserves the identification written in equation \eqref{1.6a}. For our purposes, we note that the generator of special conformal transformation is an interior differential operator. This can be easily seen from the definition \eqref{a210} in momentum space
\begin{eqnarray}
\left(z_\mu\frac{\partial}{\partial z_\nu}-z_\nu\frac{\partial}{\partial z_\mu}\right) \frac{\partial}{\partial p_\mu}\,z^2\,f(z)
&=&z^2\,\left(z_\mu\frac{\partial}{\partial z_\nu}-z_\nu\frac{\partial}{\partial z_\mu}\right) \frac{\partial}{\partial p_\mu}\,f(z)
\end{eqnarray}
We notice that acting with this operator on an expression which is linear in $z$, we shall always get an expression which is also linear in $z$.

For the $n$-point function of spin $l$ primary operators, we introduce $n$ independent null cones, of dimension $d$,  parametrized by $n$  independent auxiliary vectors, $\epsilon_1, \ldots, \epsilon_n$. For example for 2-point function we need 
$\epsilon_1\equiv \epsilon$ and $\epsilon_2\equiv\tilde{\epsilon}$, {\it i.e.},
\begin{eqnarray}
\mathbb{K}_d^\epsilon=\Big\{\epsilon\in \mathbb{C}^d,\,(\epsilon^1)^2+\dots +(\epsilon^d)^2=0\Big\}~~,~~\mathbb{K}_d^{\tilde \epsilon}=\Big\{{\tilde \epsilon}\in \mathbb{C}^d,\,({\tilde \epsilon}^1)^2+\dots +({\tilde \epsilon}^d)^2=0\Big\}
\end{eqnarray}
This guarantees that interior differential operators acting on the homogeneous polynomials defined on one null cone do not have any action on the homogeneous polynomials defined on the other cone. 
The 2-point function can be thought as defined in the complex double cone $\mathbb{K}_d^\epsilon\otimes \mathbb{K}_d^{\tilde \epsilon}$.

In \cite{Bzowski:2013sza} the calculations of correlators was done without making use of auxiliary vectors. This requires us to work with explicit indices which makes expressions complicated for higher spins. It is instructive however to see in a simple example that both approaches give the same set of equations. Consider the case of 2-point function of spin 1 operators. In the approach of \cite{Bzowski:2013sza} the most general expression for the 2-point function of spin 1 is given by
\begin{eqnarray}
A_2^{\mu\nu}(p) =\delta^{\mu\nu}A_0(p)+ p^\mu\,p^\nu\,A_1(p)\label{hgyt6}
\end{eqnarray} 
The action of the scalar part of the special conformal Ward identity \eqref{a.sctm} on the above ansatz is given by 
 \begin{eqnarray}
  {\mathcal{K}_s^\rho}\,A_2^{\mu\nu}(p)&=&\,p^\rho \delta^{\mu\nu} \left[ { K}\,A_0(p)+ 2A_1(p)\right]-2p^\nu\delta^{\mu\rho} \left[ (d-\Delta+1) +p\,\frac{\partial}{\partial p} \right] A_1(p)\nonumber\\
 &&-2p^\mu\delta^{\nu\rho} \left[ (d-\Delta+1) +p\,\frac{\partial}{\partial p} \right] A_1(p)+\,p^\rho \,p^\mu\,p^\nu\, { K}\,A_1(p)\label{fgty6}
 \end{eqnarray}
The operator $K$ in the above expression is defined in equation \eqref{defKKK}. Next, the action of the spin part of the special conformal Ward identity \eqref{a29} on \eqref{hgyt6} is given by 
 \begin{eqnarray}
 \mathcal{K}_\epsilon^\rho\,A_2^{\mu\nu}(p)
 &=& 2\delta^{\rho\mu} p^\nu\left( \frac{1}{p} \frac{\partial}{\partial p}A_0(p) + d\,A_1(p) +p\,\frac{\partial}{\partial p} A_1(p)\right) -2 \delta^{\rho\nu}\,p^\mu \frac{1}{p}\frac{\partial}{\partial p} A_0(p)\nonumber\\
&& -2 p^\rho\left( \delta^{\mu\nu}  +p^\mu p^\nu \frac{1}{p} \frac{\partial}{\partial p}\right)A_1(p) \label{fgty7}
 \end{eqnarray}
We now add the two contributions, and note that $p^\rho\delta^{\mu\nu},p^\nu\delta^{\mu\rho},p^\mu\delta^{\nu\rho} $ and $p^\mu p^\nu p^\rho$ are independent tensor structures. Hence, we can set their coefficients to zero obtaining following 4 equations 
 \be
 K A_0 &=&0\quad;\qquad KA_1-\f{2}{p}\f{\p A_1}{\p p}=0\non\\[.2cm]
 \f{1}{p}\f{\p A_0}{\p p} +(\Delta-1)A_1&=&0\quad;\quad \f{1}{p}\f{\p A_0}{\p p}+p\f{\p A_1}{\p p}+(d-\Delta+1)A_1=0\label{ewrt65}
 \ee
Next, we consider the same 2-point function in the index free formalism we have used in this paper. In the notation of section \ref{sec22pt}, the action of the scalar and spin parts of the special conformal transformations on the spin-1 two point function are given by 
  \begin{eqnarray}
b\cdot \mathcal{K}_s A_2(p)&=& 2(b\cdot p) \left[  \, \f{w}{2}KA_0 + w\,A_1(p) +\f{1}{2}\zeta\,\xi K A_1(p)\right]-2(b\cdot\epsilon_2)\zeta\left[ d-\Delta+1 +p\frac {\partial}{\partial p} \right]A_1(p)\nonumber\\
 &&-2(b\cdot\epsilon_1)\xi\left[ d-\Delta +1+p\frac{\partial}{\partial p}\right]A_1(p)\label{fgty}
 \end{eqnarray}
 and 
 \begin{eqnarray}
 b\cdot \mathcal{K}_\epsilon A_2(p)&=& 2(b\cdot\epsilon_1)\,\xi\left[ p\frac{\partial}{\partial p} A_1(p)+dA_1(p)+\frac{1}{p}\,\frac{\partial}{\partial p} A_0(p)\right]-2(b\cdot\epsilon_3)\zeta\,\frac{1}{p}\,\frac{\partial}{\partial p} A_0(p)\nonumber\\
 &&-2(b\cdot p) \left[ \zeta \xi\,\frac{1}{p}\,\frac{\partial}{\partial p} +w\right]A_1(p)\label{fgty2}
 \end{eqnarray}
Again adding the two contributions and setting the coefficients of $(b\cdot p)w, (b\cdot p)\zeta\xi, (b\cdot \epsilon_1)\xi$ and $(b\cdot\epsilon_3)\zeta$  to zero, we obtain precisely the four equations given in \eqref{ewrt65}. These coincide with equations \eqref{2eqa}, \eqref{2eqb} and \eqref{2eqc} evaluated for spin-$1$ non-conserved operators (as expected).

\section{CFT Ward Identities}
\label{appenA}
Momentum space CFT correlators may be obtained non-perturbatively by solving the Ward identities directly in momentum space. In this section, we summarise these CFT Ward identities in $d$ flat space time dimensions in both position as well as momentum space.
\subsection{Ward Identities in Position Space }
\label{appenAa}
By imposing the invariance of correlation functions under a symmetry transformation, one can obtain the following local Ward identity
\be
\p_\mu \bigl\langle J^\mu(x) \mathcal{O}_1(x_1)\cdots \mathcal{O}_n(x_n) \bigl\rangle = -i\sum_{j=1}^n \delta^d(x-x_j)\bigl\langle \mathcal{O}_1(x_1)\cdots \delta\mathcal{O}_j(x_j)\cdots\mathcal{O}_n(x_n) \bigl\rangle\label{ward1}
\ee
where $J^\mu(x)$ denotes the Noether current associated with the invariance under the transformation $\mathcal{O}(x)\rightarrow \mathcal{O}(x)+\delta\mathcal{O}(x)$.

By specialising the transformation to be conformal transformation and integrating the local Ward identity \eqref{ward1} over $x$, we obtain the global Ward identities associated with the conformal symmetry 
\be
\mbox{Translation}\quad:\quad 0&=&\sum_{i=1}^n\f{\p}{\p x_i^\mu }\Bigl\langle \mathcal{O}_1(x_1)\cdots \mathcal{O}_n(x_n)  \Bigl\rangle\\[.1cm]
\mbox{Dilatation}\quad:\quad 0&=&\sum_{i=1}^n\biggl[\Delta_i+x^\mu_i\f{\p}{\p x_i^\mu }\biggl]\Bigl\langle \mathcal{O}_1(x_1)\cdots \mathcal{O}_n(x_n)  \Bigl\rangle\\[.1cm]
\mbox{SCT}\quad:\quad 0&=&\sum_{i=1}^n\biggl[2\Delta_ix_i^\mu+2x_i^\mu x_i^\nu \f{\p}{\p x_i^\nu }-x_i^2\f{\p}{\p x_{i\mu}}\biggl]\Bigl\langle \mathcal{O}_1(x_1)\cdots \mathcal{O}_n(x_n)  \Bigl\rangle \label{a.sct}
\ee
If the operators $\mathcal{O}_i$ transform non trivially under the Lorentz transformation, then the RHS of the special conformal Ward identity has extra terms. E.g., for a tensor operator having $r$ indices, namely $\mathcal{O}_i^{\mu_1\cdots \mu_r}$, we need to add the following term in the RHS of \eqref{a.sct} (see e.g., \cite{Bzowski:2013sza})
\be
2\sum_{i=1}^n \sum_{j=1}^{r_i}\Bigl[(x_i)_{\nu_{ij}}\delta^{\rho\mu_{ij}}-(x_i)^{\mu_{ij}}\delta^{\rho}_{\;\;\nu_{ij}}\Bigl]\Bigl\langle \mathcal{O}_1^{\mu_{11}\cdots \mu_{1r_1}}(x_1)\cdots \mathcal{O}_i^{\mu_{i1}\cdots  \nu_{ij} \cdots \mu_{ir_i}}(x_i)\cdots \mathcal{O}_n^{\mu_{n1}\cdots \mu_{nr_n}}(x_n)     \Bigl\rangle
\ee

\subsection{Ward Identities in Momentum Space }
\label{appenAb}

We define the correlators in momentum space by the Fourier transform
\be
\Bigl\langle \mathcal{O}_1({ p}_1)\cdots \mathcal{O}_n({ p}_n)  \Bigl\rangle =\int d^dx_1\cdots d^dx_{n}\ e^{-i(p_1\cdot x_1+\cdots +p_n\cdot x_n)}\Bigl\langle \mathcal{O}_1(x_1)\cdots \mathcal{O}_n(x_n)  \Bigl\rangle 
\ee 
Using translation invariance, we may show the right hand side is proportional to $\delta^d(\sum_ip_i)$, and we write
\be
(2\pi)^d\delta^d(\sum_ip_i)\llangle[\Bigl] \mathcal{O}_1({ p}_1)\cdots \mathcal{O}_n({ p}_n)  \rrangle[\Bigl] =\int d^dx_1\cdots d^dx_{n}\ e^{-i(p_1\cdot x_1+\cdots +p_n\cdot x_n)}\Bigl\langle \mathcal{O}_1(x_1)\cdots \mathcal{O}_n(x_n)  \Bigl\rangle \label{stripmomentum}
\ee
The inverse Fourier transform can be expressed as 
\be
\Bigl\langle \mathcal{O}_1(x_1)\cdots \mathcal{O}_n(x_n)  \Bigl\rangle =\int \f{d^dp_1}{(2\pi)^d}\cdots \f{d^dp_{n-1}}{(2\pi)^d}\ e^{i(p_1\cdot x_1+\cdots +p_n\cdot x_n)}\llangle[\Bigl] \mathcal{O}_1({ p}_1)\cdots \mathcal{O}_n({ p}_n) \rrangle[\Bigl] \label{usefre}
\ee 
together with the constraint $p_1+p_2+\cdots +p_n =0$. 

The expression \eqref{usefre} is useful. E.g., for the scaling transformation, using
\be
\Bigl\langle \mathcal{O}_1(\lambda x_1)\cdots \mathcal{O}_n(\lambda x_n)  \Bigl\rangle =(\lambda)^{-\Delta_1-\cdots-\Delta_n}\Bigl\langle \mathcal{O}_1(x_1)\cdots \mathcal{O}_n(x_n)  \Bigl\rangle 
\ee
we find,
\be
\llangle[\Bigl] \mathcal{O}_1(\lambda { p}_1)\cdots \mathcal{O}_n(\lambda { p}_n)  \rrangle[\Bigl] =(\lambda)^{-(n-1)d+\Delta_1+\cdots+\Delta_n}\llangle[\Bigl] \mathcal{O}_1({ p}_1)\cdots \mathcal{O}_n({ p}_n) \rrangle[\Bigl] 
\ee
We shall need the expression of Ward identities in the momentum space. Taking the Fourier transform of both sides in \eqref{ward1}, the local Ward identity becomes
\be
 \llangle[\Bigl] k_\mu J^\mu(k) \mathcal{O}_1(p_1)\cdots \mathcal{O}_n(p_n) \rrangle[\Bigl]&=& - \sum_{j=1}^n \llangle[\Bigl] \mathcal{O}_1(p_1)\cdots \delta\mathcal{O}_j(k+p_j)\cdots\mathcal{O}_n(p_n) \rrangle[\Bigl]\label{longtr}
\ee
where the condition $k+p_1+\cdots+p_n=0$ is understood.

In a similar way, the global Ward identities in momentum space can be written as 
\be
\mbox{Translation}:\quad 0&=&\sum_{i=1}^np_i^\mu\llangle[\Bigl] \mathcal{O}_1(p_1)\cdots \mathcal{O}_n(p_n) \rrangle[\Bigl] \\
\mbox{Dilatation}:\quad 0&=&\biggl[-\sum_{i=1}^{n-1}p^\mu_i\f{\p}{\p p_i^\mu }+\sum_{i=1}^n\Delta_i-(n-1)d\biggl]\llangle[\Bigl] \mathcal{O}_1(p_1)\cdots \mathcal{O}_n(p_n) \rrangle[\Bigl] \label{dilwardw}\\
\mbox{SCT}:\quad 0&=&\sum_{i=1}^{n-1}\biggl[2(\Delta_i-d)\f{\p}{\p p_i^\mu}-2p_i^\nu\f{\p}{\p p_i^\nu }\f{\p}{\p p_{i}^\mu}+p_{i\mu} \f{\p}{\p p_i^\nu }\f{\p}{\p p_{i\nu} }\biggl]\llangle[\Bigl] \mathcal{O}_1(p_1)\cdots \mathcal{O}_n(p_n) \rrangle[\Bigl] \non\\ \label{a.sctm}
\ee
Again, for the operators $\mathcal{O}_i$ having $r$ indices, the extra term in the RHS of \eqref{a.sctm} takes the form
\be
2\sum_{i=1}^{n-1} \sum_{j=1}^{r_i}\Bigl[\delta^{\mu \mu_{ij}}\f{\p}{\p p_i^{\nu_{ij}}}-\delta^{\mu}_{\;\;\nu_{ij}}\f{\p}{\p p_{i\mu_{ij}}}\Bigl]\llangle[\Bigl] \mathcal{O}_1^{\mu_{11}\cdots \mu_{1r_1}}(p_1)\cdots \mathcal{O}_i^{\mu_{i1}\cdots  \nu_{ij} \cdots \mu_{ir_i}}(p_i)\cdots \mathcal{O}_n^{\mu_{n1}\cdots \mu_{nr_n}}(p_n)     \rrangle[\Bigl]
\label{a29}
\ee
We should note that the above Ward identities involve the correlators from which momentum conserving delta functions have been stripped off. One needs to be careful about the delta function in deriving these identities. The factor $-1$ in $n-1$ of \eqref{dilwardw} arises when the derivatives act on the delta function. 

If the polarisation tensor of the spin $r$ operator is expressed as $\epsilon^{\mu_1}\cdots \epsilon^{\mu_r}$, then the operator acting on the correlator in above expression can also be expressed as 
\be
\mathcal{K}^\mu_\epsilon \equiv 2\sum_{i=1}^{n-1} S_i^{\mu\nu} \f{\p}{\p p_i^\nu} \qquad,\qquad S^{\mu\nu}_i =\epsilon_i^\mu\f{\p}{\p \epsilon_{i\nu}}-\epsilon_i^\nu\f{\p}{\p \epsilon_{i\mu}}\label{a210}
\ee
To see this, we consider $r=1$ and compute 
\be
2\Bigl(\epsilon^\mu\f{\p}{\p \epsilon_{\nu}}-\epsilon^\nu\f{\p}{\p \epsilon_{\mu}}\Bigl) \f{\p}{\p p^\nu}(\epsilon^\rho T_\rho)
&=&2\epsilon_\sigma\Bigl(\delta^{\mu\sigma}\f{\p}{\p p^\rho} -\delta^{\mu}_{\rho}\f{\p}{\p p_\sigma} \Bigl)  T^\rho
\ee
where, we have used the fact that $T_\rho$ is independent of the polarisation vector. This equation establishes the relation between \eqref{a29} and \eqref{a210} and also shows that the relative sign between two terms in \eqref{a210} is consistent with the expression in \eqref{a29}. 

\subsection{Transverse Ward identity}
\label{longitudinal}
The equations \eqref{ward1} and \eqref{longtr} express the correlator involving an insertion of a conserved current in terms of the lower point correlators without the current. The expressions are valid for any symmetry transformation. However, it is instructive to see how these relations arise due to gauge invariance of the partition function when we couple the operators to corresponding sources. 

We consider the case when the non-conserved operators have spin-1. The generating functional for the CFT correlators involving the operators $J^\mu, \mathcal{O}_1^\mu$ and $\mathcal{O}_2^\mu$ is given by
\begin{eqnarray}
&&Z[a_\mu, w^{(1)}_\mu,w_\mu^{(2)}]=\int \mathcal{D}\Phi \exp\biggl[-S-\int d^d x \left( J^\mu   a_\mu +  O_1^\mu w^{(1)}_\mu+ O_2^{\mu}w^{(2)}_\mu\right)\biggl]\label{ljh1}
\end{eqnarray}
where, $a_\mu, w^{(1)}_\mu$ and $w_\mu^{(2)} $ are the sources for $J^\mu, \mathcal{O}_1^\mu$ and $\mathcal{O}_2^\mu$ respectively. These sources have a natural interpretation from the point of view of AdS/CFT where they correspond to the boundary values of bulk fields. Under a U(1) gauge transformation they transform as 
\begin{eqnarray}
a_\mu(x)\rightarrow a_\mu(x) -\partial_\mu\lambda(x) \qquad;\qquad w_\mu^{(i)}(x)\rightarrow e^{ig_i\lambda(x)}  w_\mu^{(i)}(x)  ~~,~~i=1,2
\end{eqnarray}
where $g_i$ denote the gauge couplings of the sources $w_\mu^{(i)}$.

The partition function for the connected correlator can be obtained by taking the logarithm of \eqref{ljh1}. Demanding the invariance of this under the infinitesimal variation, we find 
\begin{eqnarray}
0&=&\f{1}{Z}\delta_\lambda Z=-\f{1}{Z}\int \mathcal{D}\Phi \exp\bigl[\cdots\bigl]\int d^dx \lambda(x)\biggl( \p_\mu J^\mu+ i g_1\mathcal{O}_1^\mu w^{(1)}_\mu+ i g_2\mathcal{O}_2^\mu w^{(2)}_\mu\biggl)\label{digt3f}
\end{eqnarray}
where, we have done an integration by parts in the first term. 

Differentiating \eqref{digt3f} with respect to $w^{(1)}_\mu$ and $w_\mu^{(2)} $, we find the desired transverse Ward identity
\begin{eqnarray}
\partial_\mu\langle J^\mu (x) O_1^\nu(x_1)\,O_2^\sigma(x_2)\rangle= -ig_1\,\delta^d(x-x_1)   \langle  O_1^\nu(x_1)\,O_2^\sigma(x_2)\rangle- ig_2\,\delta^d(x-x_2)  \langle O_1^\nu(x_1)\,O_2^\sigma(x_2)\rangle
\end{eqnarray}
The Fourier transform  to momentum space defined by equation \eqref{stripmomentum} 
gives
\begin{eqnarray}
p_{1\mu}\llangle[\Bigl] J^\mu (p_1) O_1^\nu(p_2)\,O_2^\rho(p_3)\rrangle[\Bigl]
&=&-g_1\llangle[\Bigl]  O_1^\nu (-p_3)\,O_2^\rho(p_3)\rrangle[\Bigl]-g_2\llangle[\Bigl]  O_1^\nu(p_2)\,O_2^\rho(-p_2)\rrangle[\Bigl]\label{B.128}
\end{eqnarray}
 This shows that the longitudinal component of our 3-point function is fully determined by the 2-point function of the operators $O_1^\mu$ and $O_2^\mu$.
 
\section{Decomposition of correlator}
\label{appena.3}
If a correlator involves conserved currents, its analysis can be simplified by making use of the transverse Ward identities. In particular, the transverse and trace Ward identities imply that we can focus on the transverse traceless parts of the correlators  \cite{Bzowski:2013sza}. To see how it works, we consider a simple example. Suppose, we are interested in the correlator $\langle J^\mu(p_1)\mathcal{O}(p_2)\mathcal{O}(p_3)\rangle$ where $J^\mu$ is a conserved current and $\mathcal{O}$ denote some arbitrary operators. We can define the transverse part of the current $J^\mu$ as 
\be
j^\mu = \pi^\mu_{\;\nu}J^\nu \qquad;\quad \pi^{\mu\nu} =\delta^{\mu\nu} -\f{p^\mu p^\nu}{p^2} \quad;\quad \pi_{\mu\nu}p^\mu =0 
\ee
The $j^\mu$ satisfies the transversality condition $p_\mu j^\mu=0$. The current can now be written as 
\be
J^\mu = j^\mu +\f{p^\mu}{p^2} p_\nu J^\nu
\ee
Using this, we can express the correlator as sum of the transverse and longitudinal parts as 
\be
\bigl\langle J^\mu(p_1)\mathcal{O}(p_2)\mathcal{O}(p_3)\bigl\rangle&=& \bigl\langle j^\mu(p_1)\mathcal{O}(p_2)\mathcal{O}(p_3)\bigl\rangle\;\;+\;\;\f{p^\mu}{p^2}\bigl\langle p_\nu J^\nu(p_1)\mathcal{O}(p_2)\mathcal{O}(p_3)\bigl\rangle\label{a117t}
\ee
The term involving the longitudinal piece can be reduced to $2$-point function by the local Ward identity \eqref{ward1}. This shows that we only need to focus on the transverse part of the conserved current while computing a correlator. The non transverse part can be obtained from the knowledge of the lower point function. In a similar way, if we have insertions of stress tensor $T^{\mu\nu}$, we can focus on the transverse traceless part $t_{\mu\nu}$ satisfying 
\be
\p_\mu t^{\mu\nu}= 0\qquad;\qquad t^\mu_{\;\;\mu}=0
\ee
where 
\be
t^{\mu_1\mu_2}= \Pi^{\mu_1\mu_2}_{\nu_1\nu_2}\; T^{\nu_1\nu_2}\quad;\quad \Pi^{\mu_1\mu_2}_{\nu_1\nu_2} = \f{1}{2}\left[\pi^{\mu_1}_{\;\nu_1} \pi^{\mu_2}_{\;\nu_2} + \pi^{\mu_1}_{\;\nu_2} \pi^{\mu_2}_{\;\nu_1}\right]-\f{1}{d-1}\pi^{\mu_1\mu_2}\pi_{\nu_1\nu_2}
\ee
In general, for a symmetric conserved current having spin $\ell$, the projection operator can be constructed recursively as (see, e.g., \cite{Bruno})
\be
\Pi^{\mu_1\cdots \mu_\ell}_{\nu_1\cdots\nu_\ell} &=& \pi^{\mu_\ell}_{(\nu_\ell} \Pi^{\mu_1\cdots\mu_{\ell-1}}_{\nu_1\cdots\nu_{\ell-1})} - \f{2}{d+2\ell -5} \pi_{(\nu_1\nu_2}\Pi^{\mu_1\cdots\mu_\ell}_{\nu_3\cdots\nu_\ell)}
\ee

\section{Triple K integrals}
\label{apps1}
The momentum space correlators are conveniently expressed in terms of the triple K integrals which are integrals over the product of three modified Bessel functions of the second kind. In this section, we review some basic facts about these integrals. For details and proofs, see \cite{Bzowski:2013sza, Bzowski:2015yxv }. 

\subsection{Basic Identities}
The triple K integrals are defined by
\be
I_{\alpha\{\beta_1,\beta_2,\beta_3\}}\equiv \int_0^\infty dx\, x^\alpha \prod_{j=1}^3 p_j^{\beta_j} K_{\beta_j} (xp_j)\, .
\ee
This is well defined and convergent for
\be
\alpha\;>\;\; \sum_{i=1}^3|\beta_i|-1\qquad;\qquad p_1,\,p_2,\,p_3>0\, .
\ee
The $K_{n}(p x)$ denotes the modified Bessel function of the second kind and satisfies the equation 
\be
x^2\f{d^2y}{dx^2} +x\f{dy}{dx} -(p^2x^2+n^2)y=0\, . \label{beseq}
\ee
Some useful properties of this Bessel function are
\be
K_n(x)&=&K_{-n}(x)\, ;\non\\[.2cm]
K_{n+1}(x) &=&K_{n-1}(x)+\f{2n}{x}K_{n}(x)\, ; \non\\[.2cm]
\f{\p}{\p p}\Bigl[ p^nK_n(px)  \Bigl]&=&-xp^nK_{n-1}(px)\, ; \non\\[.2cm]
\f{\p K_n(x)}{\p x} 
&=&-\f{1}{2}\Bigl(K_{n-1}(x)+K_{n+1}(x)\Bigl) \, .
\ee
For our purposes, it is convenient to introduce the following quantity
\be
J_{N\{k_j\}} =I_{\f{d}{2}-1+N\{\Delta_j-\f{d}{2}+k_j\}}
\ee
The convergence condition for these are
\be
\f{d}{2}-1+N\;>\;\; \sum_{i=1}^3\Bigl|\Delta_j-\f{d}{2}+k_j\Bigl|-1\qquad\implies \qquad N-1 > \Delta_t+k_t-2d
\ee
where, we have assumed $\Delta_i-\f{d}{2}+k_i>0$.

The derivatives of the triple K integral $J_{N\{k_i\}}$ satisfy the following identities
\be
&&\biggl[\sum_{i=1}^3p_i\f{\p}{\p p_i} +(N-\Delta_t+2d-k_t)\biggl]J_{N\{k_j\}}=0 \label{jn1}\\ [.2cm]
&&\f{\p}{\p p_i} J_{N\{k_j\}}=-p_iJ_{N+1\{k_j-\delta_{ij}\}}\label{jn2}\\[.2cm]
&&K_i J_{N\{k_j\}}=2k_iJ_{N+1\{k_j-\delta_{ji}\}}- J_{N+2\{k_j\}}\label{jn3}\\[.2cm]
&&K_{ij}J_{N\{k_\ell\}}=2k_i J_{N+1\{k_\ell-\delta_{i\ell}\}}-2k_jJ_{N+1\{k_\ell-\delta_{j\ell}\}}  \label{jn4}
\ee
where
\be
K_i &\equiv&\biggl[-\f{\p^2}{\p p_i^2}+2\Bigl(\Delta_i-\f{d+1}{2}\Bigl)\f{1}{p_i}\biggl]\quad;\qquad K_{ij}\equiv K_i - K_j
\ee
The first identity in \eqref{jn4} shows that the degree of the triple K integral $J_{N\{k_i\}}$ is $ \Delta_t-2d+k_t-N$. The last identity in \eqref{jn4} shows that $K_{ij}J_{N\{k_\ell\}}=0$ iff $k_i=k_j=0$. This is useful in solving the primary Ward identities. Some other useful identities involving the triple K integrals are 
\be
\hspace*{-.3in}(N+\Delta_t+k_t-d)J_{N\{k_1,k_2,k_3\}} &=&J_{N+1\{k_1+1,k_2,k_3\}} +J_{N+1\{k_1,k_2+1,k_3\}} +J_{N+1\{k_1,k_2,k_3+1\}}\label{wotder1} \\[.3cm]
p_i^2J_{N\{k_j\}}&=& J_{N\{k_j+2\delta_{ji}\}}-2(\Delta_i-\f{d}{2}+k_i+1)J_{N-1\{k_j+\delta_{ji}\}}\label{wotder}
\ee
These identities will be very useful in solving the Ward identities.

\subsection{Regularization}
\label{regtripleK}
It can be shown that the triple K integrals $I_{\alpha\{\beta_1,\beta_2,\beta_3\}}$ diverge if 
\be
\alpha +1 \pm \beta_1 \pm \beta_2 \pm \beta_3 =-2k \qquad,\quad k=0,1,2,\cdots \label{g31}
\ee
This can be written as 
\be
\alpha +1 +\sigma_1 \beta_1 +\sigma_2 \beta_2 +\sigma_3 \beta_3 =-2k_{\sigma_1\sigma_2\sigma_3} \qquad;\quad \sigma_i =\pm 1\quad;\quad k_{\sigma_1\sigma_2\sigma_3}=0,1,2,\cdots \label{g31sigma}
\ee
If the above condition is satisfied, we need to  regularize the integrals. This can be done by shifting the parameters of the triple K integrals as 
\be
I_{\alpha\{\beta_1,\beta_2,\beta_3\}} \rightarrow I_{\tilde\alpha\{\tilde\beta_1,\tilde\beta_2,\tilde\beta_3\}}\quad\implies\qquad J_{N\{k_1,k_2,k_3\}} \rightarrow J_{N+u\epsilon\{k_1+v_1\epsilon,k_2+v_2\epsilon,k_3+v_3\epsilon\}}
\ee
where
\be
\tilde\alpha = \alpha +u\epsilon\quad,\quad \tilde\beta_1 =\beta_1 +v_1\epsilon\quad,\quad \tilde\beta_2 =\beta_2 +v_2\epsilon\quad,\quad \tilde\beta_3 =\beta_3 +v_3\epsilon
\ee
or equivalently
\be
d\rightarrow \tilde d = d+2u\epsilon \qquad;\qquad \Delta\rightarrow \tilde\Delta_i =\Delta_i +(u+v_i)\epsilon
\ee
In general, the regularisation parameters $u$ and $v_i$ are arbitrary. However, in certain cases, there may be constraints on these parameters. E.g., for spin $\ell$ conserved currents, the regularization procedure must satisfy  $\tilde \Delta_i = \tilde d+\ell -2$ together with $ \Delta_i =  d+\ell -2$. This implies $u=v_i$. In general, one may work with arbitrary values of these parameters and take the limit $u\rightarrow v_i$ in the end. 

With the above regularization, the triple K integrals become finite. However, they still diverge as $\epsilon\rightarrow 0$. For explicit calculations, we need to know this divergent behaviour. This was determined in generality in  \cite{Bzowski:2015pba} (see also the streamlined discussion in \cite{Bzowski:2022rlz}) and we summarise the main points here.
To determine this behaviour, we note that $x\rightarrow \infty$ end of the triple K integral converges even when $\epsilon\rightarrow0$ (this happens due to the exponential suppression of modified Bessel function $K_\beta(x)$ as $x\rightarrow\infty$). Hence, all the singularities come from the $x\rightarrow 0$ end. Thus, we can write the regulated triple K integral as  
\be
I_{\tilde\alpha\{\tilde\beta_i\}} &=&\int_0^{\mu^{-1}} dx \; x^{\tilde \alpha} \prod_{j=1}^3 p_j^{\tilde\beta_j}K_{\tilde\beta_j}(p_j x)\;\;+\;\; \int_{\mu^{-1}}^\infty dx x^{\tilde\alpha} \prod_{j=1}^3 p_j^{\tilde\beta_j}K_{\tilde\beta_j}(p_j x)
\ee
where $\mu$ is an arbitrary scale. The full triple K integral  $I_{\tilde\alpha\{\tilde\beta_i\}} $ is independent of this scale. Now, for small $x$, we have following Frobenius expansion
\be
x^{\tilde \alpha} \prod_{j=1}^3 p_j^{\tilde\beta_j}K_{\tilde\beta_j}(p_j x)&=& \sum_{\{\sigma_j=\pm 1\}}\sum_{\{k_j\}=0}^\infty \Bigl(\prod_{i=1}^3\f{(-1)^{k_i}}{2^{\sigma_i\tilde\beta_i+2k_i+1}k_i!}\Gamma(-k_i-\sigma_i\beta_i)p_i^{(1+\sigma_i)\tilde\beta_i+2k_i}\Bigl)x^{\tilde\alpha +\sum_j(\sigma_j\tilde\beta_j+2k_j)}\non\\[.3cm]
&=&\sum_\eta c_\eta x^\eta\label{divgencond}
\ee
where the sum runs over all values of the $\sigma_j$ and all non negative integer values of $k_j$. Also, we have defined
\be
\eta = \tilde\alpha +\sum_j(\sigma_j\tilde\beta_j+2k_j) = -1 + 2\Bigl(-k_{\sigma_1\sigma_2\sigma_3}+\sum_jk_j\Bigl) + \epsilon\Bigl(u+\sum_j v_j\sigma_j\Bigl)
\ee
The expansion \eqref{divgencond} can more explicitly be written as (dropping the tilde)
\be
&&x^\alpha \prod_{i=1}^3p_i^{\beta_i}K_{\beta_i}(xp_i)\non\\
&=& \f{x^{\alpha-\beta_1+\beta_2+\beta_3} }{2^{3-\beta_1+\beta_2+\beta_3}}\Gamma\bigl(\beta_1\bigl)\Gamma\bigl(-\beta_2\bigl)
\Gamma\bigl(-\beta_3\bigl)p_2^{2\beta_2}p_3^{2\beta_3}\;+\; \f{x^{\alpha+\beta_1+\beta_2+\beta_3} }{2^{3+\beta_1+\beta_2+\beta_3}}\Gamma\bigl(-\beta_1\bigl)\Gamma\bigl(-\beta_2\bigl)
\Gamma\bigl(-\beta_3\bigl)p_1^{2\beta_1}p_2^{2\beta_2}p_3^{2\beta_3}\non\\[.3cm]
&&+\; \f{x^{\alpha+\beta_1+\beta_2-\beta_3} }{2^{3+\beta_1+\beta_2-\beta_3}}\Gamma\bigl(-\beta_1\bigl)\Gamma\bigl(-\beta_2\bigl)
\Gamma\bigl(\beta_3\bigl)p_1^{2\beta_1}p_2^{2\beta_2}\;+\; \f{x^{\alpha+\beta_1-\beta_2+\beta_3} }{2^{3+\beta_1-\beta_2+\beta_3}}\Gamma\bigl(-\beta_1\bigl)\Gamma\bigl(\beta_2\bigl)
\Gamma\bigl(-\beta_3\bigl)p_1^{2\beta_1}p_3^{2\beta_3}\non\\[.3cm]
&&+\; \f{x^{\alpha-\beta_1+\beta_2-\beta_3} }{2^{3-\beta_1+\beta_2-\beta_3}}\Gamma\bigl(\beta_1\bigl)\Gamma\bigl(-\beta_2\bigl)
\Gamma\bigl(\beta_3\bigl)p_2^{2\beta_2}\;+\; \f{x^{\alpha-\beta_1-\beta_2+\beta_3} }{2^{3-\beta_1-\beta_2+\beta_3}}\Gamma\bigl(\beta_1\bigl)\Gamma\bigl(\beta_2\bigl)
\Gamma\bigl(-\beta_3\bigl)p_3^{2\beta_3}\non\\[.3cm]
&&+\f{x^{\alpha-\beta_1-\beta_2-\beta_3} }{2^{3-\beta_1-\beta_2-\beta_3}}\Gamma\bigl(\beta_1\bigl)\Gamma\bigl(\beta_2\bigl)\Gamma\bigl(\beta_3\bigl)\;+\; \f{x^{\alpha+\beta_1-\beta_2-\beta_3} }{2^{3+\beta_1-\beta_2-\beta_3}}\Gamma\bigl(-\beta_1\bigl)\Gamma\bigl(\beta_2\bigl)
\Gamma\bigl(\beta_3\bigl)p_1^{2\beta_1}
\ee
Using the above expansion, the triple K integral can be written as 
\be
I_{\tilde\alpha\{\tilde\beta_i\}} &=&\sum_\eta c_\eta\f{\mu^{-\eta-1}}{\eta+1}\;\;+\;\; \int_{\mu^{-1}}^\infty dx x^{\tilde\alpha} \prod_{j=1}^3 p_j^{\tilde\beta_j}K_{\tilde\beta_j}(p_j x)\label{triyu}
\ee
We have used the fact that the lower limit $x=0$ gives a vanishing contribution \cite{Bzowski:2015pba}. Now, demanding that the singularity is independent of the scale $\mu $ gives the condition 
$k_{\sigma_1\sigma_2\sigma_3}=\sum_jk_j
$ and hence we have $\eta =-1 + w\epsilon$, where
\be
w = u+\sum_j v_j\sigma_j
\ee 
The above equations determine the singularity behaviour of the triple K integral as $\epsilon\rightarrow 0$. In general, the coefficient $c_\eta$ in \eqref{triyu} may also be divergent. If the condition \eqref{g31} is satisfied in $m$ ways, the $c_{-1+w\epsilon}$ can diverge as $\epsilon^{-m+1}$ and hence, the triple K integral can diverge as $\epsilon^{-m}$. For our purposes, we need the result when the condition \eqref{g31} is satisfied in a single way. Using, $\mu^{-w\epsilon}= 1-w\epsilon\ln\mu+\cdots$, the singularity for the case of single pole can be written as 
\be
c_{\eta} \f{\mu^{-(\eta+1)}}{\eta+1} =\f{c_{-1+w\epsilon} }{w\epsilon} +O(\ln\mu)
\ee

\subsection{Zero Momentum Limit }
\label{ZML}
In solving the secondary Ward identities, we need to take into account the linear dependence among various triple K integral before setting their individual coefficients to zero. One way to do this is to analyse the equations in the zero momentum limit \cite{Bzowski:2013sza}. In this subsection, we review the behaviour of the triple K integrals in the limit $p_3\rightarrow 0$. In this limit, the momentum conservation gives 
\be
\textbf{p}_1=-\textbf{p}_2 \qquad\implies\quad p_1=p_2\equiv p,
\ee
Also, in this limit, the Bessel functions $K_\beta(p_3x )$ behave as 
\be
p_3^{n}K_{n}(p_3x) &=& \f{2^{n-1}\Gamma(n)}{x^{n}} +O(p_3^2) + p_3^{2n} \Bigl[\f{(-1)^{n+1}}{2^n\Gamma(n+1)}x^{n}\log p_3+\;\mbox{ultralocal}\; +O(p_3^2)\Bigl]\;\;,\; n=1,2,3,\cdots\non\\[.3cm]
p_3^{\beta_3}K_{\beta_3}(p_3x) &=& \f{2^{\beta_3-1}\Gamma(\beta_3)}{x^{\beta_3}} +O(p_3^2) + p_3^{2\beta_3} \Bigl[2^{-\beta_3-1}\Gamma(-\beta_3)x^{\beta_3} +O(p_3^2)\Bigl]\;\;,\quad \beta_3\not\in \mathbb{Z}\non\\[.3cm]
K_0(p_3x) &=&-\log p_3-\log x +\log2 -\gamma_E +O(p_3^2)
\ee 
This shows that for $\beta_3>0$, the zero momentum limit of $p^{\beta_3}_3K_{\beta_3}(p_3x)$ exists. Assuming $\beta_3>0$ and using the identity
 \be
\int_0^\infty dx \; x^{\alpha-1}K_{\beta_1}(px)K_{\beta_2}(px) &=&\f{2^{\alpha-3}}{\Gamma(\alpha)p^\alpha} \prod_{a,b\in \{1,-1\}}\Gamma\Bigl(\f{\alpha+a\beta_1+b\beta_2}{2}\Bigl)\label{k2j2}
 \ee
which is valid for $\mbox{Re}(\alpha) > |\mbox{Re}\; \beta_1| +  |\mbox{Re}\; \beta_2|$ and $ \mbox{Re} (p)>0$,
we find
\be
\lim_{p_3\rightarrow0} J_{N\{k_j\}}(p_1,p_2,p_3) 
&=&\mathcal{A}(N,k_i,\Delta_i,d)\;p^{\Delta_t+k_t-N-2d}\label{zep}
\ee
where, we have defined
\be
\mathcal{A}&\equiv&  \f{2^{N+\f{d}{2}-4}\Gamma(\Delta_3-\f{d}{2}+k_3)}{\Gamma(N+d-\Delta_3-k_3)} \prod_{a,b\in \{1,-1\}}\Gamma\Bigl(\f{N+d-\Delta_3-k_3+a(\Delta_1-\f{d}{2}+k_1)+b(\Delta_2-\f{d}{2}+k_2)}{2}\Bigl)\non
\ee
The compatibility between the convergence condition $\beta_3=\Delta_3-\f{d}{2}+k_3>0$ and the unitarity bound $\Delta_3\ge d+\ell-2$ for the higher spin operators imply the following condition
\be
d+\ell -2 > \f{d}{2}-k_3\qquad\implies\quad k_3 > -\Bigl(\f{d}{2}+\ell -2 \Bigl)
\ee

\section{Some Results About CFT 3 Point Functions}
\label{app:res_3pt}

In this section, we note some useful results about the 3 point functions in CFT. 
\subsection{Number of tensor structures in $d\ge 4$}
In $d\ge 4$, an arbitrary 3-point CFT correlator involving operators of spins $\ell_1,\ell_2$ and $\ell_3$ can have following number of independent tensor structures \cite{Costa:2011mg}
\be
N(\ell_1,\ell_2,\ell_3)=\f{(\ell_1+1)(\ell_1+2)(3\ell_2-\ell_1+3)}{6}- \f{p(p+2)(2p+5)}{24} -\f{1-(-1)^p}{16}\label{gennum}
\ee
where, we have ordered the spins $\ell_1\le \ell_2\le \ell_3$ and defined $p=\mbox{max}(0,\ell_1+\ell_2-\ell_3)$. 

For the cases of our interest, we have
\be
N(1,1,1) =4\qquad;\qquad N(1,2,2) = 7 \qquad;\qquad N(0,0,1) = 1
\ee
If some of the operators are conserved, then the conservation condition eliminates some tensor structures. However, there can be enhancement in the number of tensor structures if the conformal dimension of some operators coincide. We shall consider an example below.

\subsection{Number of tensor structures in $d=3$}

In 3 dimensions, there may be additional relations between the general tensor structures. It turns out that the total number of independent tensor structures in $3d$ are \cite{Costa:2011mg}
\be
N_{3d}(\ell_1,\ell_2,\ell_3)=(2\ell_1+1)(2\ell_2+1)- p(1+p)
\ee
where, we have again ordered the spins $\ell_1\le \ell_2\le \ell_3$ and defined $p=\mbox{max}(0,\ell_1+\ell_2-\ell_3)$. 

Of these, the total number of parity even structures are
\be
N^+_{3d}(\ell_1,\ell_2,\ell_3)=2\ell_1\ell_2+\ell_1+\ell_2+1- \f{p(1+p)}{2}
\ee
and the number of parity odd structures are
\be
N^-_{3d}(\ell_1,\ell_2,\ell_3)=2\ell_1\ell_2+\ell_1+\ell_2- \f{p(1+p)}{2}.
\ee
The number of parity even structures can be expressed in terms of the general formula \eqref{gennum} as
\be
N^+_{3d}(\ell_1,\ell_2,\ell_3)=N(\ell_1,\ell_2,\ell_3)-N(\ell_1-2,\ell_2-2,\ell_3-2)
\ee
This shows that the corrrelators involving spin 2 and higher fields will have less number of tensor structures in 3 dimensions than their counterparts in higher dimensions. It also means that 3-point function of operators having spin 1 fields have same number of tensor structures in all dimensions $d\ge 3$.

\subsection{Correlators Involving Three Spin-1 Operators}

\subsubsection{Generic Structure in Position Space}
The 3-point correlators of three spin-1 fields have 4 independent tensor structures. Using the embedding formalism, it is easy to work out the form of these tensor structures \cite{Costa:2011mg}. The explicit expression of the correlator is given by
\be
&&\hspace*{-.9cm}\langle   \mathcal{O}_{\mu_1}(x_1)\mathcal{O}_{\mu_2}(x_2)\mathcal{O}_{\mu_3}(x_3) \rangle \non\\[.4cm]
 &=&\f{a_1J_{\mu_1}^{23;1}J_{\mu_2}^{31;2}J_{\mu_3}^{12;3}+ a_2 J_{\mu_3}^{12;3}\f{I_{\mu_1\mu_2}(x_{12})}{(x_{12})^2} +a_3 J_{\mu_2}^{31;2}\f{I_{\mu_3\mu_1}(x_{31})}{(x_{31})^2}+a_4 J_{\mu_1}^{23;1}\f{I_{\mu_2\mu_3}(x_{23})}{(x_{23})^2} }{|x_{12}|^{\tau_1+\tau_2-\tau_3-2}|x_{23}|^{\tau_2+\tau_3-\tau_1-2}|x_{31}|^{\tau_3+\tau_1-\tau_2-2}}
\ee
where $a_i$ are arbitrary constants, $\tau_i \equiv \Delta_i+1$ and
\be
J^{ij;k}_{\mu}(x) = \f{(x_{k}-x_i)_\mu}{(x_{k}-x_i)^2}-\f{(x_{k}-x_j)_\mu}{(x_{k}-x_j)^2} \quad,\qquad I_{\mu_1\mu_2}(x_{12})= \delta_{\mu_1\mu_2}-\f{2(x_{12})_{\mu_1}(x_{12})_{\mu_2}}{(x_{12})^2}
\ee

\subsubsection{One Conserved Current}
If one of the operator is a conserved current, we need to impose the conservation condition. {\it E.g.} if the 2nd operator is a conserved current, {\it i.e.}, $\Delta_2=d-1$, then the correlator satisfies
\be
\f{\p}{\p x_2^{\mu_2}}\langle   \mathcal{O}^{\mu_1}(x_1)J^{\mu_2}(x_2)\mathcal{O}^{\mu_3}(x_3) \rangle=0
\ee
Depending upon the conformal dimensions $\Delta_1$ and $\Delta_3$, the above equation eliminates some tensor structures. If $\Delta_1\not=\Delta_3$, two structures get eliminated and the total number of independent conformal tensor structures reduces to 2. On the other hand, if $\Delta_1=\Delta_3$, the total number of independent conformal tensor structures becomes 3. However, if the two operators are identical (so that the 3-point function is symmetric under the exchange of $x_1$ and $x_3$), then all the tensor structures get eliminated and the correlator vanishes.

\section{Regularisation of Divergent Triple K Integrals}
\label{dregu4}
For our 3-point functions, some of the triple K integrals appearing in the solution of Ward identities diverge for $\Delta_1=\Delta_3$. Hence, we need to regularise these divergences using the procedure described in appendix \ref{regtripleK}. In this section, we consider the divergent triple K integrals which appear in the 3-point functions considered in this paper and extract their divergent and finite parts. These will be needed to regularise the full 3-point functions. 

Now, as discussed in appendix \ref{regtripleK}, the regularisation is done by shifting the parameters $d$ and $\Delta_i$. In our case, $\Delta_2$ corresponds to gauge field. To preserve the gauge invariance in the regulated theory, we need to set $v_2=u$. Further, we also have $\Delta_1=\Delta_3$. To maintain this condition also in the regulated theory, we need to set $v_1=v_3$. We shall work with generic values of these regularisation parameters and take the appropriate limit in the end. 
Since the condition \eqref{g31} for the divergent triple K integrals are satisfied in a single way, it follows that these triple K integrals only have single order poles in $\epsilon$. Moreover, since the right hand side of condition \eqref{g31} vanishes, we shall have $\{k_j\}=0$ in the identity \eqref{divgencond}. Thus, using \eqref{divgencond}, the divergent part of the triple K integrals is given by 
\be
\f{\Gamma(-\sigma_1\tilde\beta_1)\Gamma(-\sigma_2\tilde\beta_2)\Gamma(-\sigma_3\tilde\beta_3)}{2^{2-\tilde\alpha}\;\epsilon\; w}p_1^{(1+\sigma_1)\tilde\beta_1}p_2^{(1+\sigma_2)\tilde\beta_2}p_3^{(1+\sigma_3)\tilde\beta_3}\label{divfrt}
\ee
where, 
\be
w= u+\sum_j \sigma_j v_j
\ee

\subsection{Divergent Integrals for $s=1$}
The triple K integrals which diverge for $\Delta_1=\Delta_3$ in the spin 1 case are $J_{1\{0,1,-1\}},J_{1\{1,0,-1\}}, J_{1\{-1,1,0\}} $ and $J_{1\{-1,0,1\}}$. For $\Delta_1=\Delta_3$, these triple K integrals satisfy the divergence condition \eqref{g31} for $k=0$. To analyse these divergent triple K integrals, we start by noting the divergence condition \eqref{g31} for them. The conditions for each integrals are 
\be
J_{1\{0,1,-1\}} \quad :\quad \alpha +1 - \beta_1 - \beta_2 + \beta_3=\Delta_3 - \Delta_1 \qquad (--+)\non\\
J_{1\{1,0,-1\}} \quad :\quad \alpha +1 - \beta_1 - \beta_2 + \beta_3=\Delta_3 - \Delta_1 \qquad (--+)\non\\
J_{1\{-1,1,0\}} \quad :\quad \alpha +1 + \beta_1 - \beta_2 - \beta_3=\Delta_1 - \Delta_3 \qquad (+--)\non\\
J_{1\{-1,0,1\}} \quad :\quad \alpha +1 + \beta_1 - \beta_2 - \beta_3=\Delta_1 - \Delta_3 \qquad (+--)
\ee
Hence, for $\Delta_1=\Delta_3$, these triple K integrals satisfy the condition \eqref{g31} for $k=0$. In general, for integer dimensions, the condition \eqref{g31} may also be satisfied for other triple K integrals and for other combinations of signs. However, in this paper, we focus on non integer dimensions.

Using \eqref{divfrt}, the regularised expressions of first two triple K integrals, namely, $J_{1\{0,1,-1\}}$ and $J_{1\{1,0,-1\}}$ can be written in a single equation as
 \be
J_{N\{a,b,c\}}&=&\frac{ \Gamma \left(\frac{d}{2}-\Delta _3-c\right) \Gamma \left(\Delta_2-\frac{d}{2}+b\right) \Gamma \left(\Delta _1-\frac{d}{2}+a\right) }{ 2^{\frac{d}{2}-\Delta _1-\Delta _2+\Delta _3+3-a-b+c} \left(u-v_1-v_2+v_3\right)}\biggl[\f{1}{\epsilon}+v_1   \left(H_{\Delta_1-\frac{d}{2}+a-1} -\gamma+\log 2\right)\non\\
&&+v_2   \left(H_{\Delta _2-\frac{d}{2}+b-1}-\gamma+\log2\right)-v_3  \Bigl( H_{-\Delta _3+\frac{d}{2}-c-1}-\gamma+\log 2-2\log p_3\Bigl)+O(\epsilon)\biggl]p_3^{-d+2 \Delta _3+2c}\non\\
\label{kcrt1}
\ee
Similarly, the regularised expressions of $J_{1\{-1,1,0\}} $ and $J_{1\{-1,0,1\}}$ are given by 
 \be
J_{N\{a,b,c\}}&=&\frac{ \Gamma \left(\frac{d}{2}-\Delta _1-a\right) \Gamma \left(\Delta_2-\frac{d}{2}+b\right) \Gamma \left(\Delta _3-\frac{d}{2}+c\right) }{ 2^{\frac{d}{2}-\Delta _3-\Delta _2+\Delta _1+3+a-b-c} \left(u+v_1-v_2-v_3\right)}\biggl[\f{1}{\epsilon}-v_1   \left(H_{-\Delta_1+\frac{d}{2}-a-1} -\gamma+\log 2-2\log p_1\right)\non\\
&&+v_2   \left(H_{\Delta _2-\frac{d}{2}+b-1}-\gamma+\log2\right)+v_3  \Bigl( H_{\Delta _3-\frac{d}{2}+c-1}-\gamma+\log 2\Bigl)+O(\epsilon)\biggl]p_1^{-d+2 \Delta _1+2a}\label{kcrt2}
\ee
The $H_n$ in the above expressions denote the Harmonic number which is related to the PolyGamma function by 
\be
H_{n-1} \;=\; \psi(n)+\gamma  \; =\;  \f{\Gamma'(n)}{\Gamma(n)} +\gamma 
\ee
with $\gamma$ being the Euler–Mascheroni constant. 

From the above expressions, we also see that the above regularised expressions for the triple K integrals of the form $J_{N\{a,b,c\}}$ are related to the triple K integrals of the form $J_{N\{c,b,a\}}$ by the interchanges of $\Delta_1\leftrightarrow\Delta_3,\; p_1 \leftrightarrow p_3,\;\; a\leftrightarrow c$ and $v_1\leftrightarrow v_3$. 

\subsection{Divergent Integrals for $s=2$}
The triple K integrals which appear in $B_1$ and are divergent are
\be
&&J_{0\{0,0,-1\}}\;,\;J_{1\{1,0,-1\}}\;,\; J_{2\{0,2,-1\}}\;,\; J_{1\{0,0,-2\}}\;,\; J_{3\{1,1,-2\}}\;,\;J_{3\{2,0,-2\}}, \non\\[.3cm]
&&J_{3\{0,2,-2\}}\;,\;J_{0\{-1,0,-2\}}\;,\;J_{1\{-1,1,-2\}}\;,\;J_{2\{-1,2,-2\}}\;,\; J_{3\{-1,3,-2\}} \label{d148u}
\ee
The condition \eqref{g31} for these triple K integrals give
\be
 \alpha +1 - \beta_1 - \beta_2 + \beta_3=\Delta_3 - \Delta_1 \qquad:\quad (--+)
\ee
Similarly, the triple K integrals which appear in $B_2$ and are divergent are
\be
&&J_{0\{-1,0,0\}}\;,\;J_{1\{-1,0,1\}}\;,\; J_{2\{-1,2,0\}}\;,\; J_{1\{-2,0,0\}}\;,\; J_{3\{-2,1,1\}}\;,\;J_{3\{-2,0,2\}}, \non\\[.3cm]
&&J_{3\{-2,2,0\}}\;,\;J_{0\{-2,0,-1\}}\;,\;J_{1\{-2,1,-1\}}\;,\;J_{2\{-2,2,-1\}}\;,\; J_{3\{-2,3,-1\}} \label{d149u}
\ee
They satisfy
\be
 \alpha +1 + \beta_1 - \beta_2 - \beta_3=\Delta_1 - \Delta_3 \qquad:\quad (+--)
\ee
Thus, for $\Delta_1=\Delta_3$, the above triple K integrals diverge and we need to regularise them. Again using \eqref{divfrt}, the divergent parts of the  triple K integrals in equation \eqref{d148u} can be easily regularised and the regulated expressions are again given by equation \eqref{kcrt1}. Similarly, the regularised divergent parts of the triple K integrals in \eqref{d149u} are given by the expression in \eqref{kcrt2}.

\section{ Equations for form factors for $s=2$}
\label{appen:s2eq}
In this section, we list the primary equations which result from the special conformal Ward identity when it acts on the 3-point function involving two spin 2 non-conserved operators and one conserved vector current. 
There are a total of 60 primary equations.  At $O(z^2)$, we have following equations
\be
(\epsilon_2\cdot\pi_2\cdot p_1)(b\cdot p_1)&:&\quad (K_{1}-K_3) A_0^{(0,0)} =0 \non\\[.3cm]
(\epsilon_2\cdot\pi_2\cdot p_1)(b\cdot p_2)&:&\quad (K_{2}-K_3)A_0^{(0,0)}+2A_1^{(1,1)}=0
\ee
At $O(z)$, we have following equations
\be
(\epsilon_2\cdot\pi_2\cdot p_1)(b\cdot p_1)&:&(K_{1}-K_3) A_1^{(0,0)}-\f{2}{p_1}\f{\p A_1^{(0,0)}}{\p p_1}+\f{2}{p_3}\f{\p A_1^{(0,0)}}{\p p_3} =0 \non\\[.3cm]
&&(K_{1}-K_3) A_1^{(0,1)}+\f{2}{p_3}\f{\p A_1^{(0,1)}}{\p p_3} =0 \non\\[.3cm]
&&(K_{1}-K_3) A_1^{(1,0)}-\f{2}{p_1}\f{\p A_1^{(1,0)}}{\p p_1} =0 \non\\[.3cm]
&&(K_{1}-K_3) A_1^{(1,1)} =0 \non\\[.5cm]
(\epsilon_2\cdot\pi_2\cdot p_1)(b\cdot p_2)&:&(K_{2}-K_3) A_1^{(0,0)}-\f{2}{p_1}\f{\p A_1^{(0,1)}}{\p p_1}+\f{2}{p_3}\f{\p (A_1^{(0,0)}+A_1^{(1,0)}) }{\p p_3} +A_2^{(1,1)}=0 \non\\[.3cm]
&&(K_{2}-K_3) A_1^{(0,1)}+\f{2}{p_3}\f{\p (A_1^{(0,1)}+A_1^{(1,1)})}{\p p_3} +A_2^{(1,2)}=0 \non\\[.3cm]
&&(K_{2}-K_3) A_1^{(1,0)} -\f{2}{p_1}\f{\p A_1^{(1,1)}}{\p p_1} +A_2^{(2,1)}=0\non\\[.3cm]
&&(K_{2}-K_3) A_1^{(1,1)} +A_2^{(2,2)}=0
\ee
\be
(\epsilon_2\cdot\pi_2\cdot \epsilon_1)(b\cdot p_1)&:&(K_{1}-K_3) B_{1,0}^{(0,0)}+\f{2}{p_3}\f{\p B_{1,0}^{(0,0)}}{\p p_3} -2A_1^{(0,1)}=0 \non\\[.3cm]
&&(K_{1}-K_3) B_{1,0}^{(1,0)}-2A_1^{(1,1)} =0 \non\\[.3cm]
(\epsilon_2\cdot\pi_2\cdot \epsilon_1)(b\cdot p_2)&:&(K_{2}-K_3) B_{1,0}^{(0,0)}+\f{2}{p_3}\f{\p (B_{1,0}^{(0,0)}+B_{1,0}^{(1,0)}) }{\p p_3} -2A_1^{(0,1)}+2B_{1,1}^{(1,1)}=0 \non\\[.3cm]
&&(K_{2}-K_3) B_{1,0}^{(1,0)}-2A_1^{(1,1)}+2B_{1;1}^{(2,1)}=0 
\ee
\be
(\epsilon_2\cdot\pi_2\cdot \epsilon_3)(b\cdot p_1)&:&(K_1-K_3)B_{2,0}^{(0,0)}-\f{2}{p_1}\f{\p B_{2,0}^{(0,0)}}{\p p_1}-2A_1^{(1,0)}=0\non\\[.4cm]
&&(K_1-K_3)B_{2,0}^{(0,1)}-2A_1^{(1,1)}\non\\[.5cm]
(\epsilon_2\cdot\pi_2\cdot \epsilon_3)(b\cdot p_2)&:&(K_2-K_3)B_{2,0}^{(0,0)}-\f{2}{p_1}\f{\p B_{2,0}^{(0,1)}}{\p p_1}+2B_{2,1}^{(1,1)}\non\\[.4cm]
&&(K_2-K_3)B_{2,0}^{(0,1)}+2B_{2,1}^{(1,2)}=0
\ee
Finally, at $O(z^0)$, we have following equations
\be
(\epsilon_2\cdot\pi_2\cdot p_1)(b\cdot p_1)&:&(K_{1}-K_3) A_2^{(0,0)}-\f{4}{p_1}\f{\p A_2^{(0,0)}}{\p p_1}+\f{4}{p_3}\f{\p A_2^{(0,0)}}{\p p_3} =0 \non\\[.3cm]
&&(K_{1}-K_3) A_2^{(0,1)}-\f{2}{p_1}\f{\p A_2^{(0,1)}}{\p p_1}+\f{4}{p_3}\f{\p A_2^{(0,1)}}{\p p_3} =0 \non\\[.3cm]
&&(K_{1}-K_3) A_2^{(0,2)} +\f{4}{p_3}\f{\p A_2^{(0,2)}}{\p p_3}=0 \non\\[.3cm]
&&(K_{1}-K_3) A_2^{(1,0)}-\f{4}{p_1}\f{\p A_2^{(1,0)}}{\p p_1}+\f{2}{p_3}\f{\p A_2^{(1,0)}}{\p p_3} =0 \non\\[.5cm]
&&(K_{1}-K_3) A_2^{(1,1)}-\f{2}{p_1}\f{\p A_2^{(1,1)}}{\p p_1}+\f{2}{p_3}\f{\p A_2^{(1,1)}}{\p p_3} =0 \non\\[.5cm]
&&(K_{1}-K_3) A_2^{(1,2)}+\f{2}{p_3}\f{\p A_2^{(1,2)}}{\p p_3} =0 \non\\[.5cm]
&&(K_{1}-K_3) A_2^{(2,0)}-\f{4}{p_1}\f{\p A_2^{(2,0)}}{\p p_1}=0 \non\\[.5cm]
&&(K_{1}-K_3) A_2^{(2,1)}-\f{2}{p_1}\f{\p A_2^{(2,1)}}{\p p_1} =0 \non\\[.5cm]
&&(K_{1}-K_3) A_2^{(2,2)} =0
\ee
\be
(\epsilon_2\cdot\pi_2\cdot p_1)(b\cdot p_2)&:&(K_{2}-K_3) A_2^{(0,0)}-\f{2}{p_1}\f{\p A_2^{(0,1)}}{\p p_1}+\f{1}{p_3}\f{\p (4 A_2^{(0,0)} +2A_2^{(1,0)})}{\p p_3} =0 \non\\[.3cm]
&&(K_{2}-K_3) A_2^{(0,1)}-\f{2}{p_1}\f{\p A_2^{(0,2)}}{\p p_1}+\f{1}{p_3}\f{\p (4A_2^{(0,1)}+2A_2^{(1,1)})}{\p p_3}=0 \non\\[.3cm]
&&(K_{2}-K_3) A_2^{(0,2)} +\f{1}{p_3}\f{\p (4A_2^{(0,2)}+2A_2^{(1,2)})}{\p p_3}=0\non\\[.3cm]
&&(K_{2}-K_3) A_2^{(1,0)} -\f{2}{p_1}\f{\p A_2^{(1,1)}}{\p p_1} +\f{2}{p_3}\f{\p (A_2^{(1,0)}+A_2^{(2,0)})}{\p p_3}=0\non\\[.3cm]
&&(K_{2}-K_3) A_2^{(1,1)} -\f{2}{p_1}\f{\p A_2^{(1,2)}}{\p p_1} +\f{2}{p_3}\f{\p (A_2^{(1,1)}+A_2^{(2,1)})}{\p p_3}=0\non\\[.3cm]
&&(K_{2}-K_3) A_2^{(1,2)}  +\f{2}{p_3}\f{\p (A_2^{(1,2)}+A_2^{(2,2)})}{\p p_3}=0\non\\[.3cm]
&&(K_{2}-K_3) A_2^{(2,0)} -\f{2}{p_1}\f{\p A_2^{(2,1)}}{\p p_1}=0\non\\[.3cm]
&&(K_{2}-K_3) A_2^{(2,1)} -\f{2}{p_1}\f{\p A_2^{(2,2)}}{\p p_1}=0\non\\[.3cm]
&&(K_{2}-K_3) A_2^{(2,2)} =0
\ee
\be
(\epsilon_2\cdot\pi_2\cdot \epsilon_1)(b\cdot p_1)&:&(K_{1}-K_3) B_{1,1}^{(0,0)}-\f{2}{p_1}\f{\p B_{1,1}^{(0,0)}}{\p p_1}+\f{4}{p_3}\f{\p B_{1,1}^{(0,0)}}{\p p_3} -A_2^{(0,1)}=0 \non\\[.3cm]
&&(K_{1}-K_3) B_{1,1}^{(0,1)}+\f{4}{p_3}\f{\p B_{1,1}^{(0,1)}}{\p p_3} -A_2^{(0,2)}=0 \non\\[.3cm]
&&(K_{1}-K_3) B_{1,1}^{(1,0)} -\f{2}{p_1}\f{\p B_{1,1}^{(1,0)}}{\p p_1}+\f{2}{p_3}\f{\p B_{1,1}^{(1,0)}}{\p p_3}-A_2^{(1,1)}=0 \non\\[.3cm]
&&(K_{1}-K_3) B_{1;1}^{(1,1)}+\f{2}{p_3}\f{\p B_{1;1}^{(1,1)}}{\p p_3}-A_2^{(1,2)} =0 \non\\[.5cm]
&&(K_{1}-K_3) B_{1;1}^{(2,0)}-\f{2}{p_1}\f{\p B_{1;1}^{(2,0)}}{\p p_1}-A_2^{(2,1)}=0 \non\\[.5cm]
&&(K_{1}-K_3) B_{1;1}^{(2,1)}- A_2^{(2,2)}=0
\ee
\be
(\epsilon_2\cdot\pi_2\cdot \epsilon_1)(b\cdot p_2)&:&(K_{2}-K_3) B_{1,1}^{(0,0)}-\f{2}{p_1}\f{\p B_{1,1}^{(0,1)}}{\p p_1}+\f{1}{p_3}\f{\p (4B_{1,1}^{(0,0)}+2B_{1;1}^{(1,0)})}{\p p_3} -A_2^{(0,1)}=0 \non\\[.3cm]
&&(K_{2}-K_3) B_{1,1}^{(0,1)}+\f{1}{p_3}\f{\p (4B_{1,1}^{(0,1)}+2B_{1,1}^{(1,1)})}{\p p_3} -A_2^{(0,2)}=0 \non\\[.3cm]
&&(K_{2}-K_3) B_{1,1}^{(1,0)} -\f{2}{p_1}\f{\p B_{1,1}^{(1,1)}}{\p p_1}+\f{2}{p_3}\f{\p (B_{1,1}^{(1,0)}+B_{1,1}^{2,0})}{\p p_3}-A_2^{(1,1)}=0 \non\\[.3cm]
&&(K_{2}-K_3) B_{1;1}^{(1,1)}+\f{2}{p_3}\f{\p (B_{1;1}^{(1,1)}+B_{1;1}^{2,1})}{\p p_3}-A_2^{(1,2)} =0 \non\\[.5cm]
&&(K_{2}-K_3) B_{1;1}^{(2,0)}-\f{2}{p_1}\f{\p B_{1;1}^{(2,1)}}{\p p_1}-A_2^{(2,1)}=0 \non\\[.5cm]
&&(K_{2}-K_3) B_{1;1}^{(2,1)}- A_2^{(2,2)}=0
\ee
\be
(\epsilon_2\cdot\pi_2\cdot \epsilon_3)(b\cdot p_1)&:&(K_{1}-K_3) B_{2,1}^{(0,0)}-\f{4}{p_1}\f{\p B_{2,1}^{(0,0)}}{\p p_1}+\f{2}{p_3}\f{\p B_{2,1}^{(0,0)}}{\p p_3} -A_2^{(1,0)}=0 \non\\[.3cm]
&&(K_{1}-K_3) B_{2,1}^{(0,1)}-\f{2}{p_1}\f{\p B_{2,1}^{(0,1)}}{\p p_1}+\f{2}{p_3}\f{\p B_{2,1}^{(0,1)}}{\p p_3} -A_2^{(1,1)}=0 \non\\[.3cm]
&&(K_{1}-K_3) B_{2,1}^{(0,2)} +\f{2}{p_3}\f{\p B_{2,1}^{(0,2)}}{\p p_3}-A_2^{(1,2)}=0 \non\\[.3cm]
&&(K_{1}-K_3) B_{2;1}^{(1,0)}-\f{4}{p_1}\f{\p B_{2;1}^{(1,0)}}{\p p_1}-A_2^{(2,0)} =0 \non\\[.5cm]
&&(K_{1}-K_3) B_{2;1}^{(1,1)}-\f{2}{p_1}\f{\p B_{2;1}^{(1,1)}}{\p p_1}-A_2^{(2,1)}=0 \non\\[.5cm]
&&(K_{1}-K_3) B_{2;1}^{(1,2)}- A_2^{(2,2)}=0
\ee
\be
(\epsilon_2\cdot\pi_2\cdot \epsilon_3)(b\cdot p_2)&:&(K_{2}-K_3) B_{2,1}^{(0,0)}-\f{2}{p_1}\f{\p B_{2,1}^{(0,1)}}{\p p_1}+\f{2}{p_3}\f{\p (B_{2;1}^{(0,0)}+B_{2;1}^{(1,0)})}{\p p_3} =0 \non\\[.3cm]
&&(K_{2}-K_3) B_{2,1}^{(0,1)}-\f{2}{p_1}\f{\p B_{2,1}^{(0,2)}}{\p p_1}+\f{2}{p_3}\f{\p (B_{2,1}^{(0,1)}+B_{2,1}^{(1,1)})}{\p p_3} =0 \non\\[.3cm]
&&(K_{2}-K_3) B_{2,1}^{(0,2)} +\f{2}{p_3}\f{\p (B_{2,1}^{(0,2)}+B_{2,1}^{1,2})}{\p p_3}=0 \non\\[.3cm]
&&(K_{2}-K_3) B_{2;1}^{(1,0)}-\f{2}{p_1}\f{\p B_{2;1}^{(1,1)}}{\p p_1} =0 \non\\[.5cm]
&&(K_{2}-K_3) B_{2;1}^{(1,1)}-\f{2}{p_1}\f{\p B_{2;1}^{(1,2)}}{\p p_1}=0 \non\\[.5cm]
&&(K_{2}-K_3) B_{2;1}^{(1,2)}=0
\ee

\section{Solution of Secondary Equations for $s=2$ and $\Delta_1\not=\Delta_3$}
\label{s=2secondary}
For $\Delta_1\not=\Delta_3$, the secondary equations give following relations between the coefficients for the spin 2 case
\be
a_2^{(2,2)}&=&\frac{4 \left(\Delta _3 a_2^{(1,2)}+\Delta _1 a_2^{(2,1)}\right)}{\left(\Delta _1-\Delta _3\right) \left(d+3 \Delta _1+3 \Delta _3+4\right)}\non\\[.5cm]
a_2^{(0,2)}&=&-\frac{4 \left(\Delta _1-\Delta _3\right) \left(-d+\Delta _1+\Delta _3+2\right) a_1^{\{1,1\}}}{2\Delta _3 \left(\Delta _3-1\right)}+\frac{2 \left(\Delta _1-1\right) \Delta _1 a_2^{\{2,0\}}}{2\Delta _3 \left(\Delta _3-1\right)}\non\\
&&+\frac{\left(\left(\Delta _1+\Delta _3\right) \left(\Delta _1 \left(2 d+3 \Delta _1+8\right)-(d-8) d-3 \Delta _3^2\right)-4 (d-2) d\right) a_2^{\{1,2\}}}{2(d+3 \Delta _1+3 \Delta _3+4) \left(\Delta _3-1\right)}\non\\
&&+\frac{\Delta _1 \left(-\left(\Delta _1+\Delta _3\right) \left(\left(d+\Delta _3\right) \left(d-3 \Delta _3-8\right)+3 \Delta _1^2\right)-4 (d-2) d\right) a_2^{\{2,1\}}}{2\Delta _3 \left(d+3 \Delta _1+3 \Delta _3+4\right) \left(\Delta _3-1\right)}\non\\[.5cm]
a_2^{(1,1)}&=& \frac{2 \left(-d+\Delta _1+\Delta _3+2\right) a_1^{(1,1)}}{\Delta _3}+\frac{(d-2) \left(\left(\Delta _1+\Delta _3\right) \left(d+\Delta _1-\Delta _3\right)+2 d\right) a_2^{(1,2)}}{\left(\Delta _1-\Delta _3\right) \left(d+3 \Delta _1+3 \Delta _3+4\right)}\non\\
&&+\f{a_2^{(2,1)}}{2 \left(\Delta _1-\Delta _3\right) \Delta _3 \left(d+3 \Delta _1+3 \Delta _3+4\right)}\Bigl[\Delta _1^2 \left((d-8) d-2 \Delta _3 \left(2 d+3 \Delta _3+2\right)\right)\non\\
&&+2 (d-2) d \left(\Delta _3+2\right) \Delta _1+\Delta _3^2 \left(d+\Delta _3\right) \left(d+3 \Delta _3+4\right)+3 \Delta _1^4\Bigl]-\frac{\left(\Delta _1-1\right) a_2^{(2,0)}}{\Delta _3}\non
\ee
\be
a_2^{(1,0)}&=& \frac{\left(\Delta _1+\Delta _3-2\right) \left(d+\Delta _1-\Delta _3\right) \left(-d+\Delta _1+\Delta _3+2\right) a_1^{(1,1)}}{\Delta _1 \Delta _3}\non\\
&&+\f{1}{4 \left(\Delta _1-\Delta _3\right) \Delta _3 \left(d+3 \Delta _1+3 \Delta _3+4\right)}\Bigl[\left(\Delta _1+\Delta _3-2\right) \left(d+\Delta _1-\Delta _3\right) \non\\
&&\left(\left(\Delta _1+\Delta _3\right) \left((d-2) \Delta _1-\Delta _3 \left(3 d+2 \Delta _3+2\right)+(d-6) d+2 \Delta _1^2\right)+4 (d-2) d\right)\Bigl]a_2^{(2,1)}\non\\
&&+\f{1}{4 \Delta _1 \left(\Delta _1-\Delta _3\right) \left(d+3 \Delta _1+3 \Delta _3+4\right)}\Bigl[\left(\Delta _1+\Delta _3-2\right) \left(d+\Delta _1-\Delta _3\right)\non\\
&& \left(\left(\Delta _1+\Delta _3\right) \left(\Delta _3 \left(-2 d+\Delta _3+2\right)+(d-6) d-\Delta _1 \left(\Delta _1+6\right)\right)+4 (d-2) d\right) \Bigl]a_2^{(1,2)}\non\\
&&-\frac{\left(\Delta _1+\Delta _3-2\right) \left(3 d+\Delta _1+\Delta _3-4\right) a_2^{(2,0)}}{4 \Delta _3}\non
\ee

\be
a_2^{(0,1)}&=&-\frac{\left(\Delta _1+\Delta _3-2\right) \left(-d+\Delta _1+\Delta _3+2\right) \left(\Delta _1 \left(3 d-2 \Delta _3-2\right)-\left(\Delta _3+2\right) \left(d-\Delta _3\right)+\Delta _1^2\right) a_1^{(1,1)}}{2 \Delta _1 \left(\Delta _3-1\right) \Delta _3}\non\\
&&-\frac{(\Delta _1+\Delta _3-2)}{8 \left(\Delta _1-\Delta _3\right) \left(\Delta _3-1\right) \Delta _3 \left(d+3 \Delta _1+3 \Delta _3+4\right)}\biggl[-8 (d-2) d^2+\Delta _3^4 \left(9 d+3 \Delta _1-14\right)\non\\
&&+\Delta _3^3 \left(-2 \Delta _1 \left(2 d+\Delta _1-6\right)+d (7 d+2)-20\right)-\Delta _3^2 \Bigl(\Delta _1 \left(6 \Delta _1 \left(3 d+\Delta _1-4\right)+d (5 d-6)+4\right)\non\\
&&+d ((d-18) d+16)\Bigl)+\Delta _3 \Bigl(\Delta _1 \left(\Delta _1 \left(4 (d-3) \Delta _1-d (11 d+10)+\Delta _1^2+20\right)+2 (d-6) (d-2) d\right)\non\\
&&-2 (d-4) d (3 d-2)\Bigl)+\Delta _1 \Bigl(\Delta _1 \left(\Delta _1 \left(\Delta _1 \left(9 d+3 \Delta _1-10\right)+(d-14) d+4\right)+d (3 (d-6) d+8)\right)\non\\
&&+2 d (5 (d-2) d+8)\Bigl)+\Delta _3^5 \biggl]a_2^{(2,1)}\non\\
&&-\frac{\Delta _1+\Delta _3-2}{8 \Delta _1 \left(\Delta _1-\Delta _3\right) \left(\Delta _3-1\right) \left(d+3 \Delta _1+3 \Delta _3+4\right)}\biggl[\Delta _1^4 \left(\Delta _3-11 d\right)+\Delta _1^3 \Bigl(-d (5 d+26)+6 \Delta _3^2\non\\
&&-4 \Delta _3+28\Bigl)+\Delta _1^2 \left((d-4) d (3 d-4)-\Delta _3 \left(2 \Delta _3 \left(-7 d+\Delta _3+2\right)+d (7 d-6)+4\right)\right)\non\\
&&+\Delta _1 \left(\Delta _3 \left(\Delta _3 \left(d (d+18)+\Delta _3 \left(4-3 \Delta _3\right)-28\right)+2 (d-6) (d-2) d\right)+2 d (5 (d-2) d+8)\right)\non\\
&&-\left(\Delta _3+2\right) \left(d-\Delta _3-2\right) \left(d-\Delta _3\right) \left(\Delta _3 \left(d-\Delta _3\right)+4 d\right)-3 \Delta _1^5\biggl]a_2^{(1,2)}\non\\
&&+\frac{\left(\Delta _1-1\right) \left(\Delta _1+\Delta _3-2\right) \left(3 d+\Delta _1+\Delta _3-4\right) a_2^{(2,0)}}{4 \left(\Delta _3-1\right) \Delta _3}\non
\ee

\be
a_2^{(0,0)}&=&\frac{(d-2) \left(\Delta _1+\Delta _3-4\right) \left(\Delta _1+\Delta _3-2\right) \left(d-\Delta _1-\Delta _3-2\right) \left(d+\Delta _1-\Delta _3\right) a_1^{(1,1)}}{2 \Delta _1 \left(\Delta _3-1\right) \Delta _3}\non\\
&&-\f{(d-2) \left(\Delta _1+\Delta _3-4\right) \left(\Delta _1+\Delta _3-2\right) \left(d+\Delta _1-\Delta _3\right)}{8 \left(\Delta _1-\Delta _3\right) \left(\Delta _3-1\right) \Delta _3 \left(d+3 \Delta _1+3 \Delta _3+4\right)}\biggl[4 (d-2) d\non\\
&&+\left(\Delta _1+\Delta _3\right) \left(\left(d+\Delta _3\right) \left(d-5 \Delta _3-6\right)+3 \Delta _1^2+2 \left(\Delta _3-1\right) \Delta _1\right) \biggl] a_2^{(2,1)}\non\\
&&-\f{(d-2) \left(\Delta _1+\Delta _3-4\right) \left(\Delta _1+\Delta _3-2\right) \left(d+\Delta _1-\Delta _3\right)}{8 \left(\Delta _1-\Delta _3\right) \left(\Delta _3-1\right) \Delta _1 \left(d+3 \Delta _1+3 \Delta _3+4\right)}\biggl[4 (d-2) d\non\\
&&+\left(\Delta _1+\Delta _3\right) \Bigl(-2 \Delta _1 \left(d-\Delta _3+5\right)+\Delta _3 \left(-2 d+\Delta _3+2\right)+(d-6) d-3 \Delta _1^2\Bigl) \biggl] a_2^{(1,2)}\non\\
&&+\frac{(d-2) \left(\Delta _1+\Delta _3-4\right) \left(\Delta _1+\Delta _3-2\right) \left(d+\Delta _1+\Delta _3-2\right) a_2^{(2,0)}}{4 \left(\Delta _3-1\right) \Delta _3}\non
\ee

\be
a_1^{(1,0)}&=&\frac{ \left(\left(\Delta _1+\Delta _3\right) \left(d+\Delta _1-\Delta _3\right)-2 d+4\right) a_1^{(1,1)}}{2\Delta _1}-\f{1}{4} \left(\Delta _1-1\right) a_2^{(2,0)}
\non\\
&&-\frac{ \left(\left(\Delta _1+\Delta _3\right) \left(d+\Delta _1-\Delta _3\right) \left(\Delta _3 \left(d+\Delta _3+2\right)+d-\Delta _1^2+2\right)-4 (d-2) d\right) a_2^{(2,1)}}{4\left(\Delta _1-\Delta _3\right) \left(d+3 \Delta _1+3 \Delta _3+4\right)}\non\\
&&-\frac{\Delta _3 \left(\left(\Delta _1+\Delta _3\right) \left(\Delta _1+\Delta _3+2\right) \left(d+\Delta _1-\Delta _3\right) \left(d+\Delta _1-\Delta _3+2\right)-8 (d-2) d\right) a_2^{(1,2)}}{8\Delta _1 \left(\Delta _1-\Delta _3\right) \left(d+3 \Delta _1+3 \Delta _3+4\right)}\non
\ee

\be
a_1^{(0,1)}&=&-\frac{\left((d-1) \Delta _3-d+\Delta _1+2\right) a_1^{(1,1)}}{\Delta _3} + \frac{\left(\Delta _1-1\right) \Delta _1 a_2^{(2,0)}}{4 \Delta _3}\non\\
&&-\f{\Delta _1}{4 \left(\Delta _1-\Delta _3\right) \Delta _3 \left(d+3 \Delta _1+3 \Delta _3+4\right)}\biggl[ d \Delta _1^3-\Delta _1^2 \left(3 d+2 \Delta _3 \left(\Delta _3+1\right)-2\right)+4 (d-2) d\non\\
&&-d \Delta _1 \left(\Delta _3 \left(d+\Delta _3+2\right)-d+6\right)-\Delta _3 \left(d-\Delta _3\right) \left(\Delta _3 \left(d+\Delta _3+2\right)+3 d-2\right) +\Delta _1^4\biggl]a_2^{(2,1)}\non\\
&&+\f{1}{8 \left(\Delta _1-\Delta _3\right) \left(d+3 \Delta _1+3 \Delta _3+4\right)}\biggl[-2 d \Delta _3^3+\Delta _3^2 \left(-2 \Delta _1 \left(d+\Delta _1+2\right)+(d-6) d+4\right)\non\\
&&+2 d \Delta _3 \left(\Delta _1 \left(d+\Delta _1+2\right)+3 d-2\right)+\Delta _1 \left(\Delta _1 \left(\Delta _1 \left(2 d+\Delta _1+4\right)+d (d+10)-4\right)-2 (d-6) d\right)\non\\
&&-8 (d-2) d+\Delta _3^4\biggl]a_2^{(1,2)}\non
\ee

\be
a_1^{(0,0)}&=&-\frac{(d-2) \left(\Delta _1+\Delta _3-2\right) \left(d+\Delta _1-\Delta _3-2\right) a_1^{(1,1)}}{2 \Delta _1}+\frac{(d-2) \left(\Delta _1-1\right) \left(\Delta _1+\Delta _3-2\right) a_2^{(2,0)}}{4 \Delta _3}\non\\
&&+\f{(d-2) \left(\Delta _1+\Delta _3-2\right)}{4 \left(\Delta _1-\Delta _3\right) \Delta _3 \left(d+3 \Delta _1+3 \Delta _3+4\right)}\biggl[ \Delta _3^2 \left(\Delta _1 \left(d+2 \Delta _1-1\right)+(d-1) d\right)\non\\
&&+\Delta _3 \left(d (d+4) \Delta _1+2 (d-2) d+\Delta _1^2\right)-\left(\Delta _1-1\right) \Delta _1^2 \left(d+\Delta _1\right)-\Delta _3^4-\Delta _3^3\biggl]a_2^{(2,1)}\non\\
&&+\f{(d-2) \left(\Delta _1+\Delta _3-2\right)}{8 \Delta_1\left(\Delta _1-\Delta _3\right) \left(d+3 \Delta _1+3 \Delta _3+4\right)}\biggl[-2 (d-1) \Delta _3^3+\Delta _3^2 \left((d-6) d-2 \Delta _1 \left(d+\Delta _1+3\right)\right)\non\\
&&+2 \Delta _3 \left((d-1) \Delta _1^2+d (d+4) \Delta _1+2 (d-2) d\right)+\Delta _1^2 \left(d+\Delta _1\right) \left(d+\Delta _1+6\right)+\Delta _3^4\biggl]a_2^{(1,2)}\non
\ee

\be
a_1^{(0,0)}&=&\frac{(d-2) \left(\left(\Delta _1-\Delta _3\right) \left(d+3 \Delta _1+3 \Delta _3+4\right) a_1^{\{1,1\}}-d \Delta _3 a_2^{\{1,2\}}-d \Delta _1 a_2^{\{2,1\}}\right)}{\left(\Delta _1-\Delta _3\right) \left(d+3 \Delta _1+3 \Delta _3+4\right)}\non
\ee

\be
b_{1;0}^{(1,0)}&=&\frac{\left(\Delta _3 \left(2 d+\Delta _1-3\right)+\left(\Delta _1-1\right) \left(d-\Delta _1-2\right)+2 \Delta _3^2\right) a_1^{(1,1)}}{\Delta _3}-\frac{\left(\Delta _1-1\right) \Delta _1 \left(\Delta _3-1\right) a_2^{(2,0)}}{4 \Delta _3}\non\\
&&+\f{\Delta _1}{4 \left(\Delta _1-\Delta _3\right) \Delta _3 \left(d+3 \Delta _1+3 \Delta _3+4\right)}\biggl[ -d (d+5) \Delta _3^3-(d+1) (3 d+2) \Delta _3^2+d (10-7 d) \Delta _3\non\\
&&+d \Delta _1^3 \left(\Delta _3+1\right)-d \Delta _1 \left(\Delta _3+1\right) \left(\Delta _3 \left(d+\Delta _3+2\right)+3 d-2\right)\non\\
&&+\Delta _1^2 \left(\Delta _3 \left(3 d-2 \left(\Delta _3-1\right) \Delta _3\right)-(d-5) d+2\right)+4 (d-2) d+\Delta _3^5+\Delta _1^4 \left(\Delta _3-2\right) \biggl]a_2^{(2,1)}\non\\
&&-\f{1}{8 \left(\Delta _1-\Delta _3\right) \left(d+3 \Delta _1+3 \Delta _3+4\right)}\biggl[\Delta _3^3 \left(d^2-2 \Delta _1 \left(d+\Delta _1+2\right)-12\right)\non\\
&&+\Delta _3 \left(\Delta _1 \left(8 d^2+\Delta _1 \left(\Delta _1 \left(2 d+\Delta _1+4\right)+d (d+4)+12\right)\right)+2 d (7 d-10)\right)\non\\
&&-(2 d+5) \Delta _3^4+\Delta _3^2 \left(2 \Delta _1 \left((d+5) \Delta _1+d (d+3)+6\right)+d (5 d+6)+4\right)\non\\
&&+\Delta _1 \left(\Delta _1+2\right) \left(d+\Delta _1\right) \left(3 d-5 \Delta _1-2\right)-8 (d-2) d+\Delta _3^5\biggl]a_2^{(1,2)}\non
\ee

\be
b_{1;0}^{(0,0)}&=& -\frac{\left(\Delta _1+\Delta _3-2\right) \left(-2 \Delta _1 \left(d+\Delta _3-1\right)+\Delta _3 \left(-2 d-3 \Delta _3+2\right)+(d-2) d+\Delta _1^2\right) a_1^{(1,1)}}{2 \Delta _3}    \non\\
&&+\frac{\Delta _1 \left(\Delta _1+\Delta _3-2\right)}{8 \left(\Delta _1-\Delta _3\right) \Delta _3 \left(d+3 \Delta _1+3 \Delta _3+4\right)}\biggl[\left(\Delta _1+\Delta _3\right) \Bigl(-d \Delta _3^3-\Delta _3^2 \Bigl(d (d+11)\non\\
&&+2 \left(\Delta _1-1\right) \Delta _1-2\Bigl)+d \Delta _3 \Bigl((d-6) d+\Delta _1 \left(\Delta _1+4\right)-12\Bigl)+\Delta _1 \Bigl(\Delta _1 \Bigl(\left(\Delta _1-2\right) \Delta _1\non\\
&&-(d-2) (d-1)\Bigl)-2 (d-4) d\Bigl)+d ((d-10) d+8)+\Delta _3^4\Bigl)+4 (d-2) d^2\biggl]a_2^{(2,1)}\non\\
&&-\frac{\Delta _1+\Delta _3-2}{16 \left(\Delta _1-\Delta _3\right) \left(d+3 \Delta _1+3 \Delta _3+4\right)}\biggl[-8 (d-2) d^2+\left(\Delta _1+\Delta _3\right) \Bigl(\Delta _1 \Bigl(\Delta _1 \bigl(\Delta _1 \left(d+\Delta _1-4\right)\non\\
&&-(d-2) d-12\bigl)-(d-4) d^2\Bigl)-(3 d+8) \Delta _3^3-\Delta _3^2 \left(\Delta _1 \left(d+2 \Delta _1-4\right)+d (2-3 d)+20\right)\non\\
&&+\Delta _3 \left(\Delta _1 \left((3 d+8) \Delta _1+2 d (d+4)+32\right)+d (8-(d-12) d)\right)-2 d ((d-10) d+8)+\Delta _3^4\Bigl)\biggl]a_2^{(1,2)}\non\\
&&-\frac{\left(\Delta _1-1\right) \Delta _1 \left(\Delta _1+\Delta _3-2\right) \left(-d+\Delta _1+\Delta _3\right) a_2^{(2,0)}}{8 \Delta _3}\non
\ee

\be
b_{1;1}^{(2,1)}&=& 2a_1^{(1,1)}+\frac{\Delta _1 \left(-d \Delta _1-\Delta _3 \left(3 d+\Delta _3\right)-4 d+\Delta _1^2\right) a_2^{(2,1)}}{2 \left(\Delta _1-\Delta _3\right) \left(d+3 \Delta _1+3 \Delta _3+4\right)}\non\\
&&-\frac{\Delta _3 \left(\Delta _1+\Delta _3+2\right) \left(d+\Delta _1-\Delta _3\right) a_2^{(1,2)}}{\left(\Delta _1-\Delta _3\right) \left(d+3 \Delta _1+3 \Delta _3+4\right)}\non
\ee

\be
b_{1;1}^{(1,1)}&=&\frac{\left(-(d-1) \Delta _1-\Delta _3 \left(d+\Delta _3-1\right)+d+\Delta _1^2-2\right) a_1^{(1,1)}}{\Delta _3 \left(\Delta _1+\Delta _3\right)}-\frac{\left(\Delta _1-1\right) \Delta _1 a_2^{(2,0)}}{4 \Delta _3 \left(\Delta _1+\Delta _3\right)}\non\\
&&-\f{\Delta_1}{4 \Delta _3 \left(\Delta _3^2-\Delta _1^2\right) \left(d+3 \Delta _1+3 \Delta _3+4\right)}\biggl[ \left(\Delta _1+\Delta _3\right) \Bigl(-\Delta _1^2 \left(d+2 \Delta _3\right)\non\\
&&+\Delta _1 \left(-\Delta _3 \left(3 d+2 \Delta _3+2\right)+(d-5) d-2\right)+\Delta _3 \left(2 \Delta _3 \left(2 d+\Delta _3+1\right)+d (2 d+3)+2\right)\non\\
&&+d (3 d-2)+2 \Delta _1^3\Bigl)-4 (d-2) d  \biggl]a_2^{(2,1)}\non\\
&&+\f{1}{8 \left(\Delta _1^2-\Delta _3^2\right) \left(d+3 \Delta _1+3 \Delta _3+4\right)}\biggl[ \left(\Delta _1+\Delta _3\right) \Bigl(\Delta _1^2 \left(5 \Delta _3-2 (d+12)\right)\non\\
&&+\Delta _1 \left(d (3 d-10)+5 \Delta _3 \left(\Delta _3+4\right)-20\right)+\Delta _3 \left(\Delta _3 \left(2 d-5 \Delta _3+4\right)+3 d (d+2)+20\right)\non\\
&&+2 d (3 d-2)-5 \Delta _1^3\Bigl)-8 (d-2) d  \biggl]a_2^{(1,2)}\non
\ee

\be
b_{1;1}^{(0,1)}
&=&\frac{\left(\Delta _1+\Delta _3-2\right) \left(-d+\Delta _1+\Delta _3+2\right) \left(2 \left(d+\Delta _3-1\right)-\left(\Delta _1+\Delta _3\right) \left(d-\Delta _1+\Delta _3\right)\right) a_1^{(1,1)}}{4 \left(\Delta _3-1\right) \Delta _3}\non\\
&&+\frac{\Delta _1 \left(\Delta _1+\Delta _3-2\right)}{16 \left(\Delta _1-\Delta _3\right) \left(\Delta _3-1\right) \Delta _3 \left(d+3 \Delta _1+3 \Delta _3+4\right)}\biggl[\Delta _1 \Bigl( \Delta _3 \Bigl(\Delta _3 \left(5 d^2+\Delta _3 \left(6 d+3 \Delta _3+4\right)+4\right)\non\\
&&-2 (d-2)^2 d\Bigl)-2 d (d (d+3)-6)\Bigl)+\Delta _3 \Bigl(\Delta _3 \left(\Delta _3 \left(-d^2+\Delta _3 \left(d+\Delta _3+2\right)+4\right)-d \left(d^2+6\right)+4\right)\non\\
&&-2 d ((d-1) d+2)\Bigl)+\Delta _1^4 \left(-3 d+\Delta _3+2\right)+\Delta _1^3 \left(-2 \Delta _3 \left(3 d+3 \Delta _3+2\right)+(d-4) d-4\right)+3 \Delta _1^5\non\\
&&-\Delta _1^2 \left(\Delta _3 \left(2 \Delta _3 \left(-d+\Delta _3+2\right)+d (4-7 d)+4\right)+d ((d-8) d+2)+4\right)+8 (d-2) (d-1) d\biggl]a_2^{(2,1)}\non\\
&&-\frac{\Delta _1+\Delta _3-2}{16 \left(\Delta _1-\Delta _3\right) \left(\Delta _3-1\right) \left(d+3 \Delta _1+3 \Delta _3+4\right)}\biggl[\Delta _3^4 \left(-3 d+3 \Delta _1-7\right)-2 \Delta _3^3 \Bigl(\Delta _1 \left(d+2 \Delta _1+6\right)\non\\
&&+5 d+2\Bigl)+\Delta _3^2 \left(\Delta _1 \left(2 \Delta _1 \left(2 d-3 \Delta _1-1\right)+d (4-3 d)+4\right)+d ((d-1) d+6)+12\right)\non\\
&&+2 \Delta _3 \left(\Delta _1 \left(\Delta _1 \left(\Delta _1 \left(d+\Delta _1+6\right)+d (7-3 d)+2\right)+d (d-2)^2\right)+d ((d-1) d+2)\right)+2 \Delta _3^5\non\\
&&+\Delta _1 \left(d+\Delta _1\right) \left(\Delta _1 \left(\Delta _1 \left(-4 d+3 \Delta _1+9\right)+(d-9) d-4\right)+2 (d (d+3)-6)\right)\non\\
&&-8 (d-2) (d-1) d\biggl]a_2^{(1,2)}+\frac{\left(\Delta _1-1\right) \Delta _1 \left(\Delta _1+\Delta _3-2\right) \left(d-\Delta _1-1\right) a_2^{(2,0)}}{8 \left(\Delta _3-1\right) \Delta _3}\non
\ee

\be
b_{1;1}^{(1,0)}&=&\frac{\left(\Delta _1+\Delta _3-2\right)}{2 \Delta _1 \Delta _3}\biggl[ -\Delta _1 \left(d^2+\left(\Delta _3-2\right) \Delta _3\right)-\left(\Delta _3-2\right) \left(d-\Delta _3-2\right) \left(d+\Delta _3\right)\non\\
&&+\Delta _1^3-\left(\Delta _3-2\right) \Delta _1^2\biggl] a_1^{(1,1)}-\frac{\left(\Delta _1-1\right) \left(\Delta _1+\Delta _3-2\right) \left(d+\Delta _3\right) a_2^{(2,0)}}{4 \Delta _3}\non\\
&&+\frac{\Delta _1+\Delta _3-2}{8 \left(\Delta _1-\Delta _3\right) \Delta _3 \left(d+3 \Delta _1+3 \Delta _3+4\right)}\biggl[ \Delta _1^3 \left((d-8) d-2 \Delta _3^2\right)+\Delta _1^5\non\\
&&+\Delta _1 \left(2 d \left(d^2+d-2\right) \Delta _3+2 d^2 (d+2)+d (4-3 d) \Delta _3^2+\Delta _3^4\right)+\Delta _1^4 \left(d+\Delta _3+2\right)\non\\
&&+\left(d+\Delta _3\right) \left(\Delta _3 \left(\Delta _3 \left(d^2+\Delta _3 \left(-2 d+\Delta _3+6\right)+4\right)+2 d (d+2)\right)-8 (d-2) d\right)\non\\
&& +\Delta _1^2 \left(d \left(d^2-12\right)-\Delta _3 \left(d (d+10)+2 \Delta _3 \left(\Delta _3+4\right)+4\right)\right)\biggl]a_2^{(2,1)}\non\\
&&+\frac{\Delta _1+\Delta _3-2}{8 \Delta _1 \left(\Delta _1-\Delta _3\right) \left(d+3 \Delta _1+3 \Delta _3+4\right)}\biggl[\Delta _3^2 \left(d^3+2 \Delta _1 \left(2 \Delta _1 \left(d+\Delta _1+2\right)-(d-2) d+4\right)\right)\non\\
&&+\Delta _3 \left(\Delta _1 \left(2 d \left(d^2+d-2\right)+\Delta _1 \left(\Delta _1 \left(-2 d+\Delta _1+4\right)-d (d+2)+4\right)\right)+2 d ((d-2) d+8)\right)\non\\
&&-8 (d-2) d^2-\Delta _3^4 \left(d+2 \Delta _1\right)-\Delta _3^3 \left(2 \Delta _1 \left(-d+\Delta _1+2\right)+d (d+2)+4\right)\non\\
&&-\Delta _1 \left(\Delta _1+2\right) \left(d+\Delta _1\right) \left(d+\Delta _1+2\right) \left(2 \Delta _1-d\right)+\Delta _3^5\biggl]a_2^{(1,2)}\non
\ee

\be
b_{1;1}^{(2,0)}&=&-\frac{\left(\Delta _1+\Delta _3-2\right) \left(-d+\Delta _1+\Delta _3+2\right) a_1^{(1,1)}}{\Delta _1}+\frac{1}{2} \left(\Delta _1-1\right) a_2^{(2,0)}\non\\
&&-\f{1}{4 \left(\Delta _1-\Delta _3\right) \left(d+3 \Delta _1+3 \Delta _3+4\right)}\biggl[\left(\Delta _1+\Delta _3\right) \Bigl(\Delta _3 \left(d^2+\Delta _3 \left(-2 d+\Delta _3+6\right)+4\right)\non\\
&&+\Delta _1 \left(-\Delta _3 \left(6 d+3 \Delta _3+4\right)+(d-4) d-4\right)+2 d (d+2)+3 \Delta _1^3-\left(\Delta _3+2\right) \Delta _1^2\Bigl)-8 (d-2) d\biggl]a_2^{(2,1)}\non\\
&&-\frac{\Delta _3 \left(\left(\Delta _1+\Delta _3\right) \left(\Delta _1+\Delta _3+2\right) \left(d-3 \Delta _1-\Delta _3+2\right) \left(d+\Delta _1-\Delta _3\right)-8 (d-2) d\right) a_2^{(1,2)}}{4 \Delta _1 \left(\Delta _1-\Delta _3\right) \left(d+3 \Delta _1+3 \Delta _3+4\right)}\non
\ee

\be
b_{1;1}^{(0,0)}&=&\frac{\left(\Delta _1+\Delta _3-4\right) \left(\Delta _1+\Delta _3-2\right)}{24 \Delta _1 \left(\Delta _3-1\right) \Delta _3}\biggl[\Delta _1^3 \left(-3 d-2 \Delta _3+20\right)-\Delta _1^2 \left((d+8) \Delta _3+(d-4) (3 d+2)\right)\non\\
&&+\Delta _1 \left(-(d+16) \Delta _3^2+4 (5 d-3) \Delta _3+d (3 (d-8) d+44)+2 \Delta _3^3-40\right)\non\\
&&+3 \left(d-\Delta _3-2\right) \left(\Delta _3 \left(\Delta _3 \left(2 d+\Delta _3-6\right)+(d-4) d\right)-2 (d-4) d\right)+3 \Delta _1^4\biggl]a_1^{(1,1)}\non\\
&&+\frac{\left(\Delta _1+\Delta _3-4\right) \left(\Delta _1+\Delta _3-2\right)}{96 \left(\Delta _1-\Delta _3\right) \left(\Delta _3-1\right) \Delta _3 \left(d+3 \Delta _1+3 \Delta _3+4\right)}\biggl[-\Delta _1^4 \left(6 d^2+\Delta _3 \left(16 d+21 \Delta _3+6\right)-40 d+60\right)\non\\
&&+3 \Delta _3 \Bigl(\Delta _3 \Bigl(\Delta _3 \left(4 \left(d^2+d-8\right)-\Delta _3 \left(-2 d (d+2)+\Delta _3 \left(\Delta _3+2\right)+20\right)\right)\non\\
&&-d (d (d (d+4)-36)+32)\Bigl)-2 (d-4) d^2 (d+2)\Bigl)+24 (d-4) (d-2) d^2\non\\
&&-4 \Delta _1^3 \left(\Delta _3 \left(\Delta _3 \left(d+\Delta _3+19\right)+d (2 d-5)+24\right)-8 (d-5) d-2\right)\non\\
&&+\Delta _1^2 \Bigl(\Delta _3 \left(\Delta _3 \left(4 (4 d+3) \Delta _3+12 (d-5) d+15 \Delta _3^2+120\right)+12 (d ((d-3) d-9)+8)\right)\non\\
&&+d (8-3 (d-8) (d-4) d)\Bigl)+2 \Delta _1 \Bigl(\Delta _3 \Bigl(\Delta _3 \Bigl(\Delta _3 \left(\Delta _3 \left(2 d+\Delta _3+21\right)+2 d (5 d-7)+48\right)\non\\
&&+2 (d-2) d (3 d-8)-4\Bigl)+d (d (-3 (d-4) d-50)+68)\Bigl)+d (d (-3 (d-10) d-64)+80)\Bigl)\non\\
&&+9 \Delta _1^6+2 \left(\Delta _3+17\right) \Delta _1^5\biggl]a_2^{(2,1)}\non\\
&&-\frac{\left(\Delta _1+\Delta _3-4\right) \left(\Delta _1+\Delta _3-2\right)}{96 \Delta _1 \left(\Delta _1-\Delta _3\right) \left(\Delta _3-1\right) \left(d+3 \Delta _1+3 \Delta _3+4\right)}\biggl[-2 \Delta _1^3 \Bigl(3 d^3+2 \Delta _3 \Bigl(\Delta _3 \left(2 d-\Delta _3+17\right)\non\\
&&+(d-1) (3 d-4)\Bigl)-104 d+68\Bigl)+\Delta _1^4 \left(4 (d-3) \Delta _3+2 d (49-6 d)-15 \Delta _3^2+60\right)\non\\
&&+\Delta _1^2 \Bigl(\Delta _3 \left(\Delta _3 \left(-4 (d-6) \Delta _3+6 (d-18) d+3 \Delta _3^2-24\right)+6 d (d (7-2 d)+2)\right)\non\\
&&+d (3 d ((d-10) d+32)-104)\Bigl)+\Delta _1^5 \left(6 d-2 \Delta _3+62\right)+2 \Delta _1 \Bigl(\Delta _3 \bigl(\Delta _3 \bigl(\Delta _3 \bigl(\Delta _3 \left(d-\Delta _3+3\right)\non\\
&&-10 d+8\bigl)+d (-3 (d-4) d-56)+68\bigl)+d (d (3 (d-4) d+50)-68)\bigl)+d (d (3 (d-10) d+64)-80)\Bigl)\non\\
&&+3 \left(d-\Delta _3-2\right) \left(\Delta _3 \left(d-\Delta _3\right)+4 d\right) \left(\Delta _3 \left(\Delta _3 \left(2 d+\Delta _3-6\right)+(d-4) d\right)-2 (d-4) d\right)+9 \Delta _1^6\biggl]a_2^{(1,2)}\non\\
&&-\frac{\left(\Delta _1-1\right) \left(\Delta _1+\Delta _3-4\right) \left(\Delta _1+\Delta _3-2\right)}{{48 \left(\Delta _3-1\right) \Delta _3}} \biggl[-3 \left(\Delta _3 \left(2 d+\Delta _3-4\right)+(d-8) d+8\right)+3 \Delta _1^2\non\\
&&+2 \left(\Delta _3+5\right) \Delta _1\biggl] a_2^{(2,0)}\non
\ee

\be
b_{2;0}^{(0,1)}&=&\frac{\left(-d \left(\Delta _1+1\right) \Delta _3-\Delta _1 \left(\Delta _1 \left(-d+\Delta _1+5\right)+5 d-8\right)+2 (d-2)+\left(\Delta _1+1\right) \Delta _3^2\right) a_1^{(1,1)}}{2 \Delta _1}\non\\
&&\frac{1}{4 \left(\Delta _1-\Delta _3\right) \left(d+3 \Delta _1+3 \Delta _3+4\right)}\biggl[-(d+1) \Delta _1^4+2 \Delta _1^3 \left(2 d+\Delta _3^2+\Delta _3-3\right)\non\\
&&+\Delta _1^2 \left(\Delta _3 \left((d+2) \Delta _3+d (d+2)+2\right)+(d+1) (d+2)\right)-4 (d-2) d-\Delta _1^5\non\\
&&+\Delta _1 \left(\Delta _3 \left(\Delta _3 \left((d-2) d-\Delta _3 \left(\Delta _3+2\right)+6\right)+2 d (3 d-2)\right)+d (9 d-14)\right)\non\\
&&+\Delta _3 \left(d-\Delta _3\right) \left(\Delta _3 \left(d+\Delta _3+2\right)+d+2\right)\biggl]a_2^{(2,1)}\;+\;\;\frac{1}{4} \left(\Delta _1-1\right){}^2 a_2^{(2,0)} \non\\
&&+\frac{\Delta _3}{8 \Delta _1 \left(\Delta _1-\Delta _3\right) \left(d+3 \Delta _1+3 \Delta _3+4\right)}\biggl[(2 d+5) \Delta _1^4+\Delta _1^3 \left(2 \Delta _3 \left(d-\Delta _3\right)+d (d+16)-8\right)\non\\
&&+\Delta _1^2 \left(2 (d+3) \Delta _3 \left(d-\Delta _3\right)+d (3 d+10)+4\right)+\left(\Delta _3-2\right) \left(d-\Delta _3-2\right) \left(\Delta _3 \left(d-\Delta _3\right)+4 d\right)\non\\
&&+\Delta _1 \left(\Delta _3 \left(d-\Delta _3\right) \left(\Delta _3 \left(d-\Delta _3\right)+12 d-8\right)+2 d (9 d-14)\right)+\Delta _1^5\biggl]a_2^{(1,2)}\non
\ee

\be
b_{2;0}^{(0,0)}&=& \frac{\left(\Delta _1+\Delta _3-2\right)}{4 \Delta _1} \biggl[\Delta _1^2 \left(-2 d+\Delta _3+8\right)+\Delta _1 \left(d (d+2)-\left(\Delta _3-4\right) \Delta _3-4\right)\non\\
&&-\left(\Delta _3+2\right) \left(d-\Delta _3-2\right) \left(d-\Delta _3\right)+\Delta _1^3\biggl] a_1^{(1,1)}\non\\
&&+\frac{\Delta _1+\Delta _3-2}{8 \left(\Delta _1-\Delta _3\right) \left(d+3 \Delta _1+3 \Delta _3+4\right)}\biggl[-\Delta _3^3 \left(d^2+\Delta _1 \left(d+2 \Delta _1-2\right)+d-6\right)+4 (d-2) d^2\non\\
&&+\Delta _3^2 \left(d ((d-8) d+4)-\Delta _1 \left(d^2+\Delta _1 \left(-d+2 \Delta _1+10\right)+d+10\right)\right)+\Delta _3^4 \left(-d+\Delta _1+6\right)\non\\
&&+\Delta _3 \left(\Delta _1 \left(\Delta _1 \left(\Delta _1 \left(d+\Delta _1-2\right)-d (d+3)-6\right)+d ((d-8) d-4)\right)+(d-4) d (3 d-2)\right)\non\\
&&+\Delta _1 \left(\Delta _1 \left(\Delta _1 \left(\Delta _1 \left(\Delta _1+4\right)-(d-2) (d+5)\right)-8 d\right)-d (d (d+6)-8)\right)+\Delta _3^5\biggl]a_2^{(2,1)}\non\\
&&-\frac{\Delta _3 \left(\Delta _1+\Delta _3-2\right)}{16 \Delta _1 \left(\Delta _1-\Delta _3\right) \left(d+3 \Delta _1+3 \Delta _3+4\right)}\biggl[-\Delta _1^3 \left(2 \Delta _3 \left(\Delta _3-2 (d+1)\right)+(d-22) d+4\right)\non\\
&&\Delta _1^2 \left(\Delta _3 \left(2 \Delta _3 \left(d-\Delta _3-6\right)+d (d+22)-4\right)-d \left(d^2-32\right)\right)+\Delta _1^4 \left(d+\Delta _3+8\right)\non\\
&&+\Delta _1 \Bigl(\Delta _3 \left(\Delta _3 \left(-4 (d+1) \Delta _3+d (5 d-14)+\Delta _3^2+4\right)-2 d ((d-8) d-4)\right)\non\\
&&+2 d (d (d+6)-8)\Bigl)-\left(\Delta _3+2\right) \left(d-\Delta _3-2\right) \left(d-\Delta _3\right) \left(\Delta _3 \left(d-\Delta _3\right)+4 d\right)+\Delta _1^5\biggl]a_2^{(1,2)}\non\\
&&-\frac{1}{8} \left(\Delta _1-1\right) \left(\Delta _1+\Delta _3-2\right) \left(-d+\Delta _1+\Delta _3\right) a_2^{(2,0)}\non
\ee

\be
b_{2;1}^{(1,2)}&=&\frac{\Delta _3 \left(3 d \Delta _1+\Delta _3 \left(d-\Delta _3\right)+4 d+\Delta _1^2\right) a_2^{(1,2)}+2 \Delta _1 \left(\Delta _1+\Delta _3+2\right) \left(d-\Delta _1+\Delta _3\right) a_2^{(2,1)}}{2 \left(\Delta _1-\Delta _3\right) \left(d+3 \Delta _1+3 \Delta _3+4\right)}\non\\
&&-2a_1^{(1,1)}\non
\ee

\be
b_{2;1}^{(1,1)}&=&-\frac{\left(d \Delta _3+\Delta _1 \left(3 d+\Delta _1-4\right)-2 d-\Delta _3^2+4\right) a_1^{(1,1)}}{2 \Delta _1 \left(\Delta _1+\Delta _3\right)} -\frac{\left(\Delta _1-1\right) a_2^{(2,0)}}{4 \left(\Delta _1+\Delta _3\right)} \non\\
&&-\f{1}{4 \left(\Delta _1^2-\Delta _3^2\right) \left(d+3 \Delta _1+3 \Delta _3+4\right)}\biggl[\Delta _1^2 \left(-2 d^2+d-2 \Delta _3 \left(\Delta _3+4\right)-10\right)\non\\
&&+\Delta _3 \left(\Delta _3 \left(-d^2+d+\Delta _3 \left(\Delta _3+8\right)+10\right)-d (d+2)\right)-(d+2) \Delta _1^3\non\\
&&+\Delta _1 \left((d+2) \Delta _3^2+d (2-3 d) \Delta _3+d (6-5 d)\right)+4 (d-2) d+\Delta _1^4\biggl]a_2^{(2,1)}\non\\
&&+\f{\Delta_3}{8 \Delta _1 \left(\Delta _1^2-\Delta _3^2\right) \left(d+3 \Delta _1+3 \Delta _3+4\right)}\biggl[\Delta _1^2 \left(2 \Delta _3 \left(d-\Delta _3\right)+d (5 d-2)+4\right)\non\\
&&+(6 d-4) \Delta _1^3+2 \Delta _1 \left((3 d-2) \Delta _3 \left(d-\Delta _3\right)+d (5 d-6)\right)+\Delta _1^4\non\\
&&+\left(\Delta _3-2\right) \left(d-\Delta _3-2\right) \left(\Delta _3 \left(d-\Delta _3\right)+4 d\right)\biggl]a_2^{(1,2)}\non
\ee

\be
b_{2;1}^{(1,0)}&=&\frac{\left(\Delta _1+\Delta _3-2\right) \left(2 d+\Delta _1-\Delta _3-2\right) \left(-d+\Delta _1+\Delta _3+2\right) a_1^{(1,1)}}{4 \Delta _1}\non\\
&&+\f{\Delta _1+\Delta _3-2}{8 \left(\Delta _1-\Delta _3\right) \left(d+3 \Delta _1+3 \Delta _3+4\right)}\biggl[-\Delta _1^2 \left(\Delta _3 \left(d+2 \Delta _3+6\right)+5 d+6\right)\non\\
&&+\Delta _1 \left(d (2-3 d) \Delta _3+d ((d-5) d+2)+4 \Delta _3^2\right)+4 (d-2) (d-1) d+\Delta _1^4-4 \Delta _1^3\non\\
&&+\Delta _3 \left(\Delta _3 \left(\Delta _3 \left(d+\Delta _3+6\right)+d (7-3 d)+6\right)+d ((d-9) d+10)\right)\biggl]a_2^{(2,1)}\non\\
&&-\f{\Delta _3 \left(\Delta _1+\Delta _3-2\right)}{16 \Delta _1 \left(\Delta _1-\Delta _3\right) \left(d+3 \Delta _1+3 \Delta _3+4\right)}\biggl[\Delta _1^2 \left(4 (d-1) \Delta _3+d (d+10)-2 \Delta _3^2-4\right)\non\\
&&-2 d \Delta _1 \left(\Delta _3 \left(-3 d+2 \Delta _3+2\right)+(d-5) d+2\right)+4 d \Delta _1^3+\Delta _1^4\non\\
&&-\left(d-\Delta _3-2\right) \left(2 d-\Delta _3-2\right) \left(\Delta _3 \left(d-\Delta _3\right)+4 d\right)\biggl]a_2^{(1,2)}\non\\
&&-\frac{1}{8} \left(\Delta _1+\Delta _3-2\right) \left(d-\Delta _3-1\right) a_2^{(2,0)}\non
\ee

\be
b_{2;1}^{(0,1)}&=&\frac{\left(\Delta _1+\Delta _3-2\right)}{2 \Delta _1 \Delta _3} \biggl[\Delta _3^2 \left(d+2 \Delta _1-2\right)-2 \left(\Delta _1-1\right) \left(-d+\Delta _1+2\right) \left(d+\Delta _1\right)\non\\
&&+\Delta _3 \left(-d \Delta _1+2 d+\Delta _1^2\right)-\Delta _3^3\biggl] a_1^{(1,1)}+\frac{\left(\Delta _1-1\right) \left(\Delta _1+\Delta _3-2\right) \left(d+\Delta _1\right) a_2^{(2,0)}}{4 \Delta _3}\non\\
&&-\frac{\Delta _1+\Delta _3-2}{4 \left(\Delta _1-\Delta _3\right) \Delta _3 \left(d+3 \Delta _1+3 \Delta _3+4\right)}\biggl[\Delta _3^3 \left(d^2+2 \Delta _1 \left(\Delta _1+3\right)-4\right)\non\\
&&-\Delta _3 \left(\Delta _1 \left(\Delta _1 \left(2 \left(d^2+d-2\right)+\Delta _1 \left(\Delta _1+6\right)\right)+d \left(-d^2+d-2\right)\right)+(d-6) d^2\right)\non\\
&&-\Delta _3^2 \left(\Delta _1 \left(\Delta _1 \left(d+4 \Delta _1-4\right)+(d-3) d-2\right)+d (2-5 d)\right)+2 \left(\Delta _1-2\right) \Delta _3^4\non\\
&&+\left(\Delta _1-1\right) \left(d+\Delta _1\right) \left(\Delta _1 \left(\Delta _1 \left(-d+2 \Delta _1+2\right)+(d-6) d\right)+4 (d-2) d\right)-\Delta _3^5\biggl]a_2^{(2,1)}\non\\
&&-\frac{\Delta _1+\Delta _3-2}{8 \Delta _1 \left(\Delta _1-\Delta _3\right) \left(d+3 \Delta _1+3 \Delta _3+4\right)}\biggl[\Delta _3^3 \left((d-8) d-2 \Delta _1^2\right)+\Delta _3^5\non\\
&&\Delta _3 \left(-2 (d-6) d^2+d (4-3 d) \Delta _1^2+2 d ((d-1) d+2) \Delta _1+\Delta _1^4\right)\non\\
&&+2 \Delta _3^2 \left(\Delta _1 \left(2 \Delta _1 \left(2 d+\Delta _1\right)+3 d-2\right)+2 d (2 d-3)\right)-2 \Delta _3^4 \left(d+\Delta _1-1\right)\non\\
&&-2 \left(\Delta _1-1\right) \left(d+\Delta _1\right) \left(\Delta _1 \left(\Delta _1 \left(2 d+\Delta _1+2\right)-(d-6) d\right)-4 (d-2) d\right)\biggl]a_2^{(1,2)}\non
\ee

\be
b_{2;1}^{(0,2)}&=&-\frac{2 \left(\Delta _1-1\right) \left(-d+\Delta _1+\Delta _3+2\right) a_1^{(1,1)}}{\Delta _3} +\frac{\left(\Delta _1-1\right) \Delta _1 a_2^{(2,0)}}{2 \Delta _3}\non\\
&&-\frac{\left(\Delta _1-1\right) \Delta _1}{2 \left(\Delta _1-\Delta _3\right) \Delta _3 \left(d+3 \Delta _1+3 \Delta _3+4\right)} \biggl[4 (d-2) d\non\\
&&+\left(\Delta _1+\Delta _3\right) \left(-(d-2) \Delta _1-\Delta _3 \left(3 d+2 \Delta _3+2\right)+(d-6) d+2 \Delta _1^2\right)\biggl] a_2^{(2,1)}\non\\
&&+\f{\Delta _1-1}{2 \left(\Delta _1-\Delta _3\right) \left(d+3 \Delta _1+3 \Delta _3+4\right)}\biggl[-4 (d-2) d\non\\
&&\left(\Delta _1+\Delta _3\right) \left(2 (d+1) \Delta _1+\Delta _3 \left(2 d-\Delta _3-2\right)-(d-6) d+\Delta _1^2\right)\biggl]a_2^{(1,2)}\non
\ee

\be
b_{2;1}^{(0,0)}&=&-\frac{\left(\Delta _1+\Delta _3-4\right) \left(\Delta _1+\Delta _3-2\right)}{12 \Delta _1 \Delta _3} \biggl[\Delta _3 \left(\Delta _1 \left(3 d-2 \Delta _1-4\right)+d (3 d-14)+4\right)\non\\
&&+\Delta _3^2 \left(-2 d-3 \Delta _1+6\right)-3 \left(d-\Delta _1-2\right) \left(d+\Delta _1-4\right) \left(d+\Delta _1\right)+2 \Delta _3^3\biggl]a_1^{(1,1)}\non\\
&&-\frac{\left(\Delta _1+\Delta _3-4\right) \left(\Delta _1+\Delta _3-2\right)}{48 \left(\Delta _1-\Delta _3\right) \Delta _3 \left(d+3 \Delta _1+3 \Delta _3+4\right)}\biggl[\Delta _3^4 \left(9 d+6 \Delta _1-26\right)-4 \Delta _3^5\non\\
&&\Delta _3 \Bigl(\Delta _1 \left(2 d \left(-3 d^2+d+30\right)-\Delta _1 \left(2 \Delta _1 \left(d+2 \Delta _1+14\right)+d (7 d+16)-28\right)\right)\non\\
&&+d (d (3 (d-10) d+80)-16)\Bigl)+\Delta _3^3 \left(2 \Delta _1 \left(d+4 \Delta _1+14\right)+d (d+20)-28\right)\non\\
&&-\Delta _3^2 \left(\Delta _1 \left(2 \Delta _1 \left(9 d+6 \Delta _1-22\right)+d (9 d-34)\right)+d (d (9 d-40)+4)\right)\non\\
&&+3 \left(d+\Delta _1\right) \left(\Delta _1 \left(\Delta _1 \left(\Delta _1 \left(d+2 \Delta _1-6\right)-4 d\right)+d ((d-6) d+16)\right)+4 (d-4) (d-2) d\right)\biggl] a_2^{(2,1)}\non\\
&&+\frac{\left(\Delta _1+\Delta _3-4\right) \left(\Delta _1+\Delta _3-2\right)}{48 \Delta _1 \left(\Delta _1-\Delta _3\right) \left(d+3 \Delta _1+3 \Delta _3+4\right)}\biggl[2 \Delta _1^3 \left(3 \left(2 d^2+d-8\right)-\Delta _3 \left(3 \Delta _3+2\right)\right)\non\\
&&+\Delta _1^2 \left(\Delta _3 \left(4 \Delta _3 \left(-4 d+\Delta _3+3\right)+d (11 d-32)+4\right)+6 d (5 d-16)\right)\non\\
&&+\Delta _1 \Bigl(\Delta _3 \left(\Delta _3 \left(-2 d (3 d+5)+\Delta _3 \left(3 \Delta _3+4\right)+48\right)+2 d (d+3) (3 d-10)\right)\non\\
&&-3 d ((d-4) d (d+2)+32)\Bigl)+2 \Delta _1^4 \left(6 d-\Delta _3-3\right)+3 \Delta _1^5\non\\
&&-\left(d-\Delta _3-2\right) \left(\Delta _3 \left(d-\Delta _3\right)+4 d\right) \left(3 (d-4) d+2 \Delta _3 \left(\Delta _3+1\right)\right)\biggl] a_2^{(1,2)}\non\\
&&+\frac{\left(\Delta _1+\Delta _3-4\right) \left(\Delta _1+\Delta _3-2\right)}{48 \Delta _3} \biggl[3 \left(\Delta _1 \left(2 d+\Delta _1-4\right)+(d-8) d+8\right)-3 \Delta _3^2\non\\
&&-2 \left(\Delta _1+5\right) \Delta _3\biggl] a_2^{(2,0)}\non
\ee

\end{document}